\documentclass[aps,prb,twocolumn,amsmath,amssymb,superscriptaddress]{revtex4-1}

\usepackage{amsfonts}
\usepackage{amsbsy}
\usepackage{amsmath}
\usepackage{bm}
\usepackage{braket}
\usepackage{color}
\usepackage{graphicx}
\usepackage[colorlinks,urlcolor=magenta,linkcolor=magenta,anchorcolor=blue,citecolor=magenta]{hyperref}
\usepackage{subfigure}
\newcommand{\ttau}{\tilde\tau}
\newcommand{\tphi}{\tilde\phi}
\newcommand{\tPhi}{\tilde\Phi}
\newcommand{\tA}{\tilde A}
\newcommand{\tB}{\tilde B}
\newcommand{\ba}{\bar{a}}
\newcommand{\bb}{\bar{b}}

\begin{document}
\title{Phases and  Quantum Phase Transitions in an Anisotropic Ferromagnetic Kitaev-Heisenberg-$\ \Gamma$ Magnet}
\author{Animesh Nanda}
\email{animesh.nanda@icts.res.in}
\affiliation{International Centre for Theoretical Sciences, Tata Institute of Fundamental Research, Bangalore 560012, India}
\author{Kusum Dhochak}
\email{kdhochak@iitpkd.ac.in}
\affiliation{Department of Physics, Indian Institute of Technology, Palakkad 678557, India}
\author{Subhro Bhattacharjee}
\email{subhro@icts.res.in}
\affiliation{International Centre for Theoretical Sciences, Tata Institute of Fundamental Research, Bangalore 560012, India}
\date{\today}
\begin{abstract}
We study the spin-$1/2$ ferromagnetic Heisenberg-Kitaev-$\Gamma$ model in the anisotropic (Toric code) limit to reveal the nature of the quantum phase transition between the gapped $Z_2$ quantum spin liquid and a spin ordered phase (driven by Heisenberg interactions) as well as a trivial paramagnet (driven by pseudo-dipolar interactions, $\Gamma$). The transitions are obtained by a simultaneous condensation of the Ising electric and magnetic charges-- the fractionalized excitations of the $Z_2$ quantum spin liquid. Both these transitions can be continuous and are examples of deconfined quantum critical points. Crucial to our calculations are the symmetry implementations on the soft electric and magnetic modes that become critical. In particular, we find strong constraints on the structure of the critical theory arising from time reversal and lattice translation symmetries with the latter acting as an anyon permutation symmetry that endows the critical theory with a manifestly self-dual structure. We find that the transition between the quantum spin liquid and the spin-ordered phase belongs to a self-dual modified Abelian Higgs field theory while that between the spin liquid and the trivial paramagnet belongs to a self-dual $Z_2$ gauge theory. We also study the effect of an external  Zeeman field to show an interesting similarity between the polarised paramagnet obtained due to the Zeeman field and the trivial paramagnet driven the pseudo-dipolar interactions.  Interestingly, both the spin liquid and the spin ordered phases have easily identifiable counterparts in the isotropic limit and the present calculations may shed insights into the corresponding transitions in the material relevant isotropic limit.
\end{abstract}
\maketitle

\section{Introduction}
Recent research of spin-orbit coupled frustrated magnets have led to the discovery of a new class of candidate quantum spin liquid (QSL) materials.\cite{witczak2014correlated} Interestingly, in a subset of such magnets which ultimately order (at very low temperatures), the low temperature properties bear unconventional experimental signatures akin to  fractionalized excitations\cite{anderson1973resonating,anderson1987resonating,wen2002quantum,lee2006doping,PhysRevLett.86.1881,balents2010spin,lee2008end,savary2016quantum,wen2017colloquium,broholm2019quantum} expected in a QSL.   A framework to describe these properties start by positing that these systems are proximate to quantum phase transition between a spin ordered phase and a QSL, albeit just on the ordered side. The finite temperature properties of such a {\it proximate QSL} phase then may account for, among others, the neutron scattering of honeycomb lattice magnet $\alpha$-RuCl$_3$\cite{PhysRevB.93.134423,Banerjee2016,Banerjee2017,trebst2017kitaev,PhysRevB.97.134432} and rare-earth pyrochlore Yb$_2$Ti$_2$O$_7$.\cite{PhysRevLett.103.227202,PhysRevX.1.021002,PhysRevB.93.064406,PhysRevLett.119.057203,PhysRevLett.119.127201}

The case of $\alpha$-RuCl$_3$ is particularly interesting where a collinear Zig-Zag spin order is stabilized below $T\sim7$ K.\cite{PhysRevB.93.134423,Banerjee2016,Banerjee2017,trebst2017kitaev,PhysRevB.97.134432} However, recent neutron scattering experiments reveal that unusually intense diffused spin excitations resembling that of the two-particle fractionalized spinon continuum of a QSL survive well above the spin ordering temperature.\cite{PhysRevB.93.134423,Banerjee2016,Banerjee2017} Further, in an in-plane Zeeman field, the spin order gives away to a field induced partially-polarised paramagnet\cite{PhysRevLett.120.117204,banerjee2018excitations} with unusual spin dynamics\cite{PhysRevB.96.241107,PhysRevLett.119.227201} and quantized thermal-Hall conductivity.\cite{kasahara2018majorana} This has led to the suggestion the the zero Zeeman field Zig-Zag order in this material occurring below $7$ K\cite{PhysRevB.93.134423} is fragile and proximate to a $Z_2$ QSL with ultra short-ranged spin correlations\cite{PhysRevLett.98.247201}-- which supports fractionalized Majorana excitations and $Z_2$ fluxes.\cite{kitaev2006anyons}

Within the proximate-spin liquid scenario, therefore, the quantum phase transition between the Zig-Zag spin ordered phase and the $Z_2$ QSL then affects the low temperature physics of $\alpha-$RuCl$_3$. On generic grounds, such transitions\cite{senthil2006quantum} cannot be captured within the conventional order parameter based description.\cite{chaikin1995principles}   Further,  the $Z_2$ QSL is separated from a trivial paramagnet (one without topological order and fractionalised excitations) through a different and distinct quantum phase transition. In case of this latter transition an order parameter based description is unavailable.  If the transitions are continuous-- as is pertinent to the present work-- the correct critical theory has to essentially account for the fractionalisation  and topological order\cite{PhysRevB.78.155134,freedman2004class} in the $Z_2$ QSL in addition to the any possible spin order. Several examples of such {\it deconfined critical points}\cite{senthil2004deconfined,senthil2004quantum} are known.

\begin{figure}
\begin{center}
\subfigure[]{
\includegraphics[scale=.355]{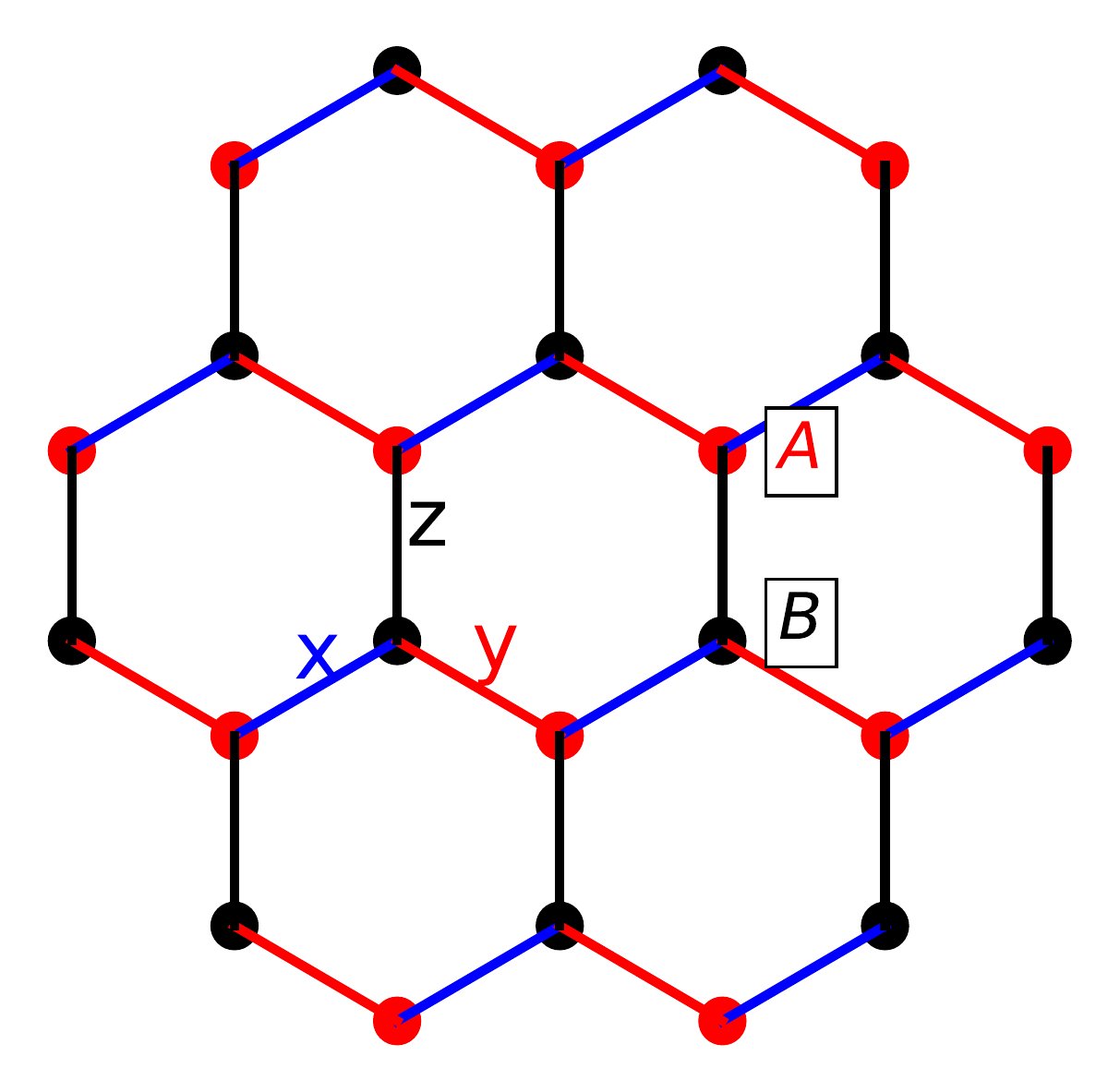}
\label{fig_kitaev}
}
\subfigure[]{
\includegraphics[scale=0.355]{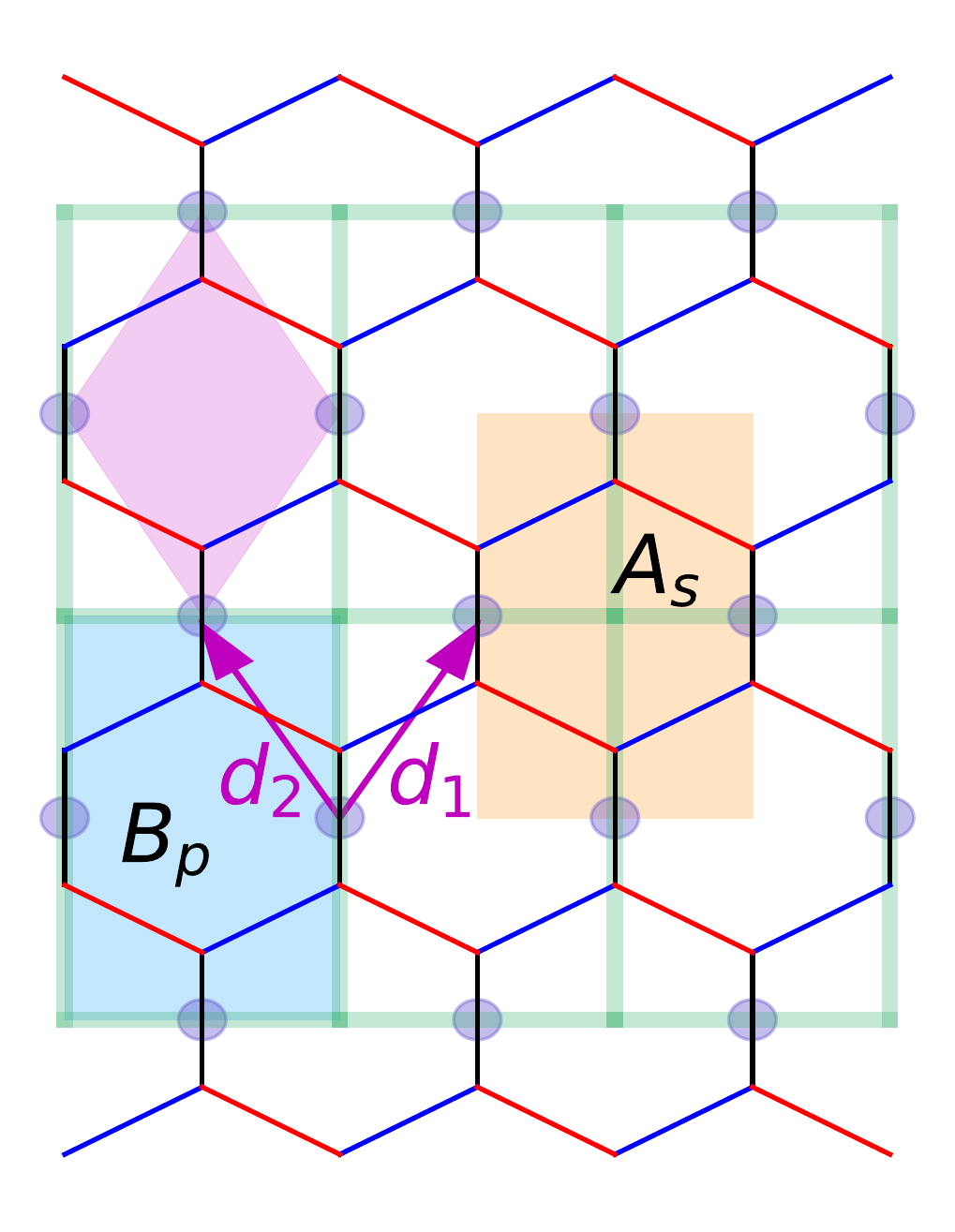}
\label{fig_toric}
}
\end{center}
\caption{{\bf (a)} The $\sigma$ spins on the honeycomb lattice (with sub-lattice $A$ and $B$). The $x,y,z$ bonds are shown in blue, red and black respectively. {\bf (b)} The low energy doublet in the anisotropic limit, the $\tau$ spins (see text), are shown in gray filled circles on the links of the light green square lattice which are also the location of the centres of the $z$-bonds of the honeycomb lattice. $\bf d_1$ and $\bf d_2$ generate translations of underlying honeycomb lattice. The plaquettes (light magenta) in the rhombic lattice become usual plaquettes (cyan) and stars (light orange) in the square lattice.}
\end{figure}

The minimal spin Hamiltonian that can capture the above physics of $\alpha$-RuCl$_3$ is given by the so-called Heisenberg-Kitaev-Pseudodipolar ($JK\Gamma$) Hamiltonian\cite{PhysRevLett.102.017205,PhysRevLett.105.027204,PhysRevLett.112.077204,Hermanns2018}
\begin{align}
H=&J\sum_{\langle p,q\rangle}\boldsymbol{\sigma_p}\cdot\boldsymbol{\sigma_{q}}+\sum_{\langle p,q\rangle \alpha}\left[\Gamma_{\alpha}\left[\sigma_{p}^{\beta}\sigma_{q}^{\gamma}+\sigma_{q}^{\beta}\sigma_{p}^{\gamma}\right]-K_{\alpha}\sigma_{p}^{\alpha}\sigma_{q}^{\alpha}\right]
\label{eq_hamiltonian}
\end{align}
where $\ \alpha=x,y,z $ refers to the $x,y,z$ bonds of the honeycomb lattice respectively (see Fig. \ref{fig_kitaev}) and $\sigma^{\alpha}_p$ denotes Pauli matrices representing spin-$1/2$ operator on the site of the honeycomb lattice. $\langle pq\rangle$ refers to nearest neighbours while $\langle pq\rangle \alpha$ refers to nearest neighbours along $\alpha$-bonds. Note that for a given $\alpha$; $\beta,\gamma(=x,y,z)\neq\alpha$. Remarkably, in addition to $\alpha$-RuCl$_3$, the Hamiltonian in Eq. \ref{eq_hamiltonian} can effectively describe the magnetic properties of several other strong spin-orbit coupled magnets on honeycomb lattice\cite{PhysRevLett.102.017205,PhysRevLett.105.027204,PhysRevLett.112.077204} that include honeycomb iridates \cite{PhysRevLett.105.027204,PhysRevB.82.064412,PhysRevB.83.220403,PhysRevLett.108.127203,PhysRevLett.108.127204,PhysRevB.85.180403} as well as three-dimensional harmonic iridates.\cite{PhysRevB.90.205116,PhysRevLett.113.197201,trebst2017kitaev,nussinov2013compass,PhysRevB.89.014414,lee2014heisenberg}  The material relevant isotropic limit ($K_x =K_y =K_z $) has a rich phase diagram including a direct phase transition between the QSL and collinear spin ordered phases.\cite{PhysRevLett.112.077204,PhysRevLett.105.027204,PhysRevB.84.100406,PhysRevB.83.245104,PhysRevB.92.020405,PhysRevB.86.224417,PhysRevLett.112.077204,PhysRevLett.109.187201,PhysRevB.97.075126,lee2019magnetic} 

An interesting and somewhat easier (for the present purpose) limit of Eq. \ref{eq_hamiltonian} occurs when one of the Kitaev couplings, $K_z$ (say) is much larger than all other couplings, {\it i.e.}, $\ |K_z| >> |J|, |K_y|, |K_x|, |\Gamma|$. In this Toric code\cite{kitaev2003fault} limit the QSL survives for $J,\Gamma=0$, albeit as a gapped $Z_2$ QSL with low energy bosonic Ising electric ($e$) and magnetic ($m$) charges while the Majorana fermion has a large gap of the order $\sim |K_z|$.\cite{kitaev2003fault,kitaev2006anyons} On increasing $J$ and/or $\Gamma$ the QSL must give way to other phases. What are these other phases and what is the nature of such phase transition are then questions of interest by themselves since any such description  must incorporate the non-trivial topological order and fractionalized $e$ and $m$ excitations of the QSL. \cite{PhysRevLett.98.070602,PhysRevB.86.214414,PhysRevB.89.235122,PhysRevB.91.134419} Also the understanding of such phase transitions in the anisotropic limit may provide us useful insights to the nature of the phase transition in the isotropic limit and thereby shed light on the finite temperature properties of candidates such as RuCl$_3$ to ascertain the validity of the {\it promximate QSL} scenario.

In this paper, with the twin motivations above, we study the phases and phase transitions in the anisotropic limit of Eq. \ref{eq_hamiltonian} and show that not only the QSL, albeit gapped, survives in the above anisotropic limit, but so do the neighbouring spin ordered phases.   Ref. \onlinecite{PhysRevB.90.035113} and \onlinecite{PhysRevB.99.115104} considered an approach  similar to ours. The starting point of Ref. \onlinecite{PhysRevB.90.035113} is slightly different version (Heisenberg exchange was also taken to be anisotropic) of the anisotropic limit for the Hamiltonian in Eq. \ref{eq_hamiltonian} with $\Gamma=0$. There, in the classical limit, all the magnetic states survive the anisotropy, and more interestingly even for quantum $S=1/2$ systems, numerical results suggest that the transition between the (now) gapped QSL and the spin orders, same as in the isotropic limit, exist.  Ref. \onlinecite{PhysRevB.99.115104}, on the other hand, considered Eq. \ref{eq_hamiltonian} for $J=0$ and and derived the effective low energy Hamiltonian through strong coupling approaches. Some of their conclusions-- such as the transverse-field Ising model results at lowest order (Eq. \ref{eq_gammak_tfim})-- agree with our effective microscopic Hamiltonian. However these above works did not systematically analyse the theory of the phase transitions. 

In this work, we substantially extend the formulation of the anisotropic problem incorporating both Heisenberg ($J$) and pseudo-dipolar ($\Gamma$) interactions to the ferromagnetic Kitaev magnet ($K_\alpha>0$ in Eq. \ref{eq_hamiltonian}). We use a combination of strong coupling expansion, numerical diagonalisation and field theoretic calculations to study the quantum phase transitions between the QSL and various spin ordered as well as paramagnetic phases by explicitly deriving a candidate critical theory for the possible deconfined quantum critical points. Our finding are summarised in the phase diagram in Fig. \ref{fig_pd}. In a following work,\cite{animesh_afm} we shall discuss the physics of antiferromagnetic Kitaev model ($K_\alpha<0$) and its proximate phases. We shall find crucial difference between the two cases along with interesting similarities. 

One of our central findings is the non-trivial implementation of the microscopic symmetries on the low energy degree of freedom. While this is quite generic to strongly correlated systems, it is even more rich in spin-orbit coupled systems where lattice and spin symmetries become intertwined. In the present case non-trivial implementation of two particular symmetries-- time reversal, $\mathcal{T}$, and lattice translation, ${\bf T_{d_1}}, {\bf T_{d_2}}$ (see Fig. \ref{fig_toric}) severely constrains the structure of the low energy theory in a novel way. In case of time-reversal symmetry, we find that while the $\sigma_i^\alpha$ spins in Eq. \ref{eq_hamiltonian} form usual Kramers doublet under time reversal ($\mathcal{T}^2=-1$)-- as is relevant for the candidate materials, the effective low energy degrees of freedom in the anisotropic limit (Eq. \ref{eq_tau_def}) is a non-Kramers doublet ($\mathcal{T}^2=+1$). The translations, ${\bf T_{d_1}}$ and ${\bf T_{d_2}}$, on the other hand, interchanges the sites and plaquettes of the underlying square lattice. This results in unconventional symmetry properties for the low energy $e$ and $m$ excitations of the QSL which transforms projectively under various symmetries.\cite{wen2002quantum,PhysRevB.87.104406} The non-Kramers nature of the low energy doublets determine-- (a) how the system couples to an external Zeeman field, and, (b) the nature of time-reversal partners $e$ and $m$ modes that become soft at the magnetic transition in such a {\it non-Kramers QSL}.\cite{PhysRevB.88.174405} The translations on the other hand, interchanges the $e$ and $m$ charges resulting in a so called {\it anyon permutation symmetry}.\cite{PhysRevB.90.115118,PhysRevB.87.104406} A profound consequence of this permutation is that the $e$ and $m$ soft modes of the QSL have the same mass resulting in placing the system along a self-dual line in the gauge-matter phase diagram of the type studied by Fradkin and Shenker in Ref. \onlinecite{PhysRevD.19.3682}. The above symmetry implementation on the soft modes then heavily constrains the nature of the critical point and hence the deconfined phase transition is protected by them. While this is certainly a feature of all symmetry breaking phase transitions, we find that in particular the anyon permutation symmetry protects the nature of the deconfined critical point between the QSL and the spin-ordered phase as well as the QSL and the trivial paramagnet. We place our results in context of the existing knowledge about this similar critical points. 

Considering the length and our multi-stage analysis of the problem, here we provide a summary of the central results obtained in this work along with the general outline of the rest of the paper.
 
\subsection{Outline and summary of the central results}
 
 In the anisotropic limit the low energy degrees of freedom for the ferromagnetic Kitaev model are given by non-Kramers doublets, $\tau^z_i$ (Eq. \ref{eq_tau_def}) sitting on the $z$-bonds of the honeycomb lattice or alternately the bonds of the square lattice as shown in Fig. \ref{fig_toric}. Due to the non-Kramers nature, only $\tau^z$ is odd under time reversal (Eq. \ref{eq_tausym}) and can couple to the Zeeman field at the linear order. 
  
The effective Hamiltonian for the $\tau$-spins, obtained within the $1/|K_z|$ degenerate perturbation theory (Eq. \ref{eq_deg pert}) describes the interaction between the $\tau$-spins (Eq. \ref{full hamiltonian}). This Hamiltonian captures the gapped $Z_2$ QSL as well as other magnetically ordered and trivial paramagnetic phases. This is easily seen by considering various limits of the effective Hamiltonian which shows the gapped $Z_2$ QSL (in $J\sim\Gamma\sim 0$ limit) gives way to either a magnetically ordered phase due to Heisenberg coupling $J$ (in $\Gamma\sim K\sim 0$ limit) or a trivial paramagnetic phase due to pseudo-dipolar coupling $\Gamma$ (in $J\sim K\sim 0$ limit). The schematic phase diagram is given by Fig. \ref{fig_pd}.  We perform exact diagonalisation calculations on finite spin clusters containing $12-32$ $\tau$-spins to further confirm the expectation for the phase diagram. While severely limited in system size, our numerical phase diagram-- based on the analysis of the spectral gap, fidelity susceptibility peaks, topological entanglement entropy and two point correlation functions provide encouraging agreement (Fig. \ref{fig_pd_numerics}) with the schematic phase diagram--reiterating the possibility of a direct phase transition between the $Z_2$ QSL and the spin-ordered state and between the $Z_2$ QSL and a trivial paramagnet apart from a $3D-Z_2$ transition between the spin ordered state and the trivial paramagnet. 
 
 A canonical way to understand the emergence of short-range entangled (with or without spontaneous symmetry breaking) phases from a QSL is in terms of the condensation of the deconfined excitations of the QSL which in this case are $Z_2$ $e$ and $m$ charges. While this formulation indeed is very powerful and lead us to understand the nature of the deconfined quantum phase transitions out of the $Z_2$ QSL, interestingly we provide an alternate insight towards understanding of the $Z_2$ QSL through proliferation of the selective domain walls of the magnetically ordered phase with a specific sign structure (Eq. \ref{eq_domgs}) as opposed to random proliferation of the domain walls in the trivial paramagnet (Eq. \ref{eq_parags}). 
  
To describe the phase transitions, it is important to understand the nature of exciations of the $Z_2$ QSL-- the $e$ and $m$ charges.  This is best done by expressing the microscopic interactions in terms of the well-known mapping to the $Z_2$ gauge theory with $e$ and $m$ charges (Eq. \ref{eq_tauz}). These bosonic $Z_2$ charges are conserved modulo 2 and see each other as source of $\pi$ flux (mutual semions). Further, they transform under projective representations of the symmetry group. In particular, we find non-trivial implementation of the time reversal symmetry (Eq. \ref{eq_gauge_trans_tr}) and translation (Eq. \ref{eq_gauge_trans_transl}) under ${\bf T_{d_1}}$ and ${\bf T_{d_2}}$ (Fig. \ref{fig_toric}) on the gauge charges. The latter leads to the permutation symmetry $e\leftrightarrow m$-- an example of an {\it anyon permutation symmetry}.  

Both the time reversal and the anyon permutation symmetry severely constrains the structure of the critical theory. Indeed for the Heisenberg perturbations the time-reversal partner soft modes of the $e$ and $m$ sectors are given by (Eqs. \ref{eq_esm} and \ref{eq_msm}) whose structure is schematically shown in Figs. \ref{fig_sm1} and \ref{fig_sm2}. These soft modes transform under the symmetries as a pair of complex bosons, $\Phi_e, \Phi_m$ (Eq. \ref{eq_ecom} and \ref{eq_mcom}) upto a quartic term that reduces the symmetry to $Z_4$ from $U_e(1)\times U_m(1)$. Crucially however, anyon permutation $\Phi_e\leftrightarrow\Phi_m$ leads to a self-dual structure of the critical theory. The mutual semionic statistics between the $\Phi_e$ and $\Phi_m$ soft modes is implemented within a mutual $U(1)\times U(1)$ Chern Simons (CS) theory resulting in a self dual $3D$ Euclidean action (Eq. \ref{eq_softct}) with Lagrangian 
\begin{align*}
\mathcal{L}=&|(\partial_\mu-i A_\mu)\Phi_e|^2+ |(\partial_\mu-i B_\mu)\Phi_m|^2\nonumber\\
&+ u(|\Phi_e|^2+|\Phi_m|^2)+v(|\Phi_e|^4+|\Phi_m|^4)\nonumber\\
&-\lambda\left[(\Phi_e)^4+(\Phi_e^*)^4 + (\Phi_m)^4+(\Phi_m^*)^4\right]\nonumber\\
&+w\left[(\Phi_e\Phi_m)^2+(\Phi_e\Phi_m^*)^2+{\rm c.c.}\right]\nonumber\\
&+\frac{i}{\pi}\epsilon^{\mu\nu\lambda} A_\mu\partial_\nu B_\lambda
\end{align*}

For $u>0$ $\Phi_e$ and $\Phi_m$ are gapped and the low energy effective action is given by the last term-- the mutual CS action. This phase is nothing but the $Z_2$ QSL with gapped $e$ and $m$ charges. The phase, $u<0$, on the other hand is characterised by finite collinear spin order characterised by the gauge invariant order parameters given by Eq. \ref{eq_magord2} that breaks time reversal symmetry. Using particle-vortex duality we can map the above action to a modified Abelian Higgs model (MAHM) which, at $u=0$ describes the transition. We note that the $U_e(1)\times U_m(1)$ breaking anisotropy terms may be irrelevant at the critical point but relevant in the spin-ordered phases. An external Zeeman field lifts the symmetry of the two time reversal partners by allowing a second order term (Eq. \ref{eq_trzeesoft}) of the form $-h_z\left[(\Phi_e)^2+(\Phi_e^*)^2+(\Phi_m)^2+(\Phi_m^*)^2\right]$. While the $Z_2$ QSL remains intact the spin-ordered phase gets affects and is now continuously connected to the polarised phase (for $J<0$) or undergoes a spin-flop transtion into a polarised phase for $J\sim |h_z|>0$. 

For the pseudo-dipolar, $\Gamma$, perturbations similarly we get a pair of complex scalar modes, $(\tPhi_e, \tPhi_m)$ (Eq. \ref{eq_tphie} and \ref{eq_tphim}), which now are time reversal invariant. The PSG of the soft modes allow for a second order term (Eq. \ref{eq_gammasc}) similar to the Zeeman case. However now the Higgs phase correspond to a time reversal symmetric trivial paramagnet. In fact we find a continuous interpolation of the soft modes driven by the Zeeman perturbation and the pseudo-dipolar perturbations by identifying the residual symmetries when both these terms are simultaneously present (Eq. \ref{eq_zeegamma}). The second-order anisotropic term acts like a pairing term in a superconductor and reduces the gauge group down to $Z_2$ from $U(1)$ at the critical point. The transition therefore belongs to $Z_2$ gauge theory on the self-dual line.
 
The rest of the paper is organised as follows.  In Section \ref{sec_hamil}, we start with the description of the anisotropic limit of the Hamiltonian given in Eq. \ref{eq_hamiltonian} and identify the low energy degrees of freedom as well as their symmetry properties. The effective low energy Hamiltonian for the $\tau$ spins, derived using degenerate perturbation theory upto fourth order, is presented in Section \ref{section_FM}. Numerical exact diagonalisation results on finite spin clusters, as presented in Section \ref{sec_numerics}. In Sec. \ref{sec_gauge_th} a $Z_2$ lattice gauge theory capturing $e$ and $m$ excitations of the $Z_2$ QSL is introduced and their symmetry transformations are analysed.  In Sec. \ref{sec_phase_tran_tc_heisenberg}, we derive the critical theory for the transition driven by the Heisenberg interaction, $J$ and show that the transition indeed occurs through the condensation of $e$ and $m$ charges. Here we also study the effect of an external Zeeman field. In Sec. \ref{sec_gammact}, we discuss the transition between the QSL and the trivial paramagnet driven by the pseudo-dipolar interactions. Finally we summarise our results in Sec. \ref{sec_summ}. Various details are given in the appendices.
 
\section{Generalized Heisenberg-Kitaev Model : Anisotropic Toric code limit}
\label{sec_hamil}

The gapped $Z_2$ QSL stabilised in the anisotropic\cite{kitaev2003fault} limit is the starting point of our analysis. It is obtained by neglecting the Heisenberg ($J$) and the pseudo-dipolar ($\Gamma$) couplings in the Hamiltonian in Eq. \ref{eq_hamiltonian} and considering one of the three Kitaev coupling to be much larger than the other two. \cite{kitaev2006anyons} Depending on which Kitaev coupling we choose we get three equivalent gapped $Z_2$ QSLs whose properties are related by appropriately rotating the underlying honeycomb lattice by $\pm2\pi/3$ about the center of the hexagon. For the rest of the paper we shall take the Kitaev couplings on the $z$-bonds of Fig. \ref{fig_kitaev} to be stronger than that of the $x$ and $y$ bonds. 

In presence of $J$ and $\Gamma$, our analysis of the anisotropic limit starts with derivation of the correct low energy effective Hamiltonian from Eq. \ref{eq_hamiltonian} in the limit of $\ |K_z| >> |J|, |K_x|, |K_y|, |\Gamma_\alpha|$.  To this end we write the Hamiltonian in Eq. \ref{eq_hamiltonian} as\cite{kitaev2006anyons,PhysRevB.91.134419}
\begin{equation}
\mathcal{H}=\mathcal{H}_{0}+\mathcal{V},
\label{eq_pert_hamiltonian}
\end{equation}
where 
\begin{equation}
\mathcal{H}_{0}=-(K_z-J)\sum_{\langle i,j\rangle,z}\sigma_i^z\sigma_j^z
\label{eq_exact_part}
\end{equation}
and $\mathcal{V}$ stands for the rest of the terms in Eq. \ref{eq_hamiltonian} which can be treated as perturbation in this limit.  For $\mathcal{V}=0$ the systems breaks up into isolated bonds and each bond has two ground states. The nature of these ground states depends crucially on the sign of $K_z$. 

For $K_z>0$, {\it i.e.} the {\it ferromagnetic} case, the two spins  participating in the bond are both parallel to each other.  Let us denote these states  in the $\sigma^z$ basis by\cite{kitaev2006anyons}
\begin{align}
|\uparrow\uparrow\rangle\equiv|+\rangle,~~~~~~~~~|\downarrow\downarrow\rangle\equiv|-\rangle
\label{eq_tau}
\end{align}
where the first (second) spin belongs to sub-lattice $A (B)$ of Fig. \ref{fig_kitaev}. The two excited states are given by
\begin{align}
|\uparrow\downarrow\rangle,~~~~~~|\downarrow\uparrow\rangle
\label{eq_afmtau}
\end{align}
where the excitation energy is $2K_z$. For the $K_z<0$, {\it i.e.} the {\it antiferromagnetic} case, the role of the two sets of doublet is reversed. As mentioned above, in this paper, we will concern ourselves with the ferromagnetic case while in a follow up work\cite{animesh_afm} we shall treat the antiferromagnetic case , $K_z<0$.

We now define $\tau^z$ operators for each $z$-bond to capture the ground state manifold, $\ \tau^z\ket{\pm}=\pm\ket{\pm} $ for both the cases of $\ K_z $. In terms of the underlying $\sigma$ spins, 
\begin{align}\label{eq_tau_def}
\tau^z=\begin{array}{c}
(\sigma_A^z+\sigma^z_B)/2
\end{array}
\end{align}
where the subscripts $A$ and $B$ label the two spins belonging to the two different sublattices participating in a particular $z$-bond (Fig. \ref{fig_kitaev}). If there are $N_z$ number of $z$ bonds then there are $2N_z$, $\sigma$-spins and hence $N_z$, $\tau$-spins. The $\tau$-spin span a rhombic lattice with $\bold{d_1}\ \&\ \bold{d_2}$ as the lattice vectors, as shown in Fig. \ref{fig_toric} (and also Fig. \ref{fig_hcomb_symm_2d_tau} in Appendix \ref{appen_symm}).

The ground state of $H_0$ is clearly $2^{N_z}$-fold degenerate. Depending on the various coupling parameters in $\mathcal{V}$, it breaks this degeneracy either by selecting an ordered ground state through quantum {\it order-by-disorder}\cite{refId0} or through {\it disorder-by-disorder}\cite{moessner2001ising} to a QSL by macroscopic superposition of the states within the degenerate manifold leading to long-range quantum entanglement. We wish to understand the nature of such ordered or disordered phases along with the nature of possible intervening quantum phase transitions. 

The effective low energy Hamiltonian below the $\sim K_z$ scale can then be gotten using a the strong coupling expansion in $1/|K_z|$ from the perturbation series 
\begin{align}
\mathcal{H}_{eff}=\mathcal{P}\left[\mathcal{V}+\mathcal{V}\mathcal{G}\mathcal{V}+\cdots\right]\mathcal{P}
\label{eq_deg pert}
\end{align}
where $\mathcal{P}$ is the projector on the ground-state manifold of $\mathcal{H}_0$ and $\mathcal{G}=(1-\mathcal{P})\frac{1}{(E-\mathcal{H}_0)}(1-\mathcal{P})$ is the propagator in the excited manifold.

Before  describing our strong-coupling calculations, however, it is useful to understand the action of the various symmetries on the $\tau^\alpha$ spins which will form an essential ingredient in our analysis. 

\subsection{Symmetries of the low energy doublet}

The lattice points of the rhombic lattice on whose sites the $\tau$-spins reside (see Fig. \ref{fig_toric} and also Fig. \ref{fig_hcomb_symm_2d_tau} in the Appendix \ref{appen_symm}) are given by
\begin{align}
i\equiv (i_1,i_2)=i_1{\bf d_1}+i_2{\bf d_2},
\end{align}
with the two diagonal translation vectors ${\bf d_1}~\&~{\bf d_2}$ of the rhombic lattice as shown in Fig. \ref{fig_toric}. Alternatively we can choose a Cartesian coordinate system (given by $\hat{\bf x}={\bf d_1}-{\bf d_2}$ and $\hat{\bf y}={\bf d_1}+{\bf d_2}$) with a two site-basis to describe the spins. We shall alternatively use both these descriptions whenever suitable.

Starting from the symmetries of the isotropic system (Eq. \ref{eq_hamiltonian}) on the honeycomb lattice (see Appendix \ref{appen_symm}) and focussing on the anisotropic limit, we find the following generators of symmetries for the anisotropic limit : 
\begin{itemize}
\item Time reversal, $\mathcal{T}$.
\item Lattice translations in the honeycomb plane, $T_{d_1}$ and $T_{d_1}$. Under translation $T_{d_1}:(i_1,i_2)\rightarrow(i_1+1,i_2)$ and $T_{d_2}:(i_1,i_2)\rightarrow(i_1,i_2+1)$.
\item Reflection about $z$-bond of the honeycomb lattice, $\sigma_v$ for which we have $\sigma_v:(i_1,i_2)\rightarrow(-i_2,-i_1)$.
\item $\pi$-rotation about the z-bond, $C_{2z}$ which gives  $C_{2z}:(i_1,i_2)\rightarrow(i_2,i_1)$.
\end{itemize}
Note that due to spin-orbit coupling, the spin quantization axes and the real space are coupled and we choose the same convention as You {\it et. al.} in Ref. \onlinecite{PhysRevB.86.085145} to understand the symmetry transformations. Further, in addition to the symmetries listed above, we find it convenient to use the additional symmetry 
\begin{itemize}
\item $\pi$-rotation about the honeycomb lattice hexagon center, $R_\pi=C_{2z}\sigma_v$
\end{itemize}

The action of the above symmetry transformations on the ground state doublets are given by (see Appendix \ref{appen_symm} for details) :
\begin{align}
&\mathcal{T}:~~~~~~\{\tau^x_i,\tau^y_i,\tau^z_i\}~~~~~~ \rightarrow \{\tau^x_i,\tau^y_i,-\tau^z_i\}\nonumber\\
&T_{\bf d_j}:~~~~\{\tau^x_i,\tau^y_i,\tau^z_i\}~~~~~~\rightarrow\{\tau^x_{i+{\bf d_j}},\tau^y_{i+{\bf d_j}},\tau^z_{i+{\bf d_j}}\}\nonumber\\
&\sigma_v:~~~~~\{\tau^x, \tau^y, \tau^z\}_{(i_1,i_2)}\rightarrow\{-\tau^x, \tau^y, -\tau^z\}_{(-i_2,-i_1)}\nonumber\\
&C_{2z}:~~~\{\tau^x, \tau^y, \tau^z\}_{(i_1,i_2)}\rightarrow\{-\tau^x_{j}, \tau^y_{j}, -\tau^z_{j}\}_{(i_2,i_1)}\nonumber\\
&R_\pi :~~~~\{\tau^x, \tau^y, \tau^z\}_{(i_1,i_2)}\rightarrow\{\tau^x, \tau^y, \tau^z\}_{(-i_1,-i_2)}
\label{eq_tausym}
\end{align}
It is important to notice that, $\tau^\alpha$s are {\it non-Kramers doublets}.  Hence any on-site (time reversal odd) magnetic ordering that can be described within this limit, has to be an ordering of $\tau^z$. This also means that an external Zeeman field can only couple to $\tau^z$ at linear order as is characteristic to such non-Kramers systems. 

With the symmetries, we now start to analyze the low energy effective theories for the Hamiltonian (Eq. \ref{eq_hamiltonian}) in the anisotropic limit for the ferromagnetic ($K_\alpha>0$) case.


\section{The anisotropic limit for $K_\alpha>0$}
\label{section_FM}

For the Isotropic model with ferromagnetic Kitaev exchanges ($K_x=K_y=K_z>0$), with increasing Heisenberg coupling, $J$, the Kitaev spin liquid gives way to a ferromagnetic (for $J<0$) or a stripy spin ordered (for $J>0$) (Fig. \ref{fig_stripy_ferro}) phase. The situation with the pseudo-dipolar interactions are much less clear and recently both the possibilities of QSL and a lattice nematic has been suggested\cite{PhysRevB.97.075126,lee2019magnetic} in related models.

In the anisotropic limit, the effective Hamiltonian in the anisotropic limit is obtained through degenerate perturbation theory as outlined in Eq. \ref{eq_deg pert}.

\subsection{The effective Hamiltonian}

For $\ K_x=K_y=K>0 $, $\ \Gamma_{\alpha}=\Gamma$  (where $\alpha=x,y,z) $, we derive the effective low energy Hamiltonian for the $\tau$-spins till fourth order perturbation theory which captures the QSL, the proximate spin ordered phases as well as possible trivial paramagnets. The effective low energy Hamiltonian for the $\tau$-spins is given by
\begin{equation}
\mathcal{H}_{eff}=\mathcal{H}_{[1]}+\mathcal{H}_{[2]}+\mathcal{H}_{[3]}+\mathcal{H}_{[4]}
\label{full hamiltonian}
\end{equation}
where,
\begin{equation}\label{single}
\begin{aligned}
	& \mathcal{H}_{[1]}= \left[2\Gamma\left(1-\frac{\Gamma^2}{\Delta^2}\right)-\frac{2\Gamma^2\delta^2}{\Delta^3}\right]\sum_{i}\tau^y_i 
\end{aligned}
\end{equation}
is the {\it single spin} interaction. The index $i$ now denotes the bonds of a square lattice as shown in Fig. \ref{fig_toric}. We have used
\begin{align}
\delta=K-J~~~~{\rm and}~~~~\Delta=K_z-J
\end{align}
for clarity. Note that the linear term in $\tau^y$ in Eq. \ref{single} is time reversal invariant and is proportional to $\Gamma$ and hence is zero when $\Gamma=0$.  This term, as we shall see below, makes the $Z_2$ QSL unstable to a trivial paramagnet as $\Gamma$ is increased. 

The other terms $a=2,3,4$  in the Hamiltonian $\ \mathcal{H}_{[a]}$ involves interactions among two, three and four spins respectively.  Odd-spin terms are generically allowed due to the non-Kramers nature of the $\tau$-spins.

In writing the the higher order terms we use the convention : each plaquette of the rhombic lattice is associated with its left edge such that we denote the spin on the left edge as $\ \tau_i $ (Fig. \ref{fig_toric}). Using the definition of $\bf d_1$ and $\bf d_2$, the top most spin is then given by $\ \tau_{i+ d_1} $ while the other two spin, the one on the right and the one in the bottom are $\ \tau_{i+d_1-d_2}~$ and $~\tau_{i-d_2} $ respectively. With this, the two spin interactions are given by

\begin{widetext}
\begin{equation}
\begin{aligned}
\mathcal{H}_{[2]} = & \left[J-\frac{J\delta}{2\Delta}-\frac{J^3\delta+J\delta^3}{8\Delta^3}\right]\sum_{\langle ij\rangle}\tau_i^z\tau_j^z +\left[\frac{J^2\delta^2}{2\Delta^3}\right]\sum_{i}\tau_i^z\tau_{i+d_1-d_2}^z  -\left[\frac{5J^2\delta^2}{8\Delta^3}\right]\sum_{i}\left(\tau_i^x\tau_{i+d_1-d_2}^x + \tau_i^y\tau_{i+d_1-d_2}^y\right)\\
& -\left[\frac{J^2\delta^2}{4\Delta^3}\right]\sum_{i}\tau_{i+d_1}^z\tau_{i-d_2}^z  -\left[\frac{J^2\delta^2}{8\Delta^3}-\frac{\Gamma^2J}{\Delta^2}\right]\sum_{i}\left(\tau_{i+d_1}^z\tau_{i-d_1}^z + \tau_{i+d_2}^z\tau_{i-d_2}^z\right)
	\label{double}
\end{aligned}
\end{equation}
\end{widetext}
The leading term (proportional to $J$) is an Ising interaction which, as we shall see drives the transition from the $Z_2$ QSL to a spin ordered phase. Unlike the trivial paramagnet above, this spin-ordered phase breaks time reversal symmetry as well as lattice point groups symmetries  $\sigma_v$ and $C_{2z}$ (Eq. \ref{eq_tausym}).

The three spin interactions are given by
\begin{widetext}
\begin{equation}\label{triple}
\begin{aligned}
	 \mathcal{H}_{[3]}=&\left[\frac{\Gamma^2}{\Delta} + \frac{7\Gamma^2\delta^2-4\Gamma^4}{4\Delta^3}\right]\sum_{i}\left(\tau^z_{i+d_{1}}\tau^x_{i}\tau^z_{i-d_{1}}-\tau^z_{i+d_{2}}\tau^x_{i}\tau^z_{i-d_{2}}\right) -\left[\frac{\Gamma^2}{\Delta} - \frac{4\Gamma^4-6\Gamma^2\delta^2+J^2\Gamma^2}{4\Delta^3}\right]\sum_{i}\tau^z_{i+d_1}\tau^z_{i-d_2}(\tau^y_i+\tau^y_{i+d_1-d_2}) \\
	& +\left[\frac{\Gamma^3}{\Delta^2}-\frac{\Gamma^4+3\Gamma^2\delta^2}{2\Delta^3}\right]\sum_{i}\left(\tau^z_i\tau^x_{i+d_1-d_2} - \tau^z_{i+d_1-d_2}\tau^x_i\right)\left(\tau^z_{i+d_1}-\tau^z_{i-d_2}\right) +\left[\frac{J\Gamma^2}{\Delta^2}\right]\sum_{i}\tau^z_{i+d_1-d_2}\tau^z_i(\tau^y_{i+d_1}+\tau^y_{i-d_2})
\end{aligned}
\end{equation}
\end{widetext}
These third order terms, along with others renormlaises the energy of various excitations in both the QSL as well as the ordered phases and trivial paramagnet. However, we expect that they do not change the qualitative nature of the phase diagram. 

Finally the four  spin interactions are given by

\begin{widetext}
\begin{equation}\label{quadruple}
\begin{aligned}
	& \mathcal{H}_{[4]}=-\left[\frac{J^4+\delta^4}{16\Delta^3}\right]\sum_{i}\tau_i^y\tau_{i+d_1-d_2}^y\tau_{i+d_1}^z\tau_{i-d_2}^z  -\left[\frac{J^2\delta^2}{8\Delta^3}\right]\sum_{i}\tau_i^x\tau_{i+d_1-d_2}^x\tau_{i+d_1}^z\tau_{i-d_2}^z
\end{aligned}
\end{equation}
\end{widetext}
where the first term is nothing but the Toric code Hamiltonain (exactly solvable for $J=0$) that has a $Z_2$ QSL ground state.\cite{kitaev2003fault,kitaev2006anyons}

Thus we have the entire effective Hamiltonian consistent with the symmetries upto fourth order in perturbation theory in $1/K_z$ which incorporates the physics of all the relevant phases.

\subsection{Phases and Phase diagram}\label{subsec_ph_&_ph_diagram}
\begin{figure}
\centering
\includegraphics[scale=0.35]{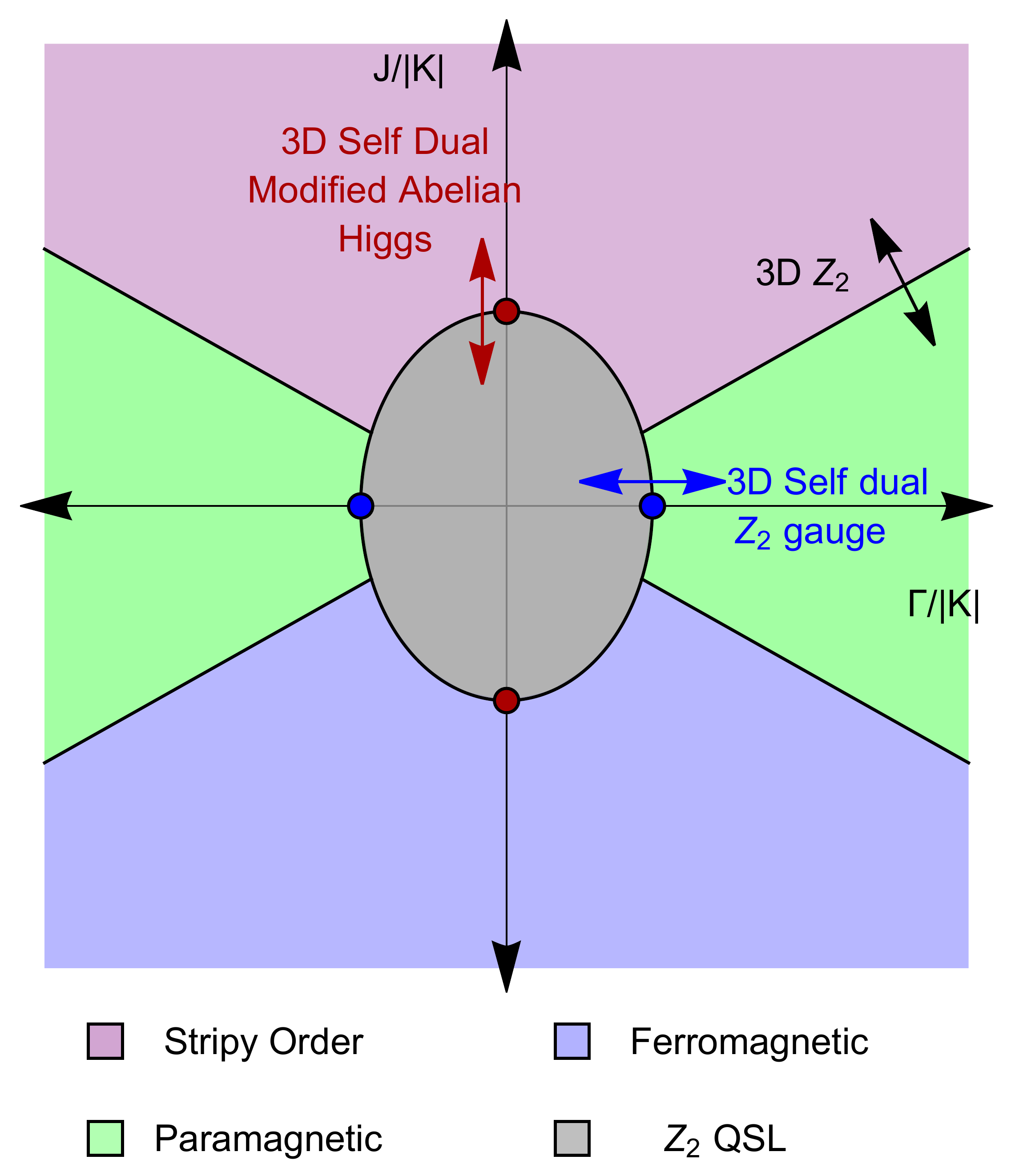}
\caption{Schematic phase diagram of the anisotropic FM Kitaev limit. At origin, i.e. $\Gamma=J=0$ is the $Z_2$ QSL, that survives the small perturbation with respect to $\Gamma/|K|,J/|K|$. However it finally gives way to the magnetically ordered phases (driven by the Heisenberg coupling, $J$) or a trivial product paramagnet (driven by the pseudo-dipolar coupling $\Gamma$). The field theoretic analysis leads to an understanding of the nature of the deconfined quantum phase transition between the QSL and the spin ordered  phase or the trivial paramagnet, in addition to the regular quantum phase transition associated with spontaneous symmetry breaking-- as mentioned in the plot above.}
\label{fig_pd}
\end{figure}

With the above effective low energy Hamiltonian (Eq. \ref{full hamiltonian}) we now study the phase diagram as a function of $J/|K|$ vs $\Gamma/|K|$. The central result of this analysis is shown in the schematic the phase diagram of Fig. \ref{fig_pd}. In the rest of this work using a combination of various field theoretic techniques and exact diagonalisation calculations on small spin clusters we substantiate the above phase diagram as well as study the possible phase transitions.

Before delving into the detailed analysis that results in the phase diagram, let us focus on the different limits to gain insights into the phase diagram. This will also allow us to understand the nature of the low energy modes near the phase transitions.
\subsubsection{{Toric code limit : $J\approx\Gamma\approx0$ and canonical representation} }
\label{para_tc_limit_fm}
In this limit the Hamiltonian in Eq. \ref{full hamiltonian} becomes 
\begin{align}
\mathcal{H}_{(J=\Gamma=0)}=-J_{TC}\sum_{i}\tau_i^y\tau_{i+d_1-d_2}^y\tau_{i+d_1}^z\tau_{i-d_2}^z
\label{eq_toric_code}
\end{align}
where $J_{TC}=\frac{K^4}{16K_z^3}$. This is exactly equivalent to the Toric code model\cite{kitaev2003fault} albeit in the {\it Wen's representation}.\cite{PhysRevLett.90.016803} While the details of this limit are well known,\cite{kitaev2003fault,kitaev2006anyons} we briefly summarise them for completion as well as to set up the notations that will be useful for our calculations.

Eq. \ref{eq_toric_code} is brought into a familiar form by the following site dependent rotation-- rotate all the spins on the horizontal bonds (Fig. \ref{fig_toric}) of the square lattice by  $U_h=\exp[i\tau^z\pi/4]$) and on the vertical bonds by $U_v=\exp\left[-i\pi(\tau^x+\tau^y+\tau^z)/(3\sqrt{3})\right]$.\cite{kitaev2006anyons}  This gives
\begin{align}
&\{\tau_i^x,\tau_i^y,\tau_i^z\}\rightarrow\{-\ttau_i^y,\ttau_i^x,\ttau_i^z\}
~~\forall i\in {\rm horizontal.~bonds}\nonumber\\
&\{\tau_i^x,\tau_i^y,\tau_i^z\}\rightarrow\{\ttau_i^y,\ttau_i^z,\ttau_i^x\}
~~\forall i\in {\rm vertical~bonds}
\label{eq_rot_ver}
\end{align}
where we denote the rotated basis by $\tilde\tau^\alpha$. Eq. \ref{eq_toric_code} then assumes the canonical Toric code form\cite{kitaev2003fault,kitaev2006anyons}
\begin{align}
\tilde{\mathcal{H}}_{J=\Gamma=0}=&-J_{TC}\left[\sum_{s} A_s+ \sum_{p} B_p\right]
\label{eq_toric_code rot}
\end{align}
where the indices $\ s,p$ denotes star and plaquette respectively on the square lattice in Fig. \ref{fig_toric} with $A_s=\prod_{i\in s}\ttau^x_i$, $B_p=\prod_{i\in p}\ttau^z_i$.\cite{kitaev2003fault,kitaev2006anyons} This stabilises a topologically ordered $Z_2$ QSL\cite{kitaev2003fault,kitaev2006anyons} with excitations being gapped bosonic $Z_2$ {\it electric} ($e$) and {\it magnetic} ($m$) charges residing on the vertices and plaquettes of the square lattice (Fig. \ref{fig_toric}) respectively. Crucially, the $e$ and $m$ charges have a mutual semionic statistics,\cite{kitaev2003fault} {\it i.e.}, they see each other as source of Aharonov-Bohm flux of $\pi$.  It is useful to remind ourselves the exact ground states wave-function of a system at this point which is given by\cite{kitaev2003fault}
\begin{align}
|\Psi^{\rm Toric}_{G.S.}\rangle=\prod_s\left(\frac{\mathbb{I}+A_s}{2}\right)|0_z\rangle
\label{eq_toricgs}
\end{align}
where
\begin{align}\label{eq_def_all_up}
|0_z\rangle=\bigotimes_i|+\rangle_i
\end{align}
represents the reference all up state in the $\ttau^z$ basis. Three other tground states on a 2-tori can be generated from the above state by operating with the following Wilson-loop operators along the two non trivial loops in the 2-tori :
\begin{equation}\label{eq_wilson_loop}
\mathcal{L}^e_{x(y)}=\prod_{i\in l_{x(y)}} \ttau^z_{i}~;~\mathcal{L}^m_{x(y)}=\prod_{i\in l^{*}_{x(y)}} \ttau^x_{i}
\end{equation}
$\mathcal{L}^e_{x(y)}$ ($\mathcal{L}^m_{x(y)}$) is product over $\ttau^z$($\ttau^x$) on the closed loop $l_{x(y)}$ ($l^*_{x(y)}$) defined on the links of the direct(dual) lattice along horizontal and vertical directions respectively. These operators have eigenvalues of $\pm 1$. The four ground states of TC model are labeled by $(\mathcal{L}^{e}_{x}=\pm1, \mathcal{L}^{e}_{x}=\pm1)$. In this notation, the ground state $|\Psi^{\rm Toric}_{G.S.}\rangle$ in Eq. \ref{eq_toricgs} is labeled as $\ket{1,1}$. The other three states are $\ket{1,-1}=\mathcal{L}^m_x|\Psi^{\rm Toric}_{G.S.}\rangle$, $\ket{-1,1}=\mathcal{L}^m_y|\Psi^{\rm Toric}_{G.S.}\rangle$ and $\ket{-1,-1}=\mathcal{L}^m_x\mathcal{L}^m_y|\Psi^{\rm Toric}_{G.S.}\rangle$.

The QSL is gapped and hence survives small Heisenberg and pseudo-dipolar perturbations as shown in Fig. \ref{fig_pd}. However due to these perturbations the $e$ and $m$ charges gain dispersion. The low energy effective description of the $Z_2$ QSL in the continuum limit is captured by a $U(1)\times U(1)$ mutual CS theory\cite{freedman2004class,PhysRevB.71.235102,PhysRevB.78.155134,xu2009global}  given by Eq. \ref{eq_u1cs} which correctly implements the semionic statistics between the gapped $e$ and $m$ excitations of the $Z_2$ QSL.

On cranking up the Heisenberg ($J$) and/or the pseudo-dipolar ($\Gamma$) couplings, however, the QSL ultimately gives way to other phases. Starting with the QSL, we can understand the possible destruction of the QSL by condensing the $e$ and $m$ charges.\cite{PhysRevB.71.125102} This leads to  different short-ranged entangled phases without or without spontaneously broken symmetries whose exact nature depend on the quantum numbers of the soft modes of the  $e$ and $m$ charges that condense. This, in turn is dictated by the energetics and the nature of the microscopic couplings, $J$ and $\Gamma$. Indeed we find that while the Heisenberg interactions, $J$, lead to a time reversal symmetry broken magnetically ordered phase, the pseudo-dipoloar term, $\Gamma$, gives rise to a trivial product paramagnet. 

\subsubsection{Heisenberg Limit : $\Gamma=K=0$}
\label{para_heisenberg_limit_fm}

Another instructive and tractable limit is when both the pseudo-dipolar and the Kitaev $x$ and $y$ exchanges, $K$, are absent. The effective Hamiltonian (Eq. \ref{full hamiltonian}) becomes
\begin{align}
\mathcal{H}_{\Gamma=K=0}=&J\sum_{\langle i,j\rangle}\tau_i^z\tau_j^z +\mathcal{O}\left[\left(J^4/\Delta^3\right)\right]
\label{eq_fm_heisenberg_min}
\end{align}

In the limit where $K_z~{(\it i.e.}~\Delta\rightarrow\infty)$ is the largest energy scale in which the above Hamiltonian is valid, the leading term is clearly given by the first term. This leads to ferromagnetic or Neel ordering for the $\tau^z$ spins depending on the sign of $J$. Higher order (in $J/\Delta$) terms though introduce fluctuations, however are expected to retain the above magnetic ordering. The same conclusion is also obtained in the limit $\Gamma=0$ and $J=K$ such that $\delta=0$.

\begin{figure}
\begin{center}
\centerline{\subfigure[]{\includegraphics[scale=.3]{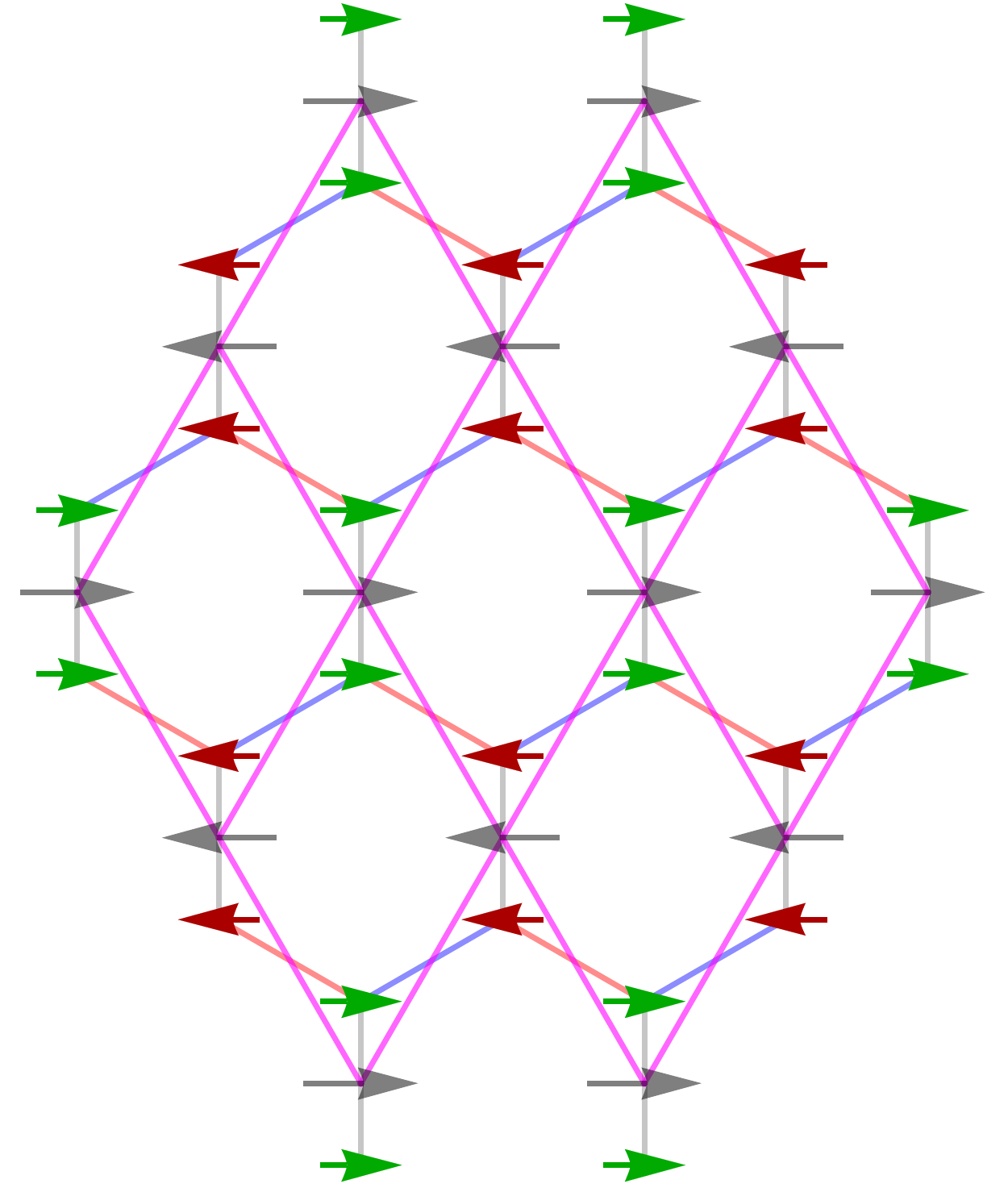}}
\quad\subfigure[]{\includegraphics[scale=0.3]{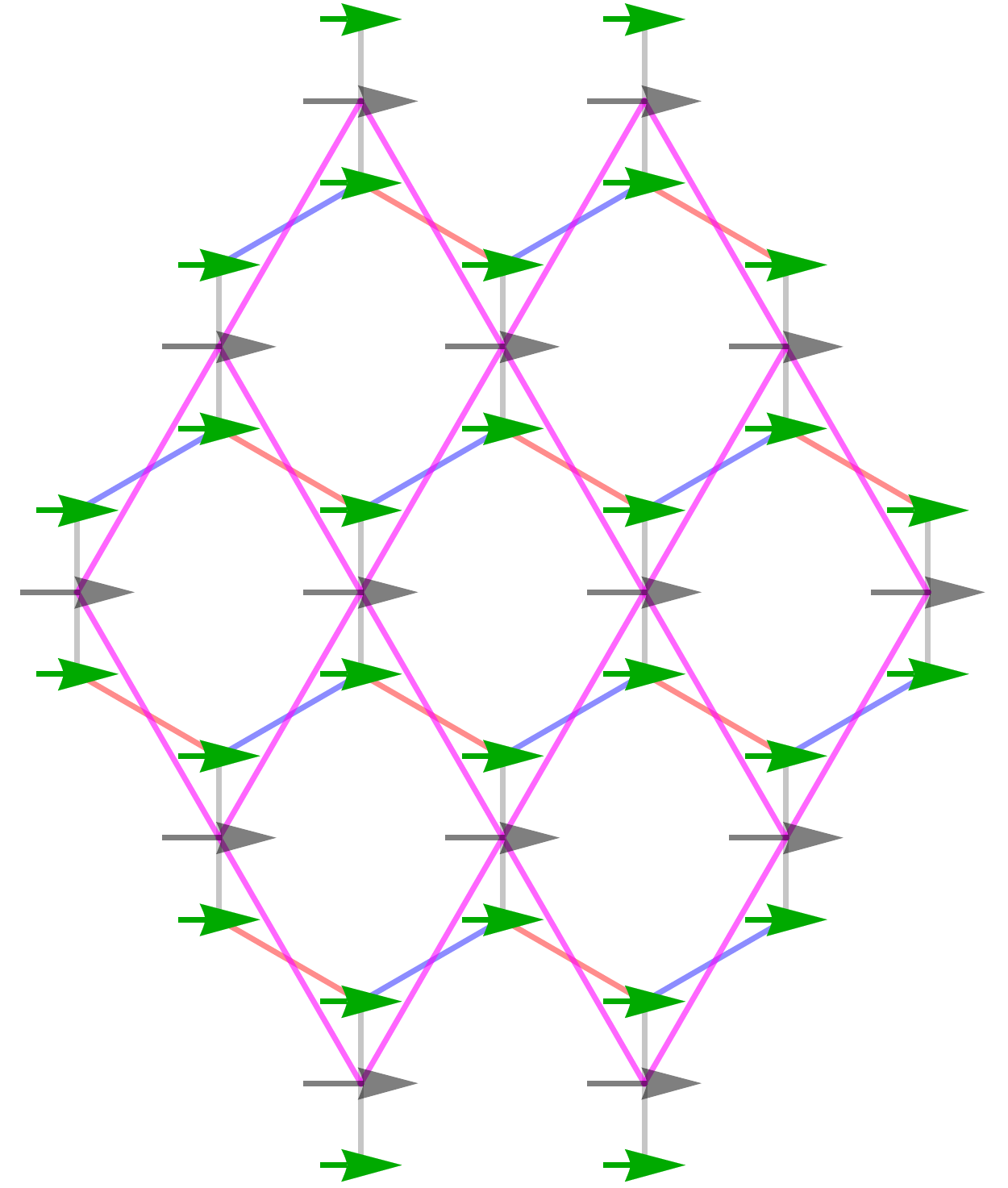}}}
\end{center}
\caption{{\bf (a) Stripy order}: For $J>0$,  Neel ordered state of the $\tau$-spins shown in grey arrows. For the $\sigma$-spin in the underlying honeycomb lattice, the magnetic ordering is shown, which is consistent Eq. \ref{eq_tau}.  This $\sigma^z$ ordering is nothing but the stripy phase. {\bf (b) Ferromagnetic order} for $J<0$: in $\tau^z$ basis all the spins point to the same direction and  equivalently for all the $\sigma^z$ spins.}
\label{fig_stripy_ferro}
\end{figure}

It is interesting to note that the Neel order (for $J>0$) in terms of the $\tau^z$ spins is actually the stripy order in terms of the original $\sigma^z$ of the underlying honeycomb lattice as shown in Fig. \ref{fig_stripy_ferro}(a). Similarly, for $J<0$, the ferromagnetic ordering in terms of $\tau^z$ transforms into a ferromagnetic ordering in terms of the underlying $\sigma^z$ as shown in Fig. \ref{fig_stripy_ferro} (b). Noticeably these are exactly the spin orders found in the immediate vicinity of the Isotropic Kitaev QSL with ferromagnetic exchanges.\cite{PhysRevLett.105.027204, PhysRevLett.112.077204} 

Hence we expect a direct transition between the Ising ferromagnet (or antiferromagnet) and the $Z_2$ QSL.\cite{PhysRevB.71.125102,PhysRevB.86.214414} To understand this transition we  re-write the minimal Hamiltonian incorporating the leading order Heisenberg perturbations in the rotated basis (Eq. \ref{eq_rot_ver}) to obtain
\begin{equation}
\begin{aligned}\label{eq_jk_min_model_rot}
& \tilde{\mathcal{H}}_{\Gamma=0} = J\sum_{\langle i,j\rangle,(i\in H; j\in V) }\ttau_i^z\ttau_j^x  - J_{TC} \left[\sum_{s} A_s+ \sum_{p} B_p\right]
\end{aligned}
\end{equation}
where $A_s$ and $B_p$ are defined below Eq. \ref{eq_toric_code rot}. We note that the perturbation by the Heisenberg term is different from that considered in Ref. \onlinecite{PhysRevLett.98.070602} of \onlinecite{PhysRevB.92.100403} as in the present case a term like $\ttau^{z(x)}_i\ttau^{z(x)}_j$ (where $i\in V$ and $j\in H$) is forbidden by time reversal.

As mentioned above, in the limit $J=0$, the ground state wave-function in the rotated basis is given by Eq. \ref{eq_toricgs}. On the other hand, for $J_{TC}=0$, when the Hamiltonian is just the first term of Eq. \ref{eq_fm_heisenberg_min}, albeit in the rotated basis, the two-fold degenerate ground states. (To be specific, let us consider $J<0$ such that the ground state in the un-rotated basis is a ferromagnet)
\begin{align}
|\Psi^\pm_{FM}\rangle=\bigotimes_{i}|\psi^\pm_i\rangle
\end{align}
where
\begin{align}
|\psi^\pm_i\rangle=\left\{\begin{array}{c}
\left\{\begin{array}{cc}
|+1_{\tilde z}\rangle & \forall i\in H\\
|+1_{\tilde x}\rangle & \forall i\in V\\
\end{array}
\right.\\
 \\
\left\{\begin{array}{cc}
|-1_{\tilde z}\rangle & \forall i\in H\\
|-1_{\tilde x}\rangle & \forall i\in V\\
\end{array}
\right.
\end{array}\right.
\end{align}
for the two time reversal partner ground states. 

Generalising the ideas of Ref. \onlinecite{PhysRevB.71.125102}, we can think about obtaining the QSL from the spin ordered state by selectively proliferating the domain walls of the latter.  Consider taking the above ferromagnetic ground state wave function in the rotated basis and project it in the zero $e$ and $m$ sector as follows 
\begin{align}
|\Psi^+\rangle=\left[\prod_s\left(\frac{1+A_s}{2}\right)\right]\left[\prod_p\left(\frac{1+B_p}{2}\right)\right]|\Psi^+_{FM}\rangle
\label{eq_projectwf}
\end{align}

\begin{figure}
\includegraphics[scale=0.4]{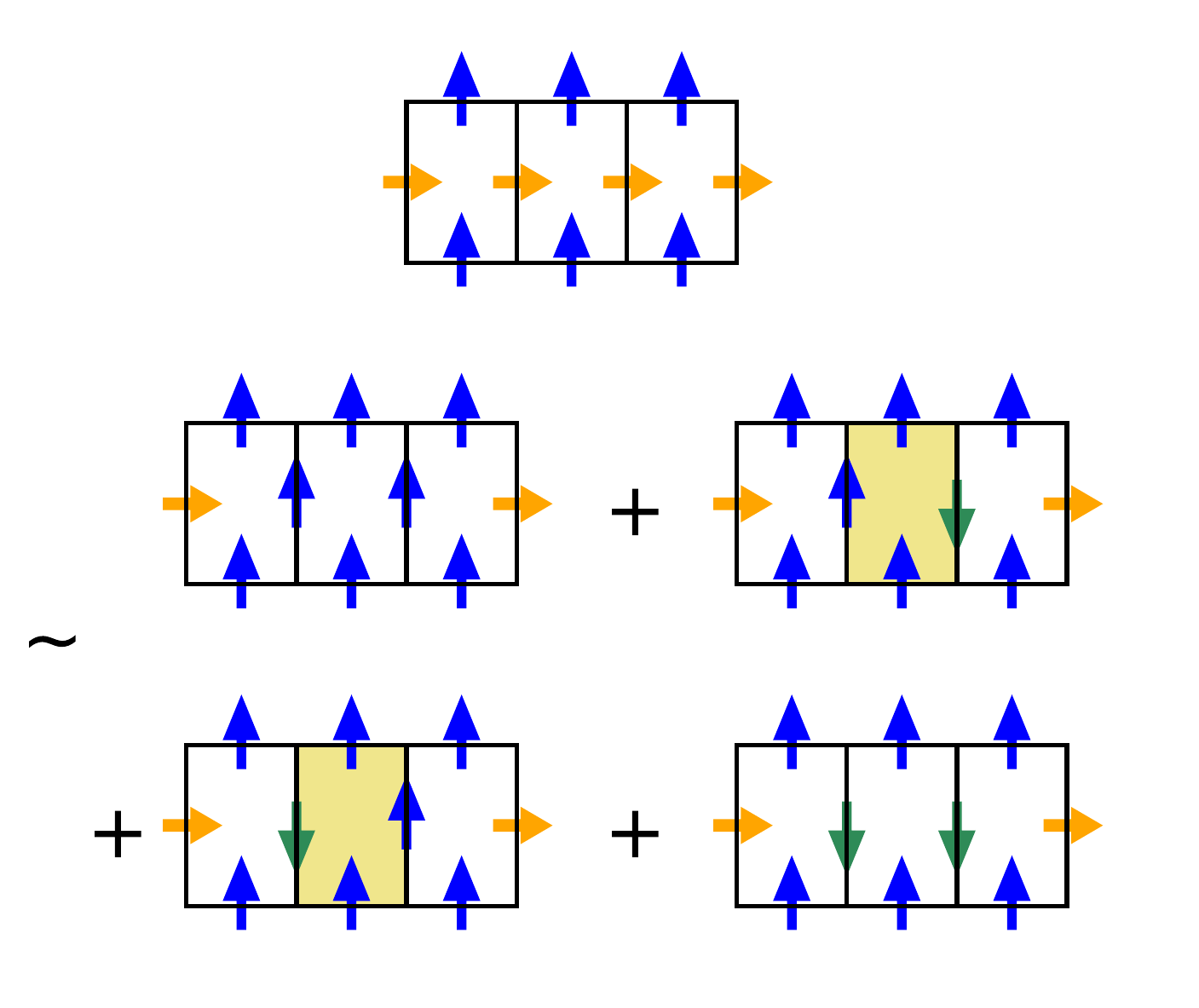}
\caption{The $|\Psi^+_{FM}\rangle$ state for one plaquette expanded in the $\ttau^z$-basis. It is clear that for two of the contributing terms there is a magnetic charge, $B_p=-1$ as marked in yellow. Blue (green) arrows stand for $\ttau^z=1(-1)$ state. Orange arrows stand for $\ttau^x=1$ state.}
\label{fig_xz}
\end{figure}

We note that the two projectors commute with each other. For a plaquette $|\Psi^+_{FM}\rangle$ is shown in Fig. \ref{fig_xz} when expanded in the $\ttau^z$-basis. It is clear that on applying $B_p$ operator to this plaquette, the amplitudes of the two contributions that has a $m$ charge ($B_p=-1$)  does not survive the projection of $\prod_p(1+B_p)/2(\equiv \mathcal{S}_B)$. Extending this argument, we conclude that the on a torus, acting $\mathcal{S}_B$ on $\ket{\psi^{+}_{FM}}$ leads to, upto normalisation, 
\begin{equation}
\prod_{x_j}\left[\prod_{p\in j^{th}\text{row}}\left(\frac{1+B_p}{2}\right)\right]|\Psi^+_{FM}\rangle\approx\prod_{x_j}\left(1+\mathcal{L}^m_{x_j}\right)\ket{0_z}
\end{equation}
where $\ket{0_z}$ is defined in Eq. \ref{eq_def_all_up}.  $\mathcal{L}^m_{x_j}$ are the horizontal Wilson loops (see Eq. \ref{eq_wilson_loop}) for each row in the square lattice, with $x_j$ being the row index. Thus it consists of closed loops of down spins on the vertical bonds running along horizontal direction along the rows.  In the above equation, the product in the right hand side is expanded to obtain
\begin{equation}
\begin{aligned}
&\left(1 + \sum_{x_j}\mathcal{L}^m_{x_j} + \sum_{x_j\neq x_k}\mathcal{L}^m_{x_j}\mathcal{L}^m_{x_k} + ...\right) |0_z\rangle\\
 =& \left(\sum_{e} \left[\mathcal{L}^m_{x_k}...\right] + \sum_{o} \left[\mathcal{L}^m_{x_k}...\right]\right)|0_z\rangle
\end{aligned}
\end{equation}
where in the last expression we have collected all the even (first summation) and the odd (second summation) powers of the $\mathcal{L}^m$ operators separately. From Eq. \ref{eq_projectwf}, it is easy to see that on application of $\prod_s(1+A_s)/2(\equiv \mathcal{S}_A)$, this leads to an equal weight superposition of the $Z_2$ QSL ground states belonging to two topological sectors, {\it i.e.},
\begin{align}
|\Psi^+\rangle\sim |1,1\rangle+|1,-1\rangle
\label{eq_top_sec_plus}
\end{align}

\begin{figure}
\centering
\subfigure[]{\includegraphics[scale=0.35]{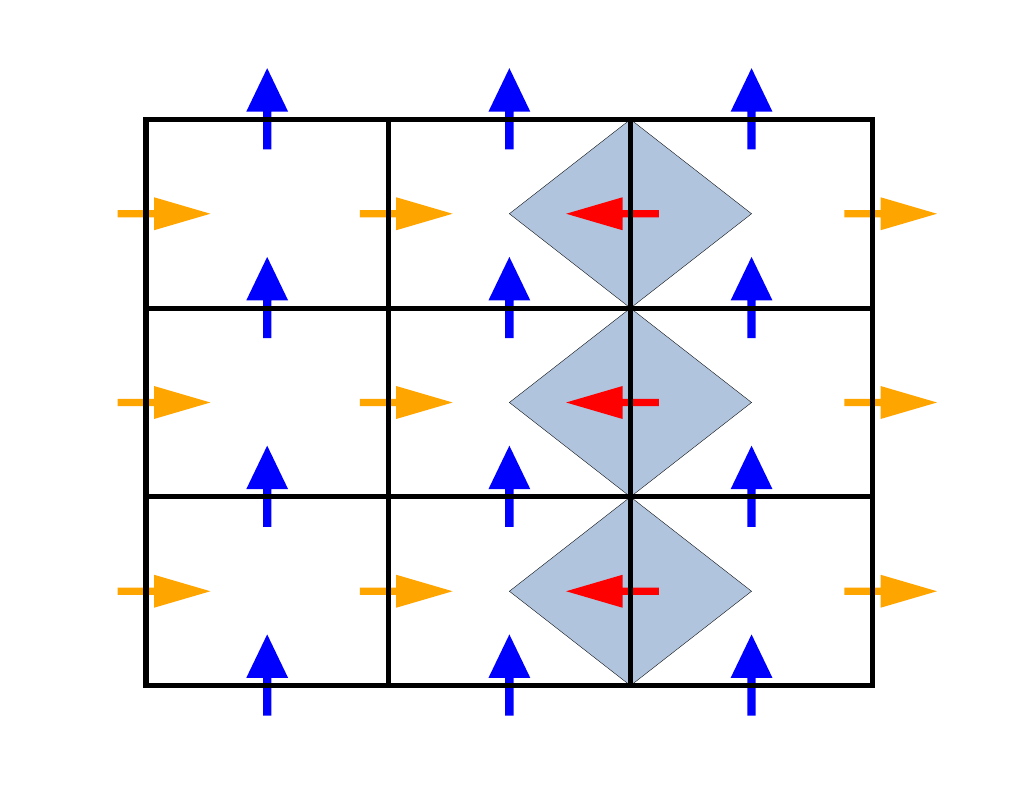}} \\
\subfigure[]{\includegraphics[scale=0.35]{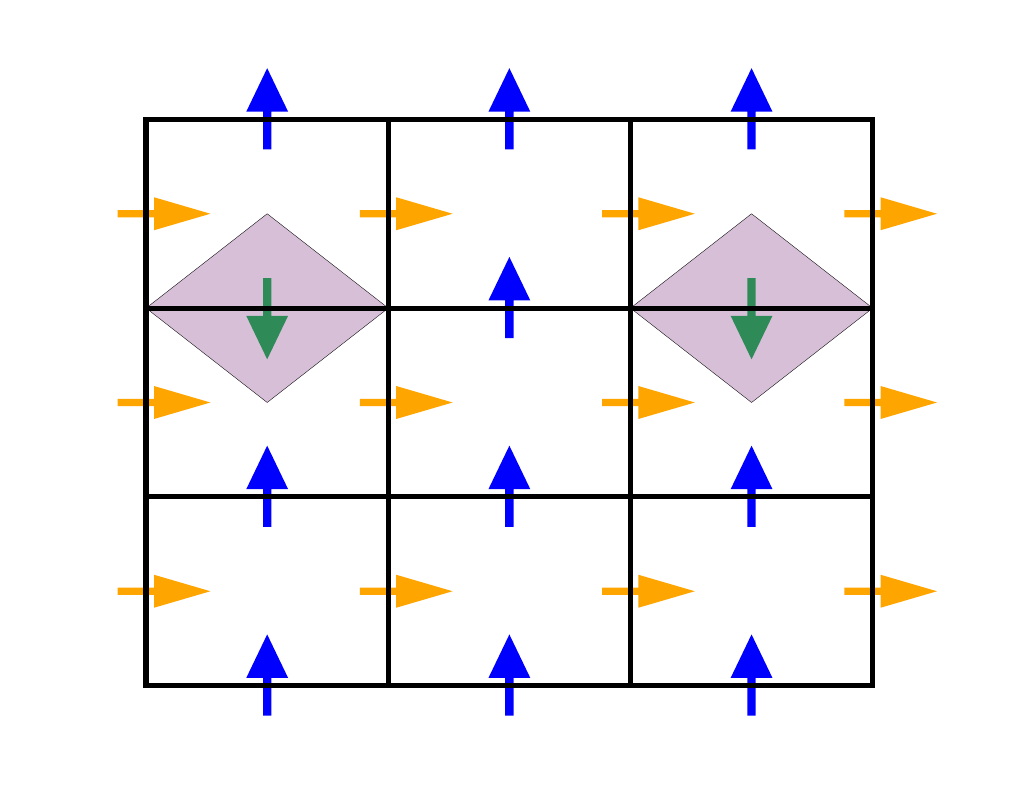}}
\subfigure[]{\includegraphics[scale=0.35]{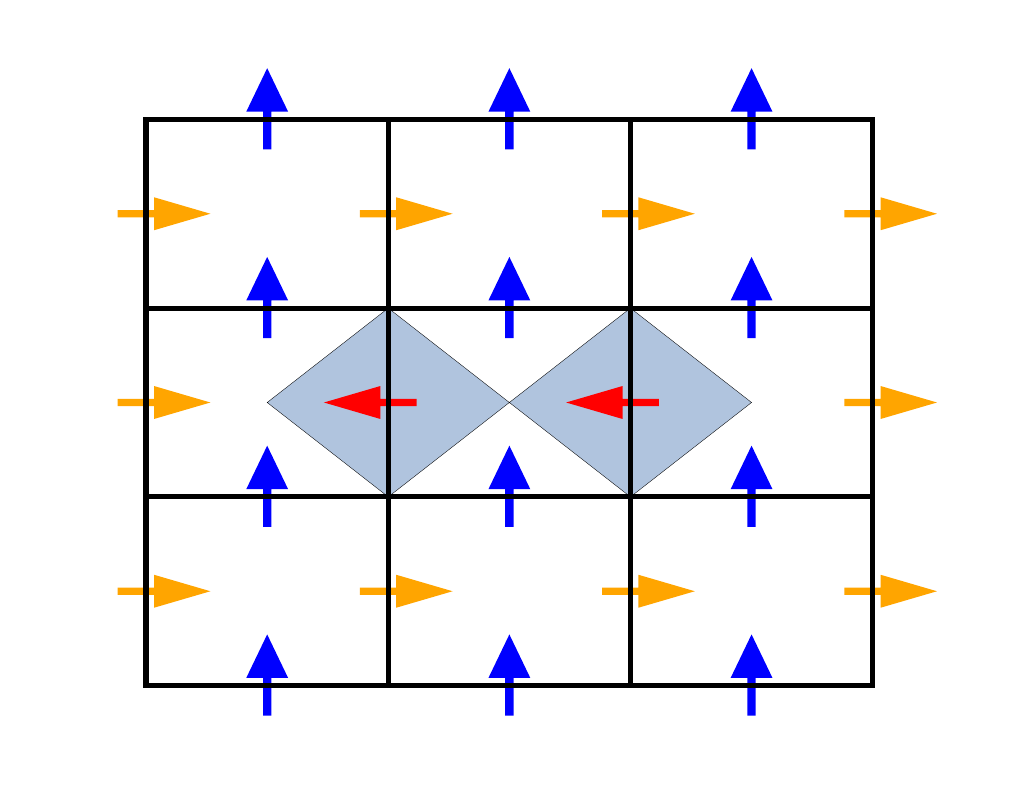}}
\subfigure[]{\includegraphics[scale=0.35]{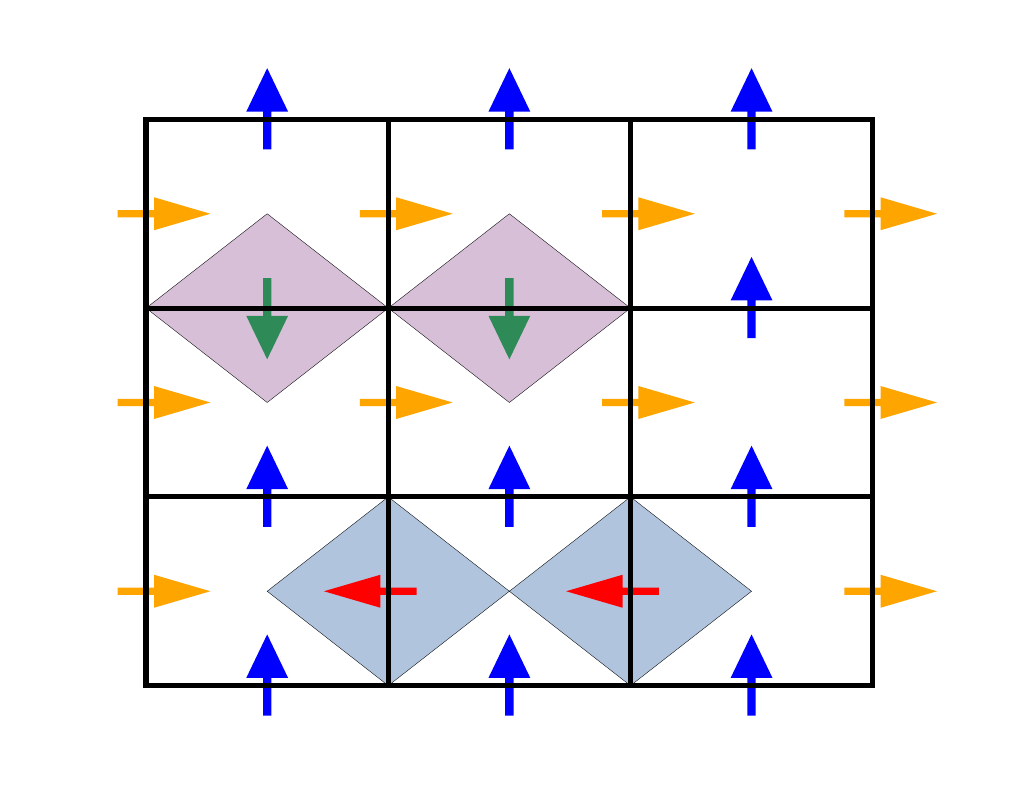}}
\subfigure[]{\includegraphics[scale=0.35]{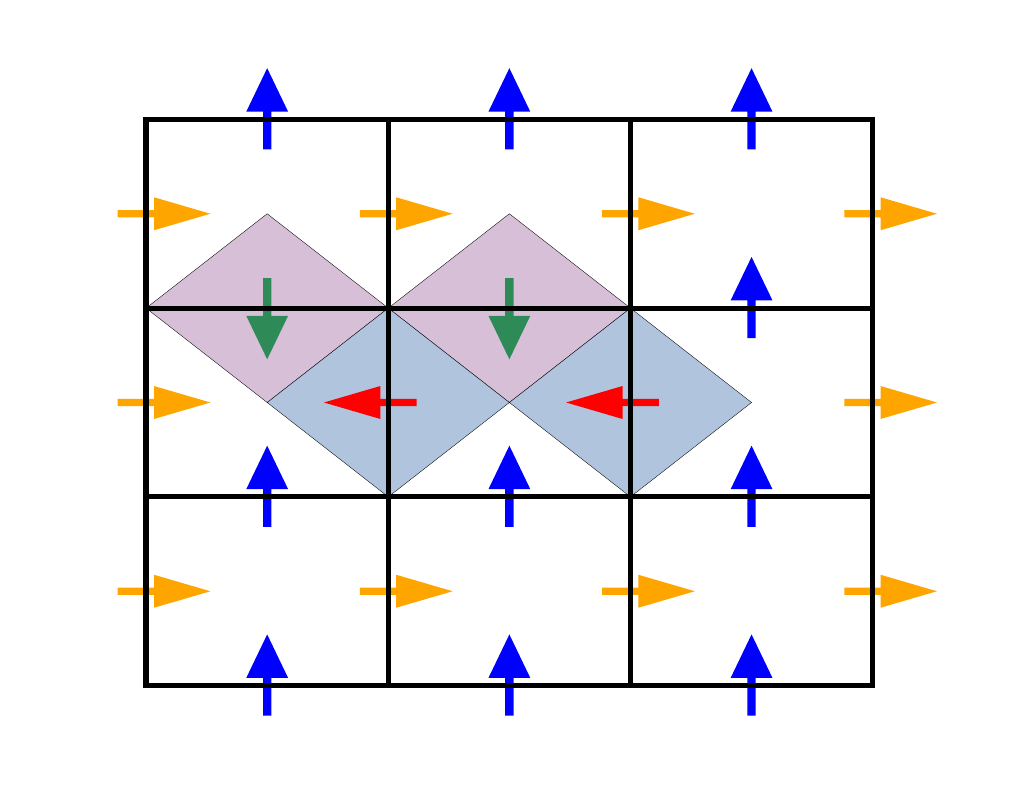}}
\caption{{\bf (a)} Domain wall created by twisted boundary condition ($\mathcal{L}^e_y$). An electric domain wall is created by the action of two neighbouring $A_s$ on $\ket{\Psi^{+}_{FM}}$ is shown in {\bf (b)},  whereas in {\bf (c)}, a magnetic domain wall is created by a single $B_p$ operator. {\bf (d)} Both domain wall with even overlap (see text). {\bf (e)} Both domain wall with odd overlap, this contribution has a relative negative sign (see text). Red arrow denote $\ttau^x=-1$, rest of the arrow definitions follow from Fig. \ref{fig_xz}}
\label{fig_adomain}
\end{figure}

Clearly from Eq. \ref{eq_top_sec_plus}, $\mathcal{L}^e_y\ket{\psi^{+}} \approx \ket{1,1}-\ket{1,-1}$, this helps us to get :
\begin{align}
 |\Psi_{G.S.}^{\rm Toric}\rangle\equiv\ket{1,1}\approx (1+\mathcal{L}^e_y)\mathcal{S}_A\mathcal{S}_B\ket{\Psi^{+}_{FM}}.
 \end{align}
The above equation connects the QSL with the spin ordered state and the operators can be interpreted in terms of the domain walls of the spin ordered state.  Expanding the right hand side of the above equation, we get
 \begin{align}
  |\Psi_{G.S.}^{\rm Toric}\rangle\sim & (1+\mathcal{L}^e_y)\left(1+(A_{s_1}\cdots A_{s_m})(B_{p_1}\cdots B_{p_n})\right)|\Psi_{FM}^+\rangle
  \end{align}
 The first term $(1+\mathcal{L}^e_y)|\Psi_{FM}^+\rangle$ is a superposition of the ordered state with periodic boundary and twisted boundary conditions (see Fig. \ref{fig_adomain}(a)) along the $x$ direction on the 2-tori for the spins on the vertical bonds (For the spins on the horizontal bonds both the states have periodic boundary conditions). Clearly the position of the twist is a choice and does not affect the observables in the QSL state. The rest of the terms on the right hand side are products of $A_s$ and $B_p$ operators and they have a straight forward interpretation in terms of the selected domain walls (defined as location of frustrated bonds) of the spin order.\cite{PhysRevB.71.125102} With the spins located on the bonds, the domain walls passes through the vertices of the square lattice of Fig. \ref{fig_toric} and has a two sub-lattice structure. As shown in Fig. \ref{fig_adomain}(b)-(d), 
the $A_s$ and the $B_p$ operators create domain walls respectively of the spin ordering on the horizontal and vertical bonds. An arbitrary product of only $A_s$ or $B_p$ operators create such domain walls of the spin order and all these contributions have an amplitude with positive sign as is explicit. For a combined set of $A_s$ and $B_p$ operators the sign is given by $(-1)^m$ where $m$ denotes the total overlap of the horizontal bonds among the participating $A_s$ and $B_p$ operators. The $A_s$ and $B_p$ in Fig. \ref{fig_adomain}(d) has zero (even) overlap on the horizontal bond compare to the single (odd) overlap in Fig. \ref{fig_adomain}(e). Thus we have
\begin{align}
|\Psi_{G.S.}^{\rm Toric}\rangle\sim |\Psi_{FM}\rangle +\sum_\alpha (-1)^{m_\alpha}|d_\alpha\rangle
\label{eq_domgs}
\end{align}
where $|d_\alpha\rangle$ denotes various domain walls states starting with the ferromagnetic state. Focusing on a single row of horizontal bonds, application of two neighbouring $A_s$ only on this row (for reference Fig. \ref{fig_adomain}(b)) leads to two disconnected domain walls. On application of further $A_s$s belonging to this row, more domain walls are either created or the ones that are already present gets transported along the chain. As a result the spin on any site on this row of horizontal bond, locally has an equal superposition of up and down spins (in $\ttau^z$-basis). This is nothing but a state where the spins on this row of horizontal bonds are polarized along $\ttau^x$ leading to the gapping out of the $e$ charge. An argument for the row of vertical bonds and the $m$ charge would lead to a similar result.  A  calculation starting from $\ket{\psi^{-}_{FM}}$ leads to equal weight superposition of the other two topological sectors of the QSL. Incidentally one can perform the above analysis starting with an all up state (in $\ttau^z$-basis) as was considered in Ref. \onlinecite{PhysRevB.71.125102}. In that case, the action of $B_p$ is trivial as the all up state is already in the zero $m$ sector resulting in a $Z_2$ QSL. Indeed the right hand side of the Eq. \ref{eq_projectwf} in that case an be interpreted in terms of the selective domain walls of the all up $\ttau^z$-ferromagnet. 

We can contrast Eq. \ref{eq_domgs} to the ground state of the trivial paramagnet obtained by arbitrarily proliferating the domain walls of the ferromagnetic state. Such a trivial paramagnet has a wave-function of the form
\begin{align}
|\Psi_{G.S.}^{\rm Trivial}\rangle\sim |\Psi_{FM}\rangle +\sum_\alpha |\tilde{d}_\alpha\rangle
\label{eq_parags}
\end{align}
 which crucially differs from Eq. \ref{eq_domgs} in nature along with the sign structure of the domain walls. Indeed $|\Psi_{G.S.}^{\rm Trivial}\rangle$ can be obtained from $|\Psi_{FM}\rangle$ by proliferating trivial domain walls using the $\ttau^z_i (\ttau^x_i)$ operators on the vertical (horizontal) bonds. Such domain wall states clearly lack the sign structure discussed above.

Inside the ferromagnetic phase, all types of domain walls are gapped. However depending on the energetics of the microscopic model there energy costs are different. Hence as a function of various coupling terms one can become energetically cheaper than the other within the ferromagnetic phase without causing a phase transition. This provides a crossover within the ferromagnetic phase similar to the U(1) case in three dimensions as discussed in Ref. \onlinecite{PhysRevB.71.125102}. In this lights, it is clear that the Toric code interaction term such as in Hamiltonian in Eq. \ref{eq_jk_min_model_rot} favours  {\it decorated} (by sign) domain walls energetically whose subsequent proliferation leads to the QSL. This also suggest that a different perturbation involving single-site spin operators can lead to a trivial paramagnet starting from the FM. This, we argue below is exactly what the $\Gamma$ term does.

\subsubsection{Pseudo-dipolar limit : $J=K=0$ :}\label{para_pseudo_dipolar_limit}

Finally, we consider the effect of only $\Gamma$ term on the $\tau$ spins. From Eq. \ref{full hamiltonian}, we put $J=K=0$, then we get 
\begin{align}\label{eq_pseudo-dipolar}
\mathcal{H}^{F}_{(J=K=0)}=& 2\Gamma\sum_{i}\tau^y_i  +\mathcal{O}\left[\left(\Gamma^2/\Delta\right)\right]
\end{align}
In the $\Gamma/\Delta\rightarrow 0$ limit, only the first term survives which is just non-interacting spins in a ``magnetic field". The ground state is a product state, $\ \ket{0;\Gamma_{\pm}} = \otimes_{j}\ket{\boldsymbol{\mp}1_y}_{j} $. In terms of $z$-basis it is defined as $\ket{\boldsymbol{\mp}_y}=\frac{\ket{+1_z}\mp i\ket{-1_z}}{\sqrt{2}}$. 

These two ground states describe a time reversal symmetric trivial paramagnetic states (one for either sign of $\Gamma$) of the form described by Eq. \ref{eq_parags}. To see this is is useful to go to the rotated basis (Eq. \ref{eq_rot_ver}) whence the first term of Eq. \ref{eq_pseudo-dipolar} becomes
\begin{align}\label{eq_gammaham_rot}
\tilde{\mathcal{H}}_{J=K=0}=2\Gamma\left[\sum_{i\in V}\ttau^z_i+\sum_{i\in H}\ttau^x_i \right] 
\end{align}
However, as discussed above, in the FM state the same operators as above create elementary trivial domain walls of the ferromagnetic state leading to a paramagnet. 

As an aside, it is interesting to note that, though explicitly broken in the anisotropic limit that we consider this work, the two above states have finite $z$-bond-spin-nematic correlations as measured from the expectation value of the operator
\begin{align}
\hat{Q}^{\alpha \beta}_{ii^{\prime}} = \left(\frac{\sigma^{\alpha}_i\sigma^{\beta}_{i^{\prime}} + \sigma^{\beta}_i\sigma^{\alpha}_{i^{\prime}}}{2} - \frac{\delta_{\alpha\beta}}{3}\boldsymbol{\sigma}_i.\boldsymbol{\sigma}_{i^{\prime}}\right)
\end{align}
We find
\begin{align}
\bra{\pm}_y\hat{Q}^{\alpha \beta}_{ii^{\prime}}\ket{\pm}_y=\begin{bmatrix}
	- \frac{1}{3} & \mp1 & 0\\
	\mp1 & - \frac{1}{3} & 0\\
	0 & 0 & \frac{2}{3}\\
\end{bmatrix}
\label{eq_nematicop}
\end{align}
which describes a nematic with principle axis along $\hat{\bf n}=[1\bar10]$ for $\ \Gamma<0 $ and is along $\hat{\bf n}=[110]$ for $\ \Gamma>0 $. However, as stated above, this does not break any symmetry of the anisotropic Hamiltonian spontaneously and hence represents a featureless paramagnetic phase with gapped excitations which is continuously connected to the product state. Indeed signatures of such a nematic phase was numerically observed in the isotropic $K-\Gamma$ model recently\cite{lee2019magnetic} where the rotational symmetry $\sigma_hC_6$ of the extended Kitaev model is spontaneously broken down by the development of the nematic order.

The above discussion of the phases sharpens the questions about  the quantum phase transitions between the $Z_2$ QSL and  the spin-ordered or a trivial paramagnetic phase as a function of $J/K$ and $\Gamma/K$  respectively as indicated in Fig. \ref{fig_pd}.  However, before moving on to the theory of quantum phase transition, we present our preliminary numerical calculations in the form of exact diagonalisations on finite spin clusters. This provide further insights into the nature of the soft $e$ and $m$ modes which then is used to construct the critical theory. 

\begin{figure*}
\centering
\subfigure[]{\includegraphics[scale=.251]{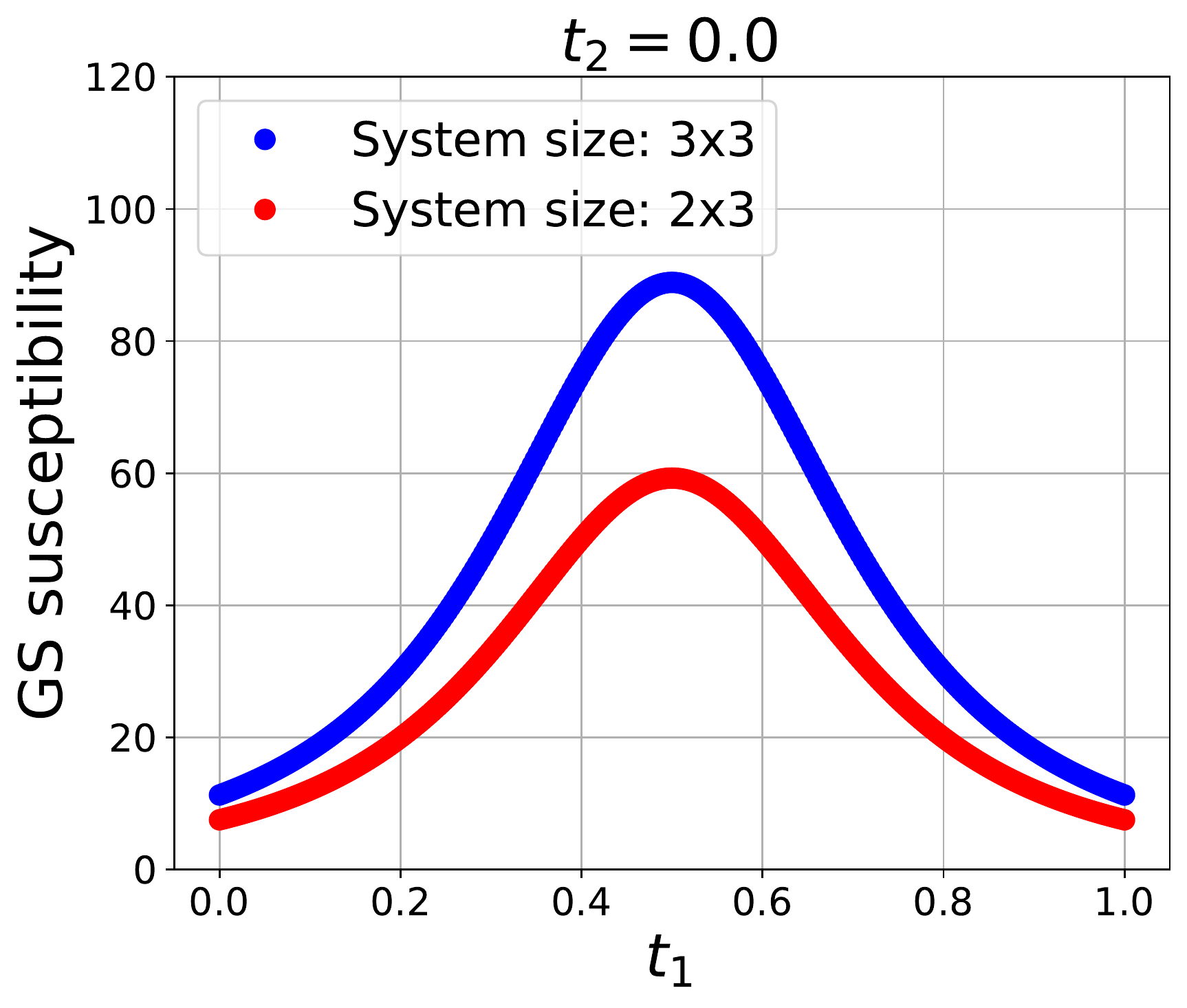}}
\subfigure[]{\includegraphics[scale=.251]{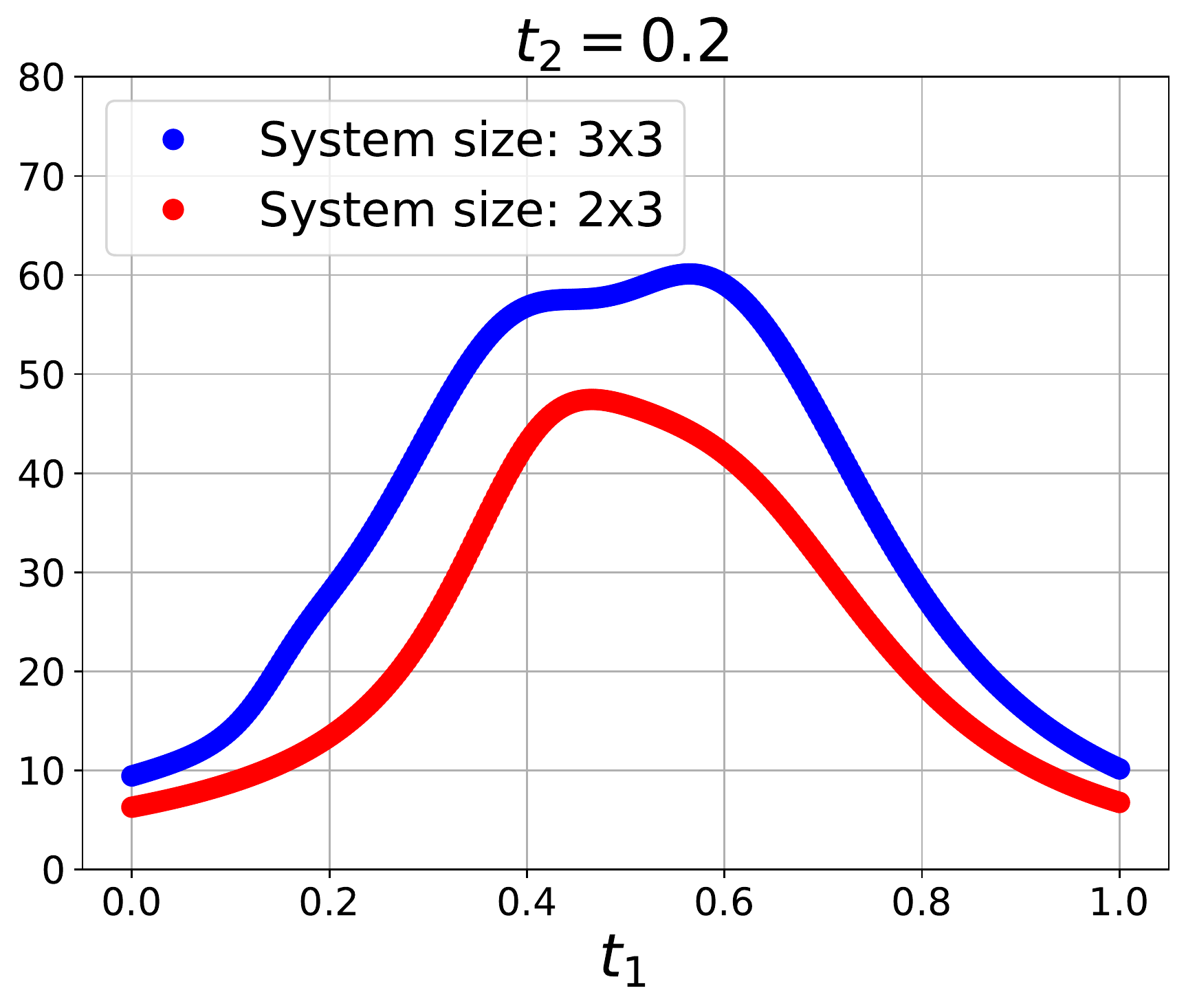}}
\subfigure[]{\includegraphics[scale=.251]{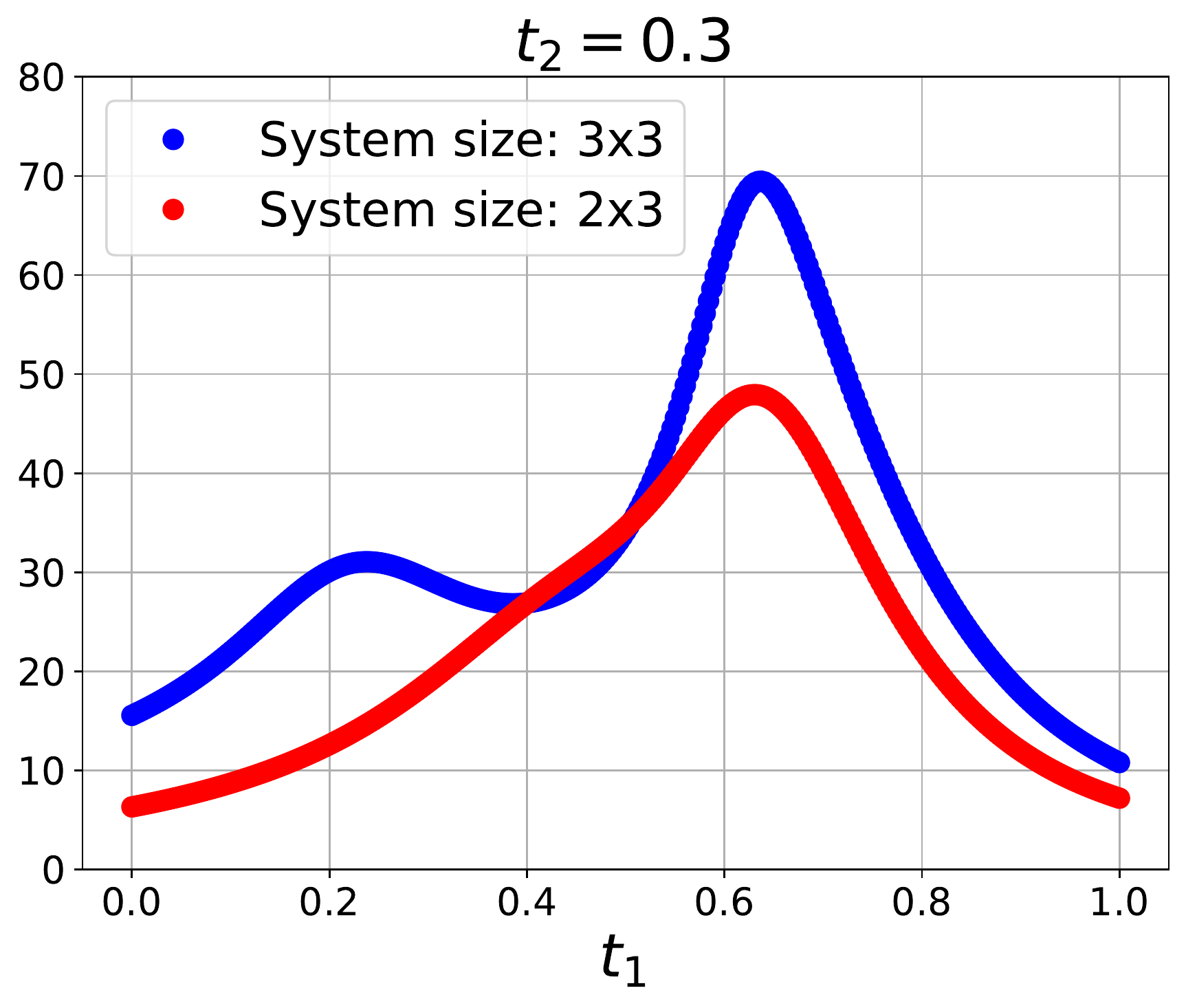}}
\subfigure[]{\includegraphics[scale=.251]{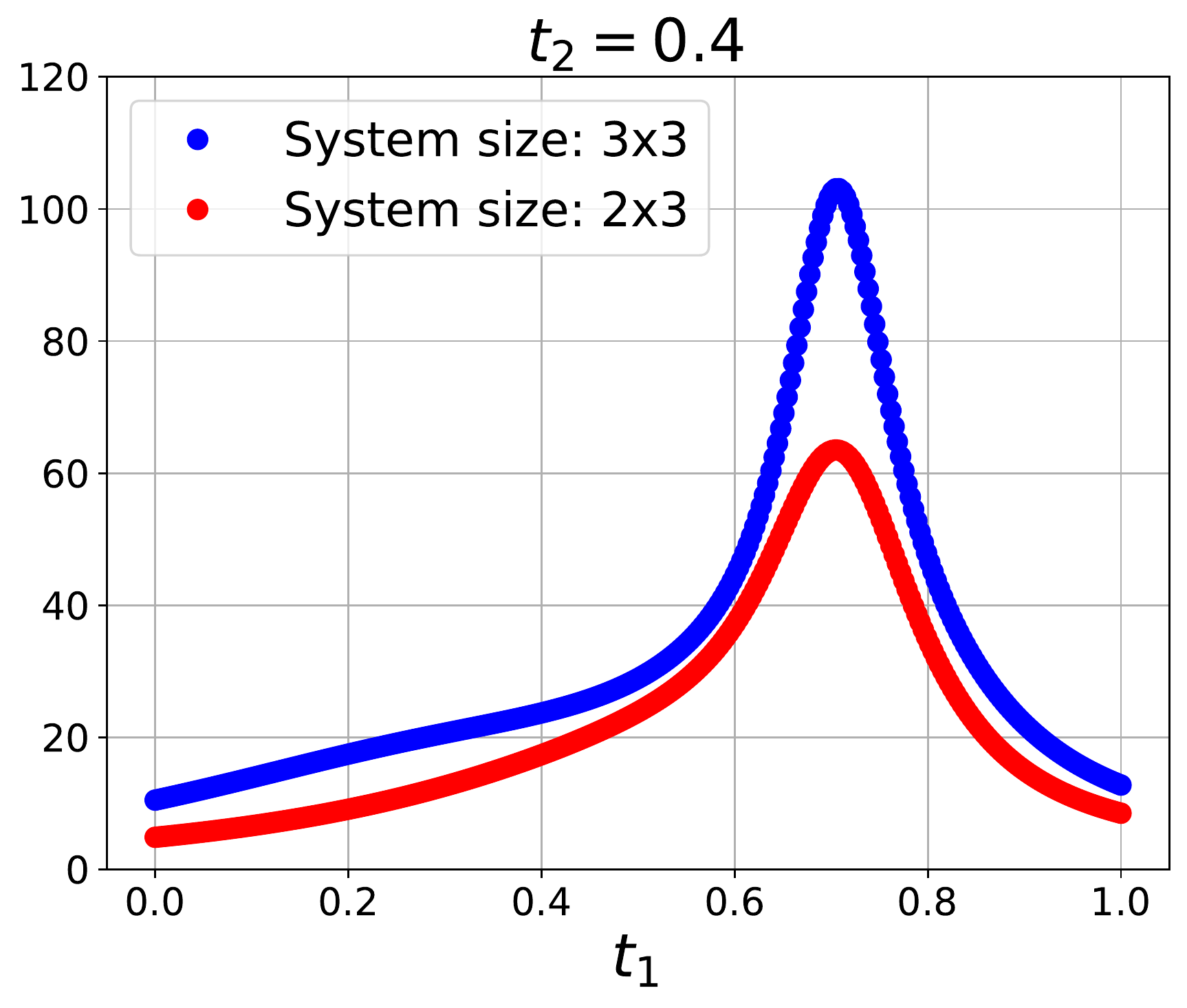}}
\subfigure[]{\includegraphics[scale=.251]{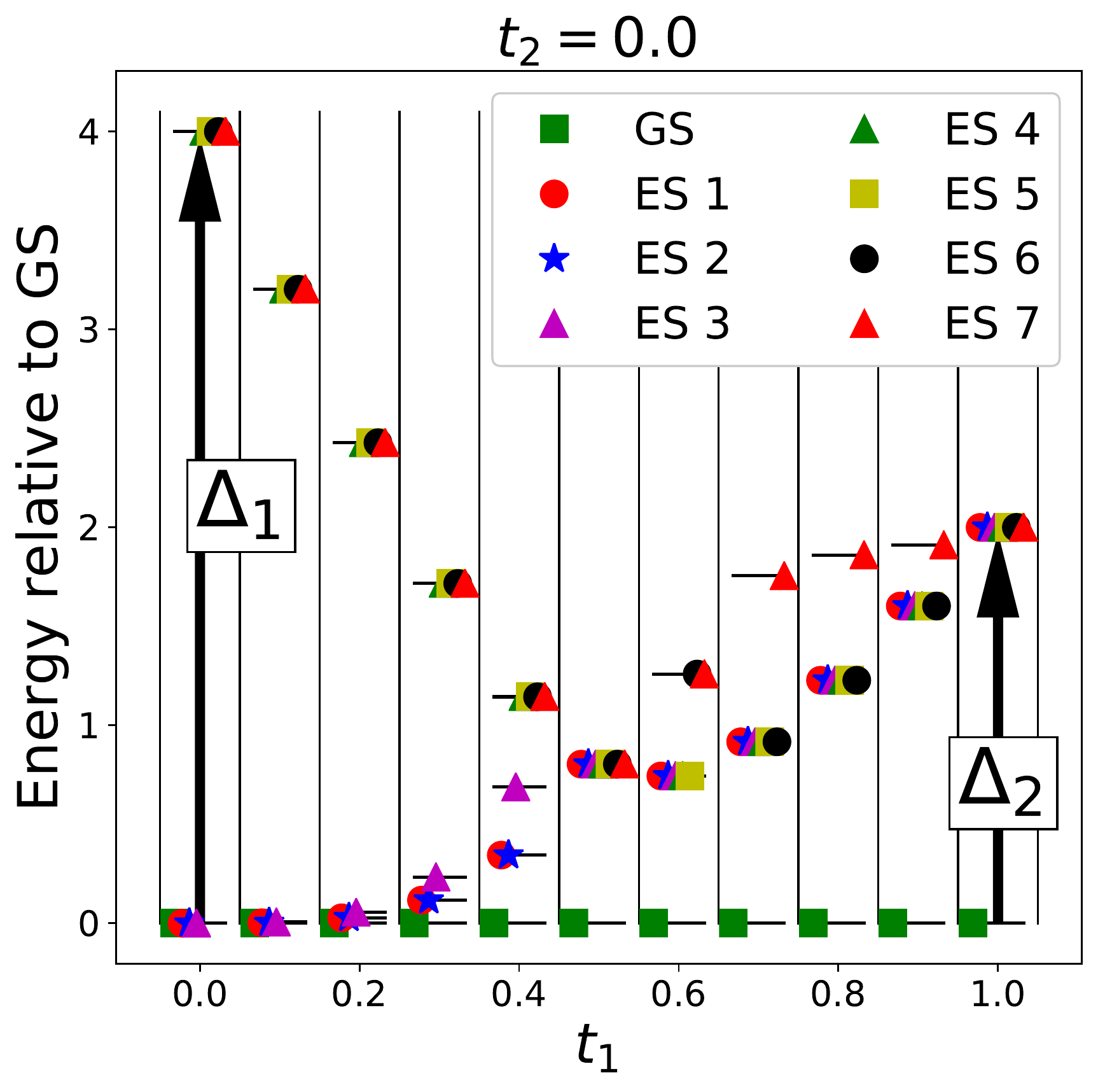}}
\subfigure[]{\includegraphics[scale=.251]{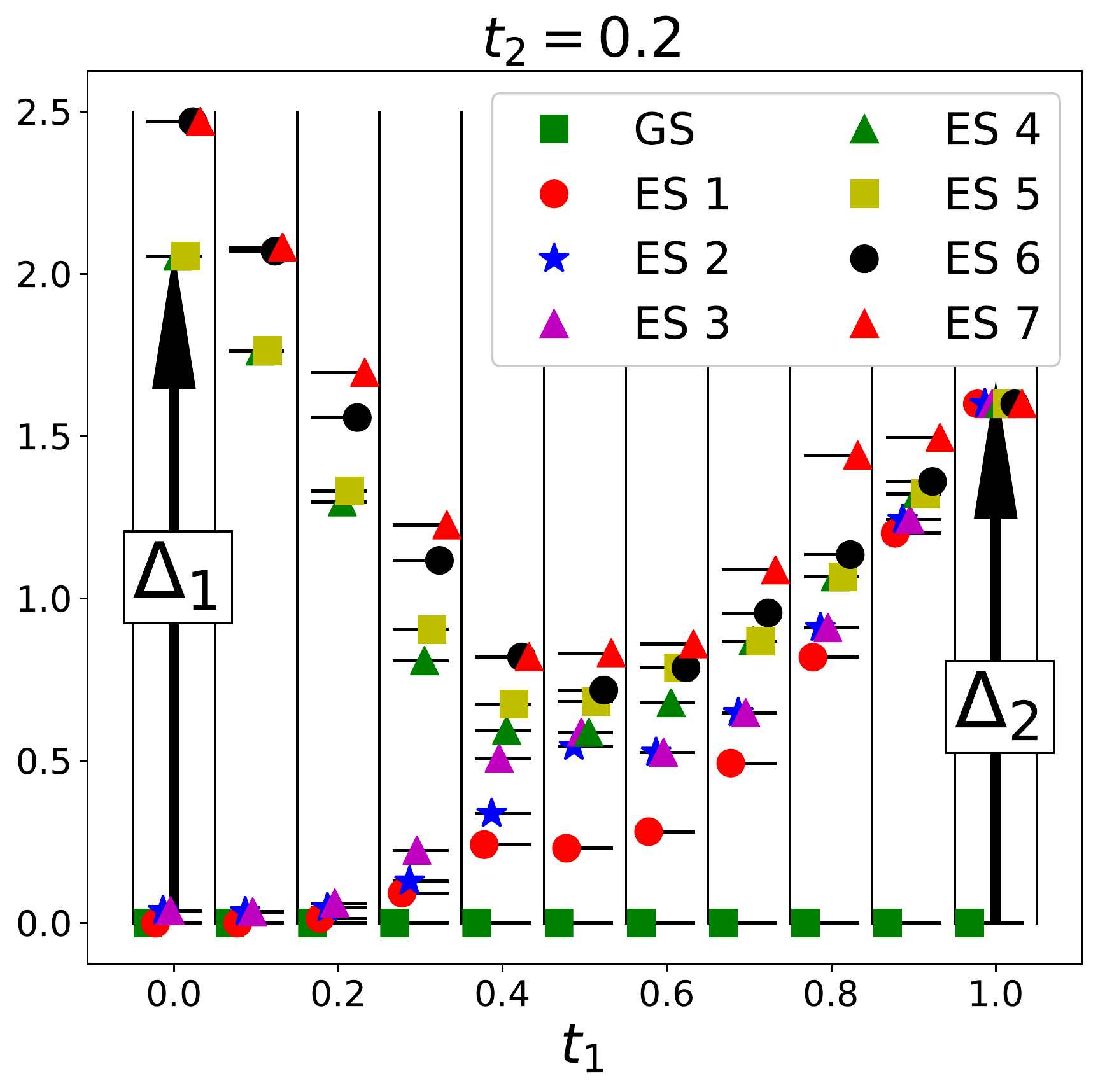}}
\subfigure[]{\includegraphics[scale=.251]{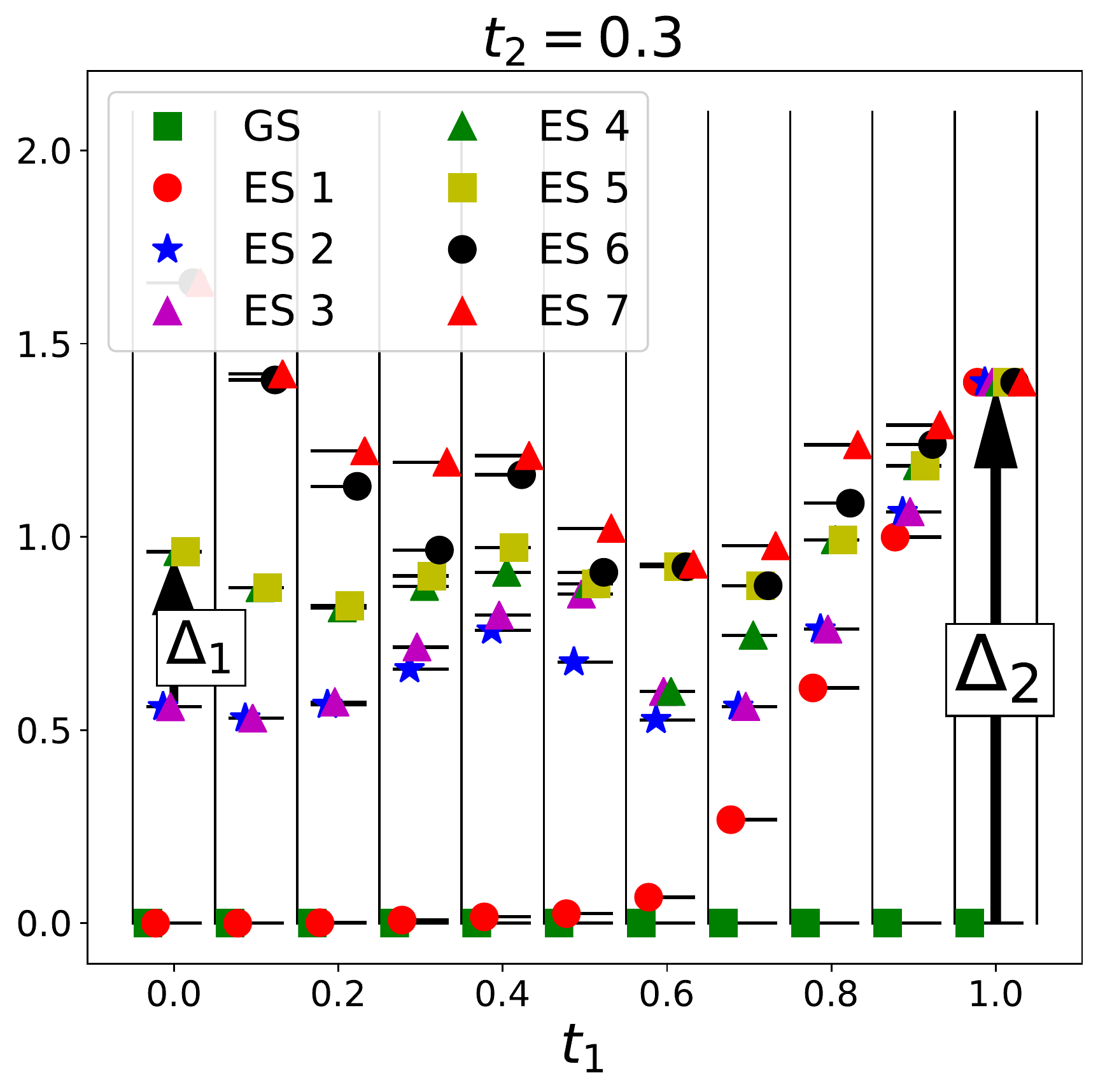}}
\subfigure[]{\includegraphics[scale=.251]{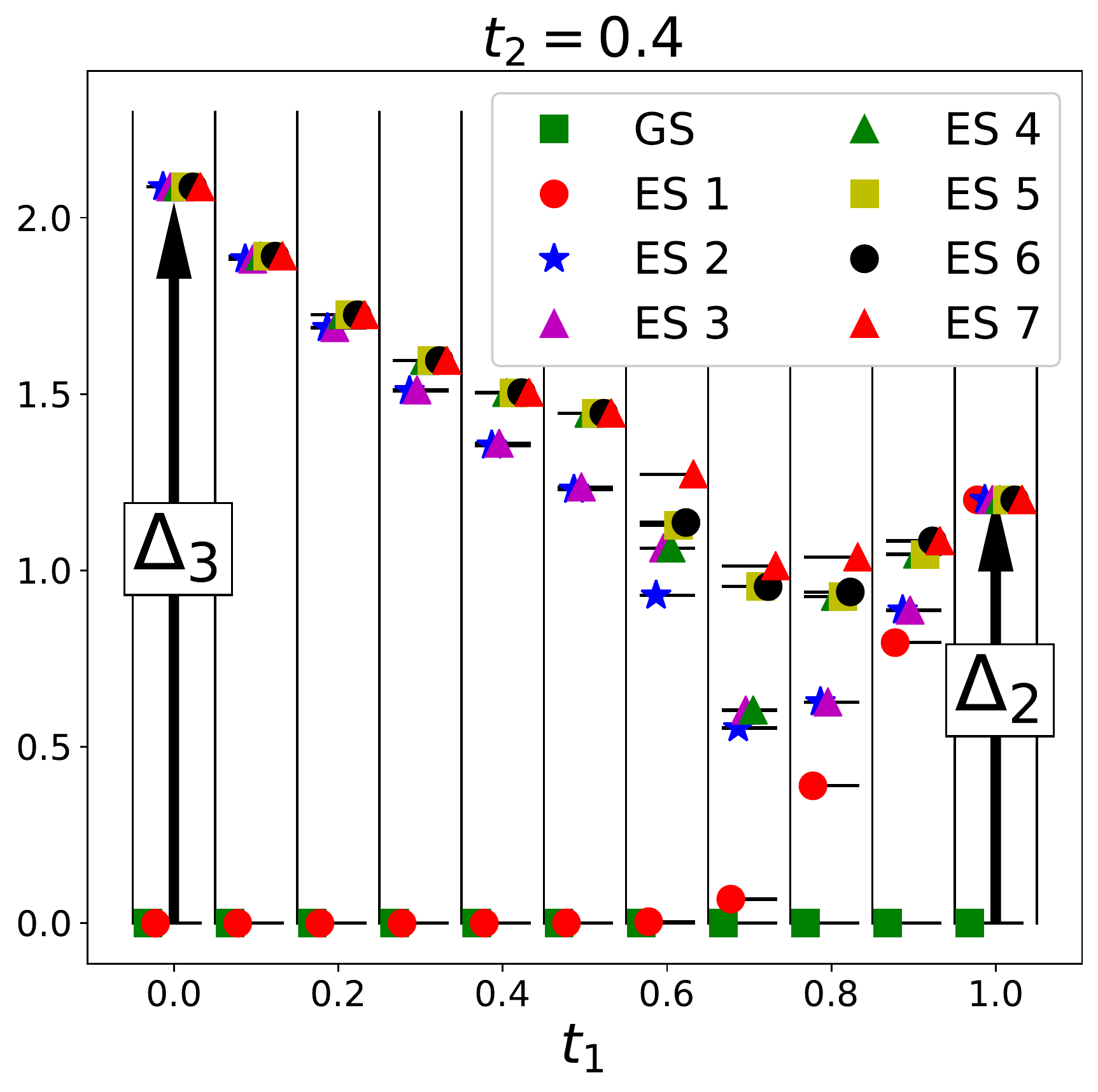}}
\caption{Peak in the absolute value of $\frac{\partial^2 E_{0}}{\partial t^2_1}$, for different $t_2$s. {\bf (a)} $t_2=0.0$ {\bf (b)} $t_2=0.2$ {\bf (c)} $t_2=0.3$ {\bf (d)} $t_2=0.4$. The blue and red dot are for the system size $3\times3$  and $2\times3$ respectively, we see the height of the peak increases as we increase the system size. In the thermodynamic limit the peaks are expected to diverge. Change of excitation gaps along $t_1$, for different $t_2$'s. {\bf (e)} $t_2=0.0$ {\bf (f)} $t_2=0.2$ {\bf (g)} $t_2=0.3$ {\bf (h)} $t_2=0.4$. GS (ES $j$) stands for ground state (jth excited state). $\Delta_1,~\Delta_2~\&~\Delta_3$ are the excitation gaps above the $Z_2$ QSL, paramagnetic (PM) and ferromagnetic (FM) GS respectively.}
\label{fig_33gs_sus_t1_dt2}
\end{figure*}

\section{Numerical Results}
\label{sec_numerics}

We perform exact diagonalisation calculations on finite spin clusters\cite{weinberg2017quspin,weinberg2019quspin}. For the present purpose, we focus on the third quadrant of the phase diagram (Fig. \ref{fig_pd}) while other details will be discussed in a follow-up work.\cite{animesh_afm} For this, we  take the minimal Hamiltonian from Eqs. \ref{eq_fm_heisenberg_min} and \ref{eq_pseudo-dipolar} which captures the leading perturbations to the QSL (Eq. \ref{eq_toric_code}) arising due to the Heisenberg and the pseudo-dipolar terms. The Hamiltonian that interpolates between the different limits is given by
\begin{equation}
\begin{aligned}
\mathcal{H}(t_1,t_2) = & - (1-t_1)(1-t_2)\mathcal{H}_{TCM}  \\
& - t_1(1-t_2)\mathcal{H}_{\Gamma} - t_2(1-t_1)\mathcal{H}_{zz}
\end{aligned}
\end{equation} 
where to compare with the couplings introduced above, we note $\mathcal{H}_{TCM} \equiv \frac{1}{J_{TC}}\mathcal{H}^{F}_{J=\Gamma=0}$, $\mathcal{H}_{\Gamma} \equiv \frac{1}{2\Gamma}\mathcal{H}^{F}_{J=K=0}$ and $\mathcal{H}_{zz} \equiv \frac{1}{J}\mathcal{H}^{F}_{\Gamma=K=0}$, defined in Eqs. \ref{eq_toric_code}, \ref{eq_fm_heisenberg_min} and \ref{eq_pseudo-dipolar} respectively and $t_1=\frac{2\Gamma}{J_{TC}+2\Gamma}~;~t_2=\frac{J}{J_{TC}+J}$.

In this parameter space, at the points $(t_1,t_2)=(0,0),(1,0),(0,1)$ the $\mathcal{H}(t_1,t_2)$ becomes $\mathcal{H}_{TCM}$, $\mathcal{H}_{\Gamma}$ and $\mathcal{H}_{zz}$ respectively.  We perform exact diagonalization (ED) for $2\times3,~3\times3,~5\times2,~4\times3,~5\times3~$ and $4\times4$ clusters with periodic boundary conditions (PBC) such that they contain $12-32$ spins. 

To make an estimate of the phase boundaries in the system, we calculate ground state fidelity susceptibility and spectral gap. The numerical results for representative parameter sets, as discussed below, are plotted in Fig. \ref{fig_33gs_sus_t1_dt2}. The three different phases are then characterised by calculating the Topological entanglement entropy\cite{kitaev2006topological,levin2006detecting} that characterise the $Z_2$ QSL, the magnetisation, $\langle\tau^y\rangle$, that characterises the trivial paramagnet and the two point correlator $\langle\tau^z_i\tau^z_j\rangle$ that characterises the ferromagnet. These are then plotted in representative parameter regimes in Fig. \ref{fig_tee_mag}.  A combination of the above signatures result in the phase diagram given by Fig. \ref{fig_pd_numerics} which should then be compared with the third quadrant of the schematic phase diagram in Fig. \ref{fig_pd}. 

\begin{figure}
\centering
\subfigure[]{\includegraphics[scale=.35]{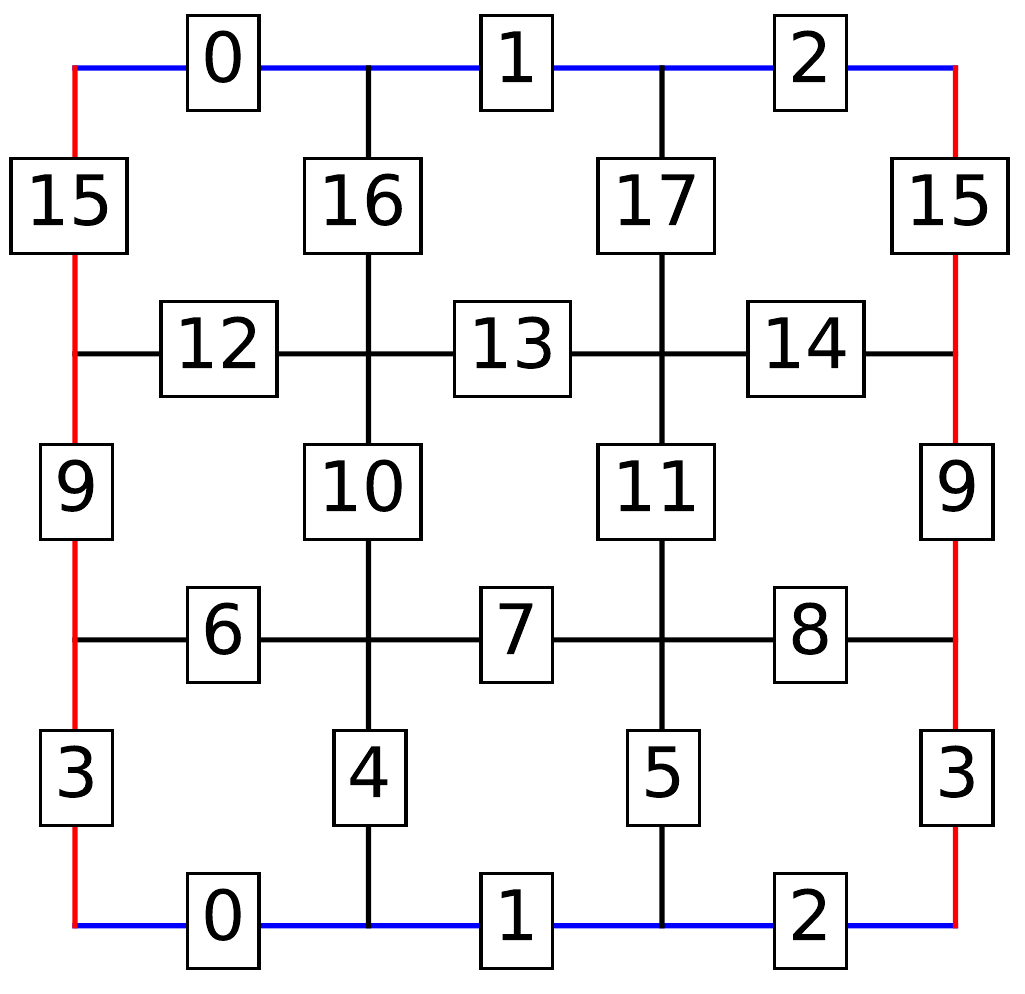}}
\subfigure[]{\includegraphics[scale=.35]{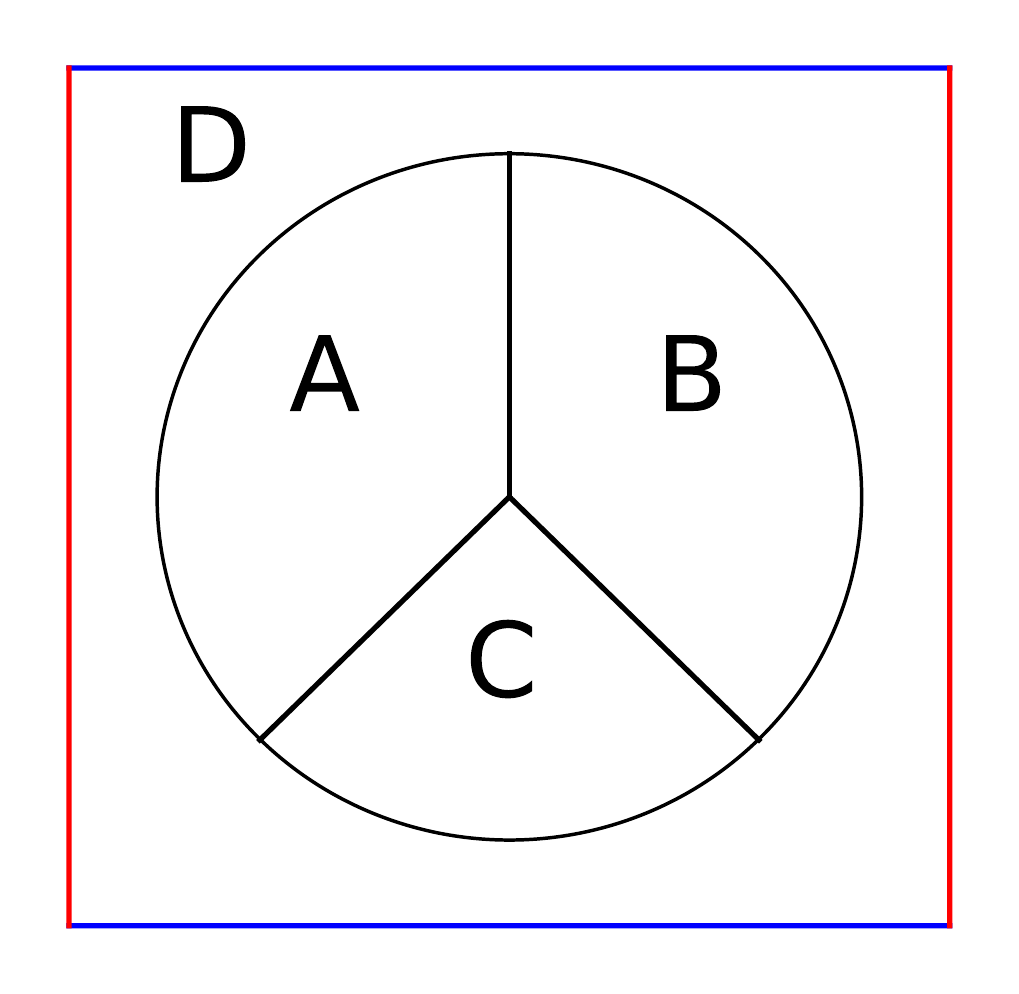}}
\caption{{\bf (a)} The $3\times 3$ cluster with 18 spins, the blue and red edges are identified due to PBC {\bf (b)} Geometry of the four sub-systems, for the calculation of the topological entanglement entropy.}
\label{fig_ed_details}
\end{figure}

\begin{figure}
\centering
\subfigure[]{\includegraphics[scale=.325]{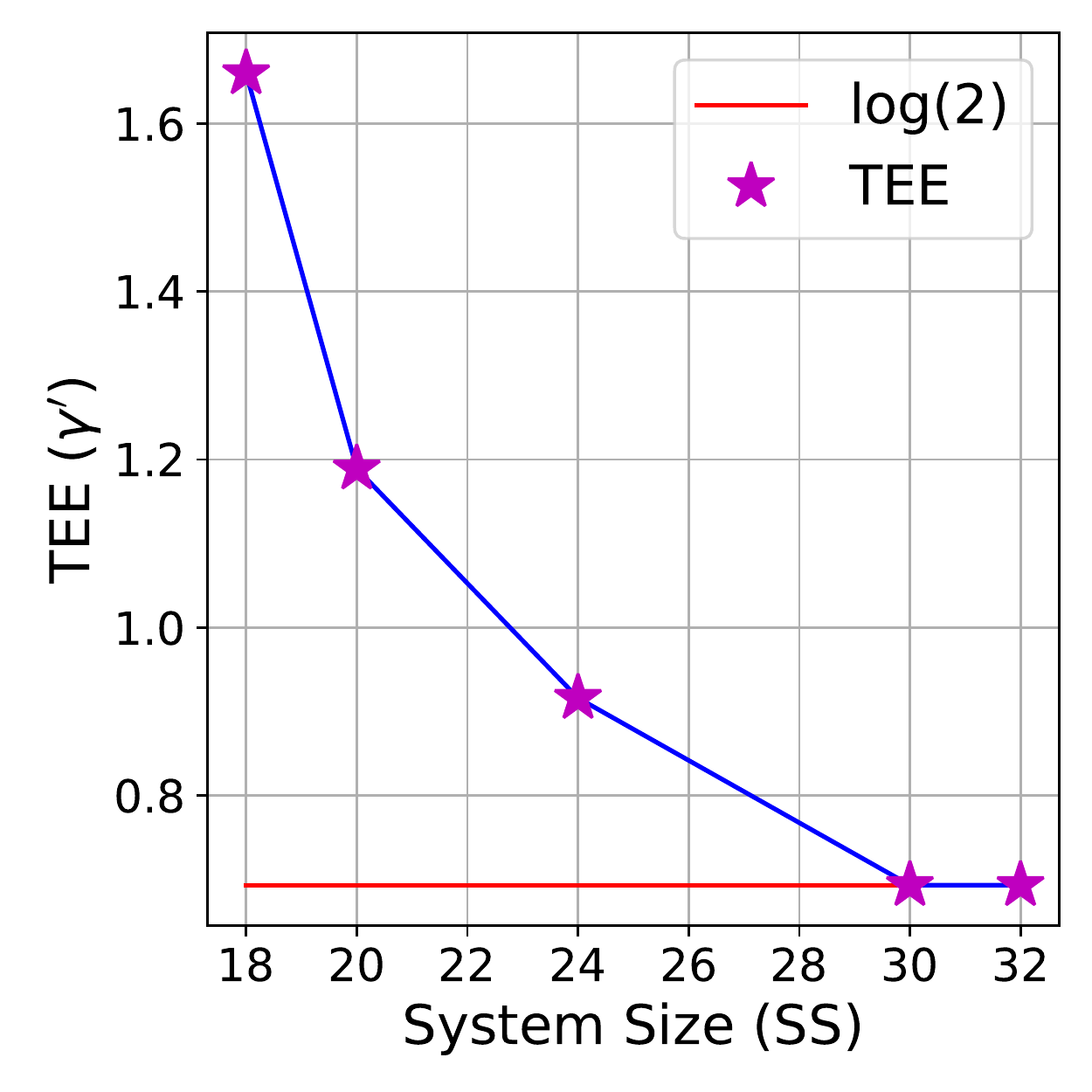}}
\subfigure[]{\includegraphics[scale=.325]{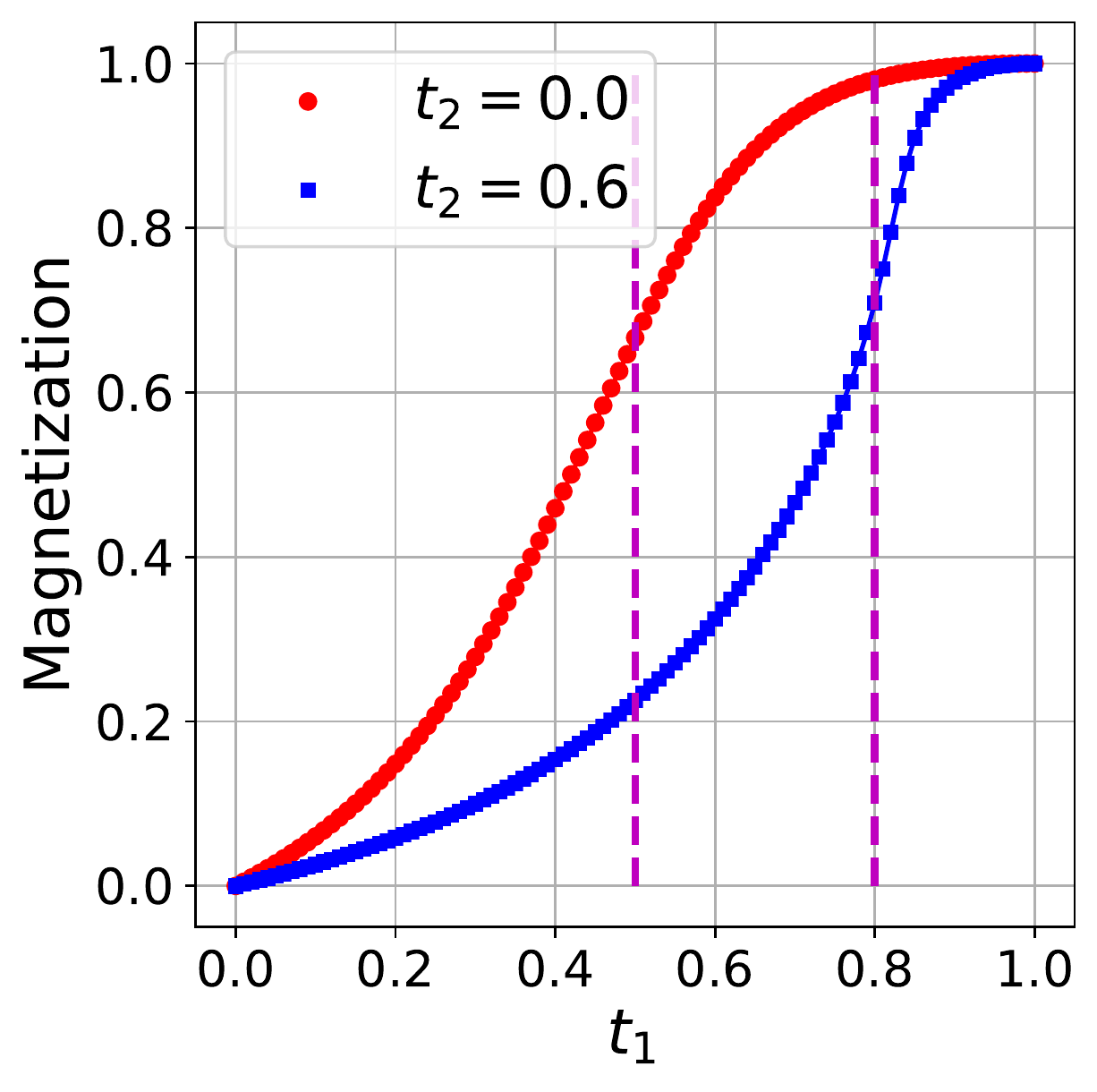}}
\subfigure[]{\includegraphics[scale=0.45]{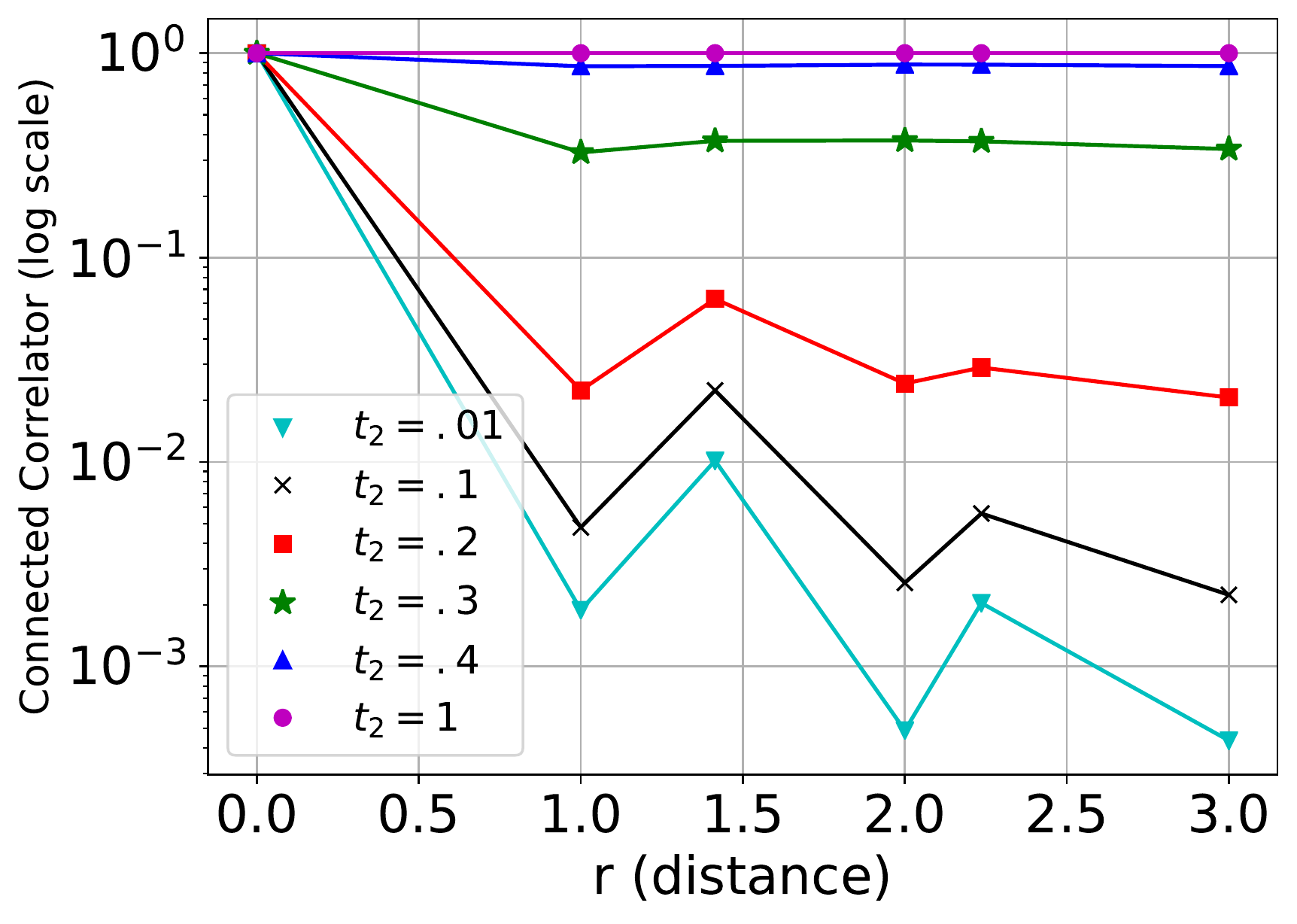}}
\caption{{\bf (a)} The topological entanglement entropy (TEE) for increasing system sizes (see text). The considered system sizes are $3\times3,~5\times2,~4\times3,~5\times3~{\rm and}~4\times4$ which have 18, 20, 24, 30 \& 32 spins respectively. Only the $3\times3$ cluster is shown in Fig. \ref{fig_ed_details}(a).  {\bf (b)} Magnetization in the $\Gamma$-direction, as function of $t_1$ for constant $t_2=0.0~\&~0.6$, along the red dashed lines in the Fig. \ref{fig_pd_numerics}. The dashed magenta line shows the phase transition points along $t_1$, obtained from the phase diagram in Fig. \ref{fig_pd_numerics} for the respective $t_2$ values. {\bf (c)} Plot of normalized correlation function as a function of distance for different values of $t_2$ (with $t_1=0$, see text for more details). Deep inside the FM phase the correlation does not decay, beyond $t_2=0.3$ the correlation decays exponentially.}
\label{fig_tee_mag}
\end{figure}

\begin{figure}
\centering
\includegraphics[scale=0.45]{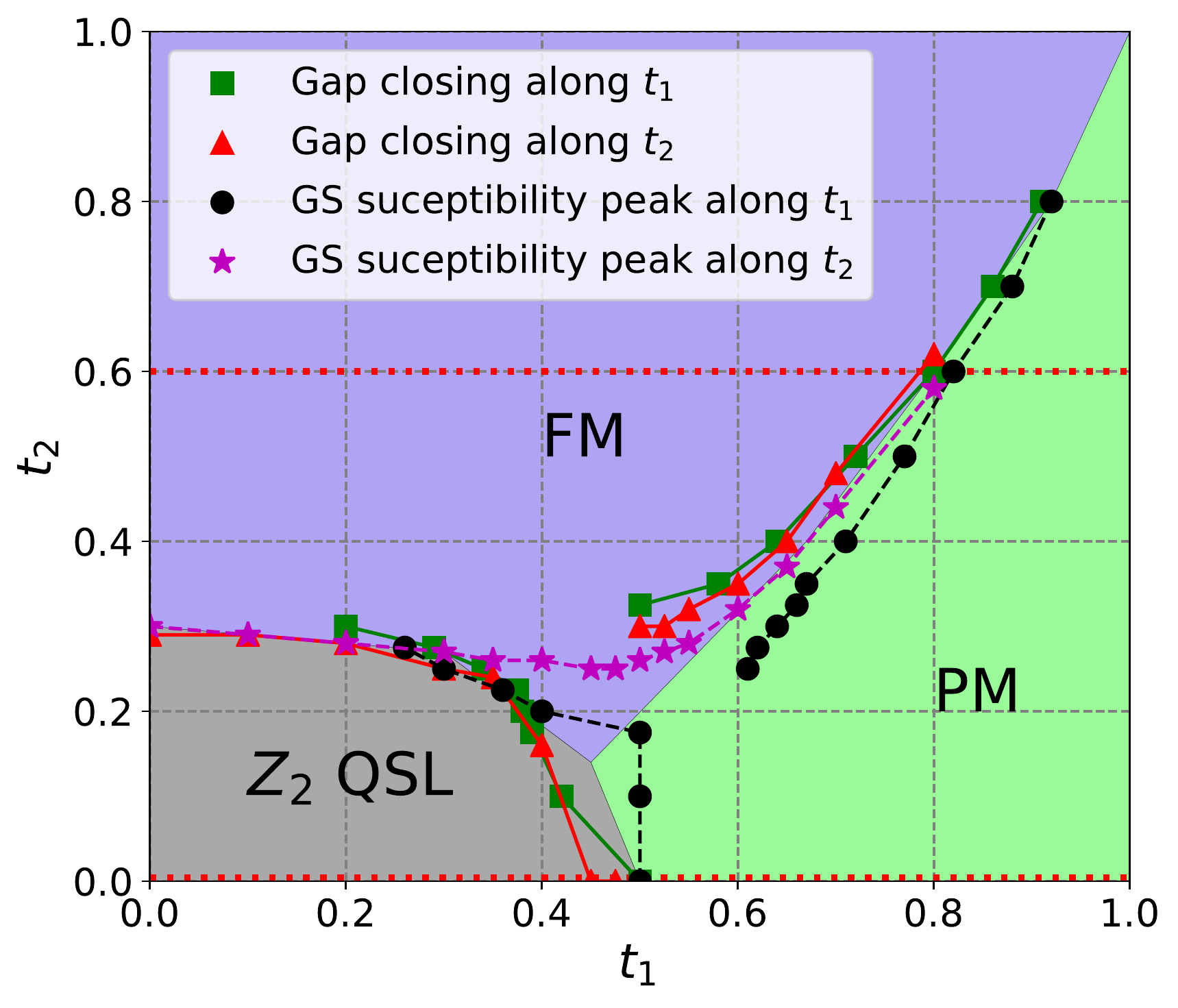}
\caption{Numerically obtained phase diagram focusing on the third quadrant of the Fig. \ref{fig_pd}. The phase transition points are obtained from the gap analysis and the GS susceptibility peak for the cluster $3\times 3$ (Fig. \ref{fig_ed_details}(a)). The magnetization plots in Fig. \ref{fig_tee_mag}(b) are along the red dashed lines: $t_2 = 0.0~\&~0.6$.}
\label{fig_pd_numerics}
\end{figure}

\paragraph{Fidelity Susceptibility :} An estimate of the phase boundaries can be obtained from the study of the response of the ground state energy  due to the change in the parameters $t_1$ and  $t_2$ through the fidelity susceptibility\cite{yu2009fidelity}: $\chi_\lambda=-\partial^2E_{GS}/\partial\lambda^2$ (with $\lambda=t_1, t_2$). In the Fig. \ref{fig_33gs_sus_t1_dt2} ((a)-(d)) we plot $|\chi_{t_1}|$ as a function of $t_1$ for four different representative values of $t_2$. The peaks, which increase with system size  indicate possible phase transitions. Similar peaks are observed in $\chi_{t_2}$ (not shown). The position of the peaks is plotted in Fig. \ref{fig_pd_numerics} which gives an estimate of the phase boundaries. 

\paragraph{Ground state degeneracy and spectral gap :} A related way to characterize the phase boundaries is obtained by tracking the closing of the spectral gap. The corresponding results are shown in Fig. \ref{fig_33gs_sus_t1_dt2} ((e)-(h)) as a function of $t_1$ for the same values of $t_2$ as $\chi_{t_1}$ figures in the upper panel. 

$t_1=0,t_2=0$ corresponds to the exactly solvable Toric code limit which has $Z_2$ QSL ground state, similarly at $(t_1=1,t_2=0)$, we have spin polarized ground state (paramagnetic phase).  In the Toric code limit the system is expected to have four fold degenerate ground state. In the spin polarized limit, there is no GS degeneracy. The gap closing gives us an estimate of the transition which is again plotted in the numerical phase diagram of Fig. \ref{fig_pd_numerics}. The general agreement of the susceptibility data and the gap data is noticeable. 

Fig. \ref{fig_33gs_sus_t1_dt2}(e) shows for $t_2=0$ the evolution of gap at different $t_1$. At $t_1=0$ we have the exactly solvable Toric code model with a $Z_2$ QSL ground state with  the gap scale is $\Delta_1$, above the four fold degenerate ground state which is exactly equal to $4$ for the pure Toric code model in accordance with the expectation. The gap closes around $t_1=0.5$ and towards $t_1=1$ another gap, $\Delta_2$ opens up, which is above the trivial spin polarized paramagnetic ground state. In the \ref{fig_33gs_sus_t1_dt2}(f) and \ref{fig_33gs_sus_t1_dt2}(g), for $t_2=0.2$ and $0.3$ respectively, the size of the gap and the closing point along $t_1$ changes significantly. In both the cases the perturbation to the Toric code model lifts the four fold degeneracy of the topologically ordered QSL ground state via finite size effects. Finally in \ref{fig_33gs_sus_t1_dt2}(h), at $t_2=0.4$, the two fold degeneracy at $t_2\sim 0$ originates from the two possible time reversal partners describing the ferromagnetic state which spontaneously breaks time reversal symmetry as discussed in the previous section. The gap above the ground state manifold is given by $\Delta_3$.  At $t_1 \approx 0.7$ this gap closes so that the system goes into paramagnetic phase signalled by the unique time reversal symmetric ground state as seen in the figure. 

Having gotten an estimate of the phase boundaries, we now turn to further characterisation of the phases.

\paragraph{Topological Entanglement Entropy :} In the $Z_2$ QSL, the entanglement entropy ($S_A$) between a sub-system ($A$), and its compliment ($\bar{A}$) follows the area law with a sub-leading topological correction given by\cite{kitaev2006topological,levin2006detecting}
\begin{equation}
S_{A}(L)=\alpha L -\tilde{\gamma}(L)~;~\tilde{\gamma}(L)=\log(2)+O(1/L)
\end{equation}
Where $\alpha$ is a non-universal constant and $L$ is the length of the boundary between $A~\&~\bar{A}$. In the limit $L\rightarrow\infty$ the TEE saturates to $\log(2)\approx 0.693$. By partitioning the whole system into 4 sub-systems in a particular way, as shown in Fig. \ref{fig_ed_details}(b), the constant part of the TEE can be extracted as\cite{kitaev2006topological,levin2006detecting}
\begin{equation}
\begin{aligned}
& -\gamma^{\prime}= S_{A}+S_{B}+S_{C}+S_{ABC}-S_{AB}-S_{BC}-S_{AC} \\
&~~~~~~~~~~~~~ = \log(2) + \sum_{\beta}O(1/L_{\beta})
\end{aligned}
\end{equation}
where $\beta$ is the different choices of combinations of the sub-system, such as $\beta = A,~AB$ and so on. In Fig. \ref{fig_tee_mag}(a) the $\gamma^{\prime}$ is shown as a function of increasing system (SS) size, which is denoted by the total number of spins in a cluster, for the higher system size TEE saturates to $\log(2)$. The clusters considered here are  $3\times3,~5\times2,~4\times3,~5\times3~\&~4\times4$ which have 18, 20, 24, 30 \& 32 spins respectively. For the smallest system with 18 spins, the sub-systems (A, B, C in Fig. \ref{fig_ed_details}(b)) has 3-4 spins, whereas for the largest system size considered here with 32 spins, the sub-systems has 7-8 spins.

\paragraph{Transverse magnetisation along $\Gamma$ :} To characterise the trivial paramagnet, we calculate the magnetisation along $\Gamma$, {\it i.e.} $\langle\tau_i^y\rangle$. For two representative values of $t_2=0.0~{\rm and}~0.6$, this has been plotted as a function of $t_1$ in Fig. \ref{fig_tee_mag}(b). In the $(t_1,t_2)$ parameter space, these are along the red dotted lines in the Fig. \ref{fig_pd_numerics}. For both the values of $t_2$, in the limit $t_1=1$, the system is in PM phase, where the magnetization saturates to 1. The magnetization decreases along with the decreasing $t_1$, eventually being zero in the limit $t_1=0$. However for two different values of $t_2$, the magnetization changes differently. From the Fig. \ref{fig_pd_numerics}, we see for $t_2=0.0~(0.6)$ the phase transition is around $t_1\approx0.5~(0.8)$, the magenta lines in Fig. \ref{fig_tee_mag}(b) denote the corresponding $t_1$ values for phase transition.

\paragraph{The two point correlator for the ferromagnetic order parameter :} To characterize the ferromagnet, connected correlator $c(r)=\langle\tau^z_i\tau^z_{i+r}\rangle-\langle\tau^z_i\rangle\langle\tau^z_{i+r}\rangle$ is used where $\langle...\rangle$ denotes the ground state expectation value. The normalised $c(r)$ is plotted as a function of distance in Fig. \ref{fig_tee_mag}(c) for different values of $t_2$ with $t_1$ being zero. This is along the $t_2$ axis of Fig. \ref{fig_pd_numerics}. In the FM phase, starting from the $t_2=1$ until $t_2=0.4$ the spins are correlated. Bellow $t_2=0.3$, the correlation falls off exponentially, however due to small system size it is difficult to extract the correlation length.

The above exact diagonalisation, is severely limited by system size. However, it has well understood limits.  The results  suggest possibility of direct transitions out of the $Z_2$ QSL into the symmetry broken ferromagnet or the symmetric trivial  paramagnet. The results are summarised in Fig. \ref{fig_pd_numerics} which is in rough agreement with the expectation of Fig. \ref{fig_pd}. 

In the rest of this work, we present our understanding of the unconventional quantum phase transitions assuming that they are continuous. To successfully describe the transition, we need to obtain a description of the non-trivial low energy excitations of the QSL and their behaviour determines the critical theory. This naturally takes the form of a gauge theory coupled with matter matter fields.

\section{Gauge theory description of the phases and phase transitions}
\label{sec_gauge_th}

As the first step towards the gauge theory description we find it convenient to separate the $e$ and $m$ charges and this is done by rotating the spins as outlined in Eq. \ref{eq_rot_ver}. 

Following usual techniques,\cite{PhysRevLett.98.070602,PhysRevB.91.134419} we introduce the Ising variables, $\mu^\alpha$, on the sites and $\rho^\alpha$ on the bonds of the square lattice (Fig. \ref{fig_toric}) as follows :
\begin{align}
\ttau_{i}^z=\mu^x_a\rho^z_{ab}\mu^x_b, \quad\quad\quad\ttau^x_{i}=\rho^x_{ab}
\label{eq_tauz}
\end{align}
with the Gauss's law constraint
\begin{align}
\prod_{b\in +_a}\rho^x_{ab}=\mu^z_a=\prod_{i\in+_a}\ttau^x_{i}
\label{gauss}
\end{align}
where $ a(\equiv(a_x,a_y))$ and $b$ denote sites the square lattice (fig. \ref{fig_toric}) joined by the bond $i$ where $\ttau_i$ sits. For N sites, there are $2N$, $\tau^\alpha$-spins sitting on the bonds. Hence the total dimension fo the Hilbert space is $2^{2N}$. In terms of the gauge theory, there are $N$ $\mu$-spins and $2N$ $Z_2$ gauge potentials $\rho$ leading to a total degree of freedom of $2^{N}\times 2^{2N}$ which form a  redundant description. However for each site there is one Gauss's law constraint (Eq. \ref{gauss}) leading to $\ \frac{2^{2N}\times2^{N}}{2^{N}}=2^{2N} $ physical degree of freedom equivalent to that of the $\tau$ spins. Thus the above mapping leads to a faithful representation.

The physical picture for the above mapping is easy to understand. The Gauss's law shows that $\mu_a^z=+1(-1)$ denotes the absence (presence) of an $e$ charge at the sites of the square lattice in Fig. \ref{fig_toric}. Therefore, $\mu^x_a$ are creation/annihilation operators for  $e$ charges at the sites and $\rho^x_{ab}$ are the electric fields of the Z$_2$ gauge theory whose flux is related to the density of the electric charges $\mu_a^z$ through the Gauss's law. Finally, from Eq. \ref{eq_tauz}, we get
\begin{align}
\prod_{i\in\Box}\ttau^z_{i}=\prod_{ab\in\Box}\rho^z_{ab}
\end{align}
This is nothing but the $m$ excitations which is now given by the lattice curl of the Z$_2$ gauge potential.

At this point it is useful to also introduce the dual gauge fields where the $m$ charges are explicit. This is obtained using the standard $Z_2$ version of the electromagnetic duality\cite{hansson2004superconductors} 
\begin{align}
\ttau^x_{i}=\tilde\mu^x_{\ba}\tilde\rho^z_{\ba\bb}\tilde\mu^x_{\bb};~~~~~~~\ttau^z_{i}=\tilde\rho^x_{\ba\bb}
\label{eq_dualmap}
\end{align}
where the $m$ charge creation operators, $\tilde\mu_{\ba}^x$ are now defined on the sites of the dual lattice, denoted by ${\bar{a}}\equiv T_{\bf d_1}({a})\equiv(\bar{a}_x,\bar{a}_y)$ and ${\bb}$ (we use the bar above the symbol to denote dual lattice sites), obtained by joining the centres of the direct square lattice of Fig. \ref{fig_toric} such that in the above expression the bond of the direct lattice denoted by $i$ is bisected by the dual bond joining the sites ${\ba}$ and ${\bb}$. The dual gauge fields, $\tilde\rho^\alpha_{\ba\bb}$, reside on the links of the dual lattice and the dual Gauss's law is given by
\begin{align}
\prod_{{\bb}\in+_{\ba}}\tilde\rho^x_{\ba\bb}=\tilde\mu^z_{\ba}=\prod_{i \in \Box_{\ba}}\ttau^z_{i}
\end{align}

Therefore in this dual representation, $\tilde\rho^x_{\ba\bb}$ is the magnetic field of the $Z_2$ gauge theory whose divergence is equal to the $m$ charge $\tilde\mu_{\ba}^z$ at the site of the dual plaquette. As previously, the dual representation along with the dual Gauss' s law span the physical Hilbert space. This is further clear by the relation between the direct and the dual degrees of freedom which is obtained by comparing Eq. \ref{eq_tauz} and \ref{eq_dualmap}, which gives 
\begin{align}
\rho^x_{ab}=\tilde\mu^x_{\ba}\tilde\rho^z_{\ba\bb}\tilde\mu^x_{\bb}
\label{eq_dualefield}
\end{align}
where the ($ab$) bond on the direct lattice bisect the dual bond $({\ba\bb})$, and
\begin{align}
\prod_{b\in+_a}\rho^x_{ab}&=\mu^z_a=\prod_{\langle {\ba\bb}\rangle \in \Box_a}\tilde\rho^z_{\ba\bb}\nonumber\\
\prod_{{\bb}\in+_{\ba}}\tilde\rho^x_{\ba\bb}&=\tilde\mu^z_{\ba}=\prod_{\langle ab\rangle\in \Box_{\ba}}\rho^z_{ab}
\end{align}
The last equation encode that while the $e$ and $m$ charges are bosons, they see each other as source of $\pi$ fluxes. In fact, these equations are actually not independent but are related to each other through duality.

We can use either of the representations discussed above. However, it is often useful to introduce both the charges explicitly, each coupled to its own gauge field and the mutual semionic statistics is then represented by a mutual $Z_2$ Chern-Simons (SC) action or, \cite{PhysRevB.62.7850,PhysRevB.84.104430} in the continuum limit,  a mutual $U(1)\times U(1)$ CS theory.\cite{freedman2004class,PhysRevB.71.235102,PhysRevB.78.155134,xu2009global}

\subsubsection{Action of the symmetries on the gauge charges and the gauge fields}

Having expressed the elementary excitations, the gauge charges, of the QSL, we now turn to the action of symmetries on them. From Eq. \ref{eq_tausym}, we get the symmetry transformations of the rotated spins $\ttau$s using Eq. \ref{eq_rot_ver} (Table \ref{table_tau_symm_fm_h}  in Appendix  \ref{appen_symm}).

\paragraph{\underline{Lattice Translations } :}\label{para_gauge_symm_tran} 
Under both the translations, along the directions ${\bf d}_1$ and ${\bf d}_2$ (see Fig. \ref{fig_toric}), the plaquettes and the vertices are interchanged. Hence the $e$ and $m$ charges are interchanged (the original square lattice and its dual gets interchanged). This is thus an example of an anyon permutation symmetry.\cite{PhysRevB.87.104406} The translation symmetry acts on the gauge degrees of freedom in the following manner.
\begin{equation}
\begin{aligned}
 T_{\bf d_j}: \begin{array}{l} \{\mu^x,\mu^z\}_{a} \rightarrow \{\tilde{\mu}^x,\tilde{\mu}^z\}_{T_{d_j}({a})} \\
  \{\tilde{\mu}^x,\tilde{\mu}^z\}_{\bar{a}} \rightarrow \{\mu^x,\mu^z\}_{T_{d_j}({\bar{a}})}  \\
\{\rho^x,\rho^z\}_{ab} \rightarrow \{\tilde{\rho}^x,\tilde{\rho}^z\}_{T_{d_j}({ab})} \\
\{\tilde{\rho}^x,\tilde{\rho}^z\}_{\bar{a}\bar{b}} \rightarrow \{\rho^x,\rho^z\}_{T_{d_j}({\bar{a}\bar{b}})}
\end{array}
\end{aligned}\label{eq_gauge_trans_transl}
\end{equation}

For translation along the cartesian axes, the lattice vectors are given by $\hat x={\bf d_1-d_2}$ and $\hat y={\bf d_1+d_2}$. Under this, the gauge charges and potentials transform as
\begin{equation}
\begin{aligned}
T_{\hat x(\hat y)} : \begin{array}{l}\{\mu^x,\mu^z\}_{a} \rightarrow \{\mu^x, \mu^z\}_{a+\hat x(\hat y)} \\
\{\tilde{\mu}^x,\tilde{\mu}^z\}_{\bar{a}} \rightarrow \{\tilde{\mu}^x,\tilde{\mu}^z\}_{\bar{a}+\hat x(\hat y)}  \\
\{\rho^x,\rho^z\}_{\bar{a}\bar{b}} \rightarrow \{\rho^x,\rho^z\}_{\bar{a}+\hat x(\hat y),\bar{b}+\hat x(\hat y)} \\
\{\tilde{\rho}^x,\tilde{\rho}^z\}_{\bar{a}\bar{b}} \rightarrow \{\tilde{\rho}^x,\tilde{\rho}^z\}_{\bar{a}+\hat x(\hat y),\bar{b}+\hat x(\hat y)}
\end{array}
\end{aligned}
\end{equation}

\paragraph{\underline{Time Reversal} :}\label{para_gauge_symm_tr} The bond dependent rotation of Eq. \ref{eq_rot_ver} imply that in the rotated basis, natural to the Toric code QSL, on the vertical bonds, the $\tau^x$ is odd under time reversal, whereas on the vertical bonds $\tau^z$ continues to remain time reversal odd. This endows the gauge charges and the gauge fields non-trivial transformation under time reversal which depends on their spatial location and is given by
\begin{align}
\mathcal{T}:  \begin{array}{l}
\{\mu^x,\mu^z\}_{a}\rightarrow \{\mu^x,\mu^z\}_{a} \\
\{\tilde\mu^x,\tilde\mu^z\}_{\bar{a}}\rightarrow \{\tilde\mu^x,\tilde\mu^z\}_{\bar{a}}  \\
\{\rho^x,\rho^z\}_{ab} \rightarrow \{(-1)^{a_y+b_y}\rho^x,(-1)^{a_x+b_x}\rho^z\}_{ab} \\
\{\tilde{\rho}^x,\tilde{\rho}^z\}_{\bar{a}\bar{b}} \rightarrow \{(-1)^{\bar{a}_y+\bar{b}_y}\tilde{\rho}^x,(-1)^{\bar{a}_x+\bar{b}_x}\tilde{\rho}^z\}_{\bar{a}\bar{b}}\\
\end{array}
\label{eq_gauge_trans_tr}
\end{align}

\paragraph{\underline{Reflections about $z$ bond, $\sigma_v$} :}
\begin{equation}
\begin{aligned}
\sigma_v: \begin{array}{l}
\{\mu^x,\mu^z\}_{a} \rightarrow \{\mu^x,\mu^z\}_{\sigma_v({a})} \\
\{\tilde{\mu}^x,\tilde{\mu}^z\}_{a} \rightarrow \{\tilde{\mu}^x,\tilde{\mu}^z\}_{\sigma_v({a})} \\
\{\rho^x,\rho^z\}_{ab} \rightarrow \{(-1)^{a_y+b_y}\rho^x,(-1)^{a_x+b_x}\rho^z\}_{\sigma_v({ab})} \\
 \{\tilde{\rho}^x,\tilde{\rho}^z\}_{\bar{a}\bar{b}} \rightarrow \{(-1)^{\bar{a}_y+\bar{b}_y}\tilde{\rho}^x,-1)^{\bar{a}_x+\bar{b}_x}\tilde{\rho}^z\}_{\sigma_v({\bar{a}\bar{b}})}\\
 \end{array}
\end{aligned}\label{eq_gauge_trans_sigv}
\end{equation}

\paragraph{\underline{$\pi$-rotation about the $z$-bond, $C_{2z}$} :}  
\begin{equation}
\begin{aligned}
C_{2z}: \begin{array}{l}\{\mu^x,\mu^z\}_{\bf a} \rightarrow \{\mu^x,\mu^z\}_{C_{2z}({a})} \\
\{\tilde{\mu}^x,\tilde{\mu}^z\}_{\bar{a}} \rightarrow \{(-1)^{\bar{a}_x}\tilde{\mu}^x,\tilde{\mu}^z\}_{C_{2z}({\bar{a}})} \\
\{\rho^x,\rho^z\}_{ab} \rightarrow \{(-1)^{a_y+b_y}\rho^x,(-1)^{a_x+b_x}\rho^z\}_{C_{2z}(ab)} \\
\{\tilde{\rho}^x,\tilde{\rho}^z\}_{\bar{a}\bar{b}} \rightarrow \{(-1)^{\bar{a}_y+\bar{a}_y}\tilde{\rho}^x,(-1)^{\bar{a}_x+\bar{b}_x}\tilde{\rho}^z\}_{C_{2z}({\bar{a}\bar{b}})}
\end{array}
\end{aligned}\label{eq_gauge_trans_c2z}
\end{equation}

\paragraph{\underline{$\pi$-rotation honeycomb lattice centre, $R_{\pi}$} :} 
\begin{equation}
\begin{aligned}
R_{\pi}: \begin{array}{l}\{\mu^x,\mu^z\}_{a} \rightarrow \{\mu^x,\mu^z\}_{R_{\pi}({a})} \\
\{\tilde{\mu}^x,\tilde{\mu}^z\}_{\bar{a}} \rightarrow \{\tilde{\mu}^x,\tilde{\mu}^z\}_{R_{\pi}({\bar{a}})} \\
\{\rho^x,\rho^z\}_{ab} \rightarrow \{\rho^x,\rho^z\}_{R_{\pi}({ab})} \\
\{\tilde{\rho}^x,\tilde{\rho}^z\}_{\bar{a}\bar{b}} \rightarrow \{\tilde{\rho}^x,\tilde{\rho}^z\}_{R_{\pi}({\bar{a}\bar{b}})}
\end{array}
\end{aligned}\label{eq_gauge_trans_rpi}
\end{equation}

With this we start to investigate the nature of the phase transition out of the Z$_2$ QSL discussed in the previous section. To this end we begin with the phase transition along the line of vertical and horizontal axes of the phase diagram in Fig. \ref{fig_pd} starting with the transition between the $Z_2$ QSL and the spin-ordered state brought about by the Heisenberg interactions and followed by the description of the transition between the QSL and the trivial paramagnet tuned by the pseudo-dipolar term. Here we note that as indicated previously, we expect that the transition between the ferromagnet and the trivial paramagnet is described by a transverse field Ising model whose transition is well understood and belongs to the well known 3d Ising universality class.

 \section{Phase transition between QSL and the spin ordered phase}\label{sec_phase_tran_tc_heisenberg}

Along the vertical axis of Fig. \ref{fig_pd} at $\Gamma=0$,  there are two competing phases-- the $Z_2$ QSL for  $J\sim 0$ and the spin ordered phase in the Heisenberg limit, $J/|K|\gg 1$. While, as we already described, the QSL can be understood in terms of selective proliferation of domain walls of the spin ordered phase, to understand the phase transition between them, it is much more convenient to start with the QSL and obtain the description of the transition in terms of the soft modes, as a function of $J$, of its excitations-- the $e$ and $m$ charges.

To the leading order in $J$ the pertinent Hamiltonian is given by Eq. \ref{eq_jk_min_model_rot} which generates the dispersion for the localised (in the exactly solvable Toric code limit) bosonic $e$ and $m$ charges eventually resulting in soft-modes which condense to give rise to the spin order as we shall show below.  We neglect the higher order terms in $J$ and later shall return to them to understand their effects.

In terms of the gauge charges of Eq. \ref{eq_tauz}, the Hamiltonian in eq. \ref{eq_jk_min_model_rot} becomes
\begin{equation}
\begin{aligned}
& \tilde{\mathcal{H}}_{\Gamma=0}^F=J\sum_{\langle ab\rangle \in H; \langle bc\rangle\in V}\left[\mu^x_a\rho^z_{ab}\mu^x_b\right]\left[\rho^x_{bc}\right] \\
& ~~~~~~~~~-J_{TC}\sum_a\mu^z_a-J_{TC}\sum_p\prod_{\langle ab\rangle\in p}\rho^z_{ab}
\label{eq_leading}
\end{aligned}
\end{equation}

The second and the third term represents the energy costs for creating $e$ and $m$ charges respectively. Indeed for $J=0$, the theory is nothing but an even Ising gauge theory\cite{sachdev2018topological} that describes the $Z_2$ QSL.

The first term, on the other hand, creates and mobilises both $e$ and $m$ charges. Of central importance for our purpose is the particular form of  hopping term-- both $e$ and $m$ charges, once created, can only disperse along the horizontal directions (with reference to Fig. \ref{fig_toric}) at this leading order of $J$. Somewhat similar effect was observed in dopped isotropic Kitaev model.\cite{PhysRevB.90.035145} The decoupling of various  horizontal electric and magnetic ``chains" lead to a dimensional reduction at this order. However, different such chains, as we shall see below, gets coupled by higher order terms. This generically leads to anisotropic kinetic energy for the $e$ and $m$ charges and hence one expects anisotropic correlation lengths.

\subsection{Gauge mean field theory}\label{subsubsec_gmft}

We start our analysis by decoupling the first term in Eq. \ref{eq_leading} within gauge mean field theory\cite{PhysRevLett.108.037202} where we systematically neglect the gauge fluctuations. A mean field decoupling of the gauge charges and the gauge fields in the $e$ and $m$ sectors for the first term in Eq. \ref{eq_leading} :  $\left[\mu^x_a\rho^z_{ab}\mu^x_b\right]\left[\rho^x_{bc}\right]\rightarrow \langle\mu^x_a\rho^z_{ab}\mu^x_b\rangle\rho^x_{bc} +\mu^x_a\rho^z_{ab}\mu^x_b\langle\rho^x_{bc}\rangle$, gives 
\begin{align}
\tilde{\mathcal{H}}_{\Gamma=0}^F\rightarrow\tilde{\mathcal{H}}_{\Gamma=0}^{\rm GMFT}=\tilde{\mathcal{H}}_{\Gamma=0}^{\rm GMFT}(e)+\tilde{\mathcal{H}}_{\Gamma=0}^{\rm GMFT}(m)
\end{align}
where
\begin{align}
\tilde{\mathcal{H}}_{\Gamma=0}^{\rm GMFT}(e)=\sum_{\langle ab\rangle\in H}J_{ab} \mu^x_a\rho^z_{ab}\mu^x_b-J_{TC}\sum_a\mu_a^z
\label{eq_egmft}
\end{align}
describes the $e$ sector with 
\begin{align}
J_{ab}= J\left[\langle\rho^x_{b,b-\hat y}\rangle+\langle\rho^x_{b,b+\hat y}\rangle+\langle\rho^x_{a,a-\hat y}\rangle+\langle\rho^x_{a,a+\hat y}\rangle\right]
\label{eq_jeff}
\end{align}
being the effective coupling and 
\begin{align}
\tilde{\mathcal{H}}_{\Gamma=0}^{\rm GMFT}(m)=\sum_{\langle \bar{a}\bar{b}\rangle\in H}J_{\bar{a}\bar{b}} \tilde\mu^x_{\bar{a}}\tilde\rho^z_{\bar{a}\bar{b}}\tilde\mu^x_{\bar{b}}-J_{TC}\sum_{\bar{a}}\tilde\mu_{\bar{a}}^z
\label{eq_mgmft}
\end{align}
describes the $m$ sector with
\begin{align}
J_{\bar{a}\bar{b}}=J\left[\langle\tilde\rho^x_{\bar{b},\bar{b}-\hat y}\rangle+\langle\tilde\rho^x_{\bar{b},\bar{b}+\hat y}\rangle+\langle\tilde\rho^x_{\bar{a},\bar{a}-\hat y}\rangle+\langle\tilde\rho^x_{\bar{a},\bar{a}+\hat y}\rangle\right]
\label{eq_jmff}
\end{align}

Clearly, at this order in $J$, the $e$ and $m$ sectors completely decouple into a series of transverse field Ising chains in the horizontal direction in Fig. \ref{fig_toric}. For the horizontal direction, we can choose a gauge 
\begin{align}
\rho^z_{a,a+\hat x}=\tilde\rho^z_{\bar{a},\bar{a}+\hat x}=+1
\label{eq_gaugechoice}
\end{align}
as these links do not cross. The $Z_2$ QSL is then the paramagnetic phase of this decoupled transverse field Ising chains where the $e$ and $m$ charges are both gapped. The Heisenberg term gives kinetic energy to both the $e$ and $m$ charges in the horizontal direction which then develops soft modes which condense to give rise to $\langle\mu^x\rangle\neq 0$  and $\langle\tilde\mu^x\rangle\neq 0$  
for the respective chains. 

\begin{figure}
\subfigure[]{
\includegraphics[scale=0.3]{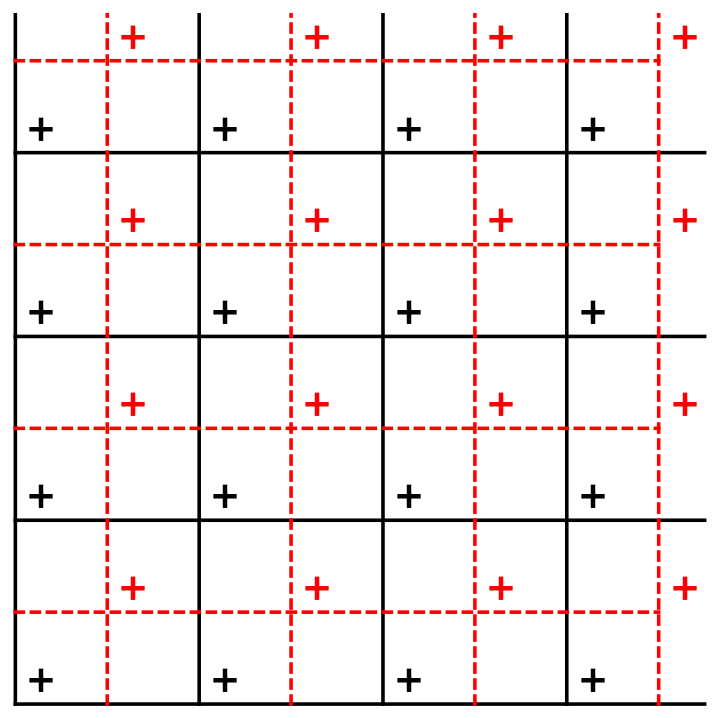}
\label{fig_sm1}
}
\subfigure[]{
\includegraphics[scale=0.3]{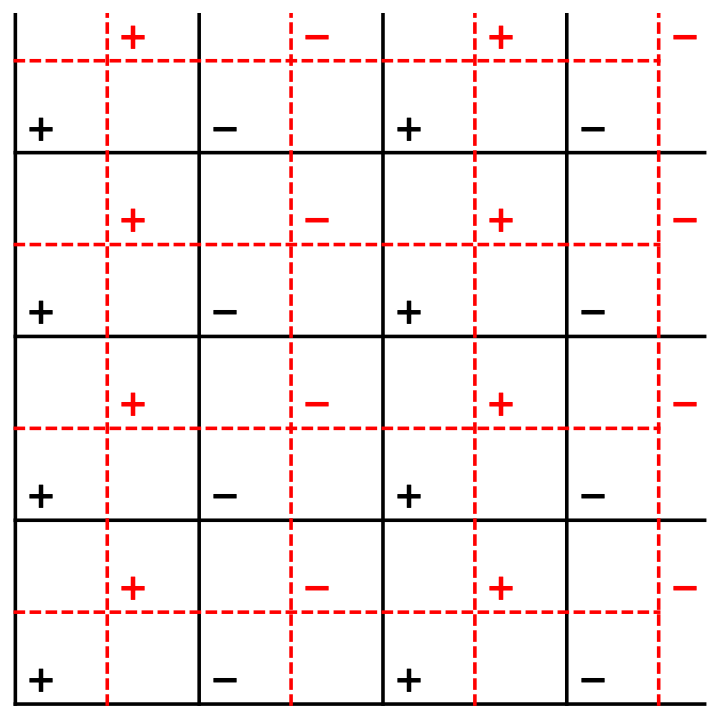}
\label{fig_sm2}
}
\caption{The electric (black) and the magnetic (red) soft modes on the direct and dual lattice respectively. The $\boldsymbol{\pm}$ denotes $\mu^x=\pm 1$ and $\tilde\mu^x=\pm 1$ respectively. Fig. (a) and (b) shows the two time reversal partners respectively, $(\hat{\nu}_e^{(1)},\hat{\nu}_e^{(2)})$ for the electric and $(\hat{\nu}_m^{(1)},\hat{\nu}_m^{(2)})$ the magnetic sectors.}
\end{figure}

For the above gauge the soft mode develops at zero momentum as shown in Fig. \ref{fig_sm1} for both the $e$ and $m$  sectors. This can be denoted by 
\begin{align}
\hat\nu_e^{(1)}=1;~~~~~~\hat\nu_m^{(1)}=1
\end{align}
for the $e$ sector on the direct lattice and $m$ sector on the dual lattice respectively.

Application of time reversal symmetry (see Eq. \ref{eq_gauge_trans_tr}) gives the time reversal partner soft mode for both the $e$ and $m$ sectors as shown in Fig. \ref{fig_sm2} which are given by
\begin{align}
\hat{\nu}_e^{(2)}=e^{i\pi x};~~~~\hat{\nu}_m^{(2)}=e^{i\pi X}
\end{align}
for the $e$ sector and $m$ sectors. The cartesian coordinates of the direct and dual lattices are given by $(x,y)$ and ($X,Y$) with $X=x+1/2$ and $Y=y+1/2$. Other symmetries do not generate any further soft modes and hence the transition out of the QSL into the spin-ordered phase is described in terms of the above soft modes.

\subsection{Soft modes}

The soft mode expansion for the $e$ sector is therefore given by\cite{lannert2001quantum,xu2009global,PhysRevB.84.104430}
\begin{align}
\Psi_e({\bf r},\tau)=\phi_e^{(1)}({\bf r},\tau)~\hat\nu_e^{(1)}+\phi_e^{(2)}({\bf r},\tau)~\hat\nu_e^{(2)}
\label{eq_esm}
\end{align}
where $(\phi_e^{(1)}({\bf r},\tau), \phi_e^{(2)}({\bf r},\tau))$ are real fields that represents amplitudes of the electric soft modes. Similarly, for the $m$ sector, the soft mode expansion is given by
\begin{align}
\Psi_m({\bf r},\tau)=\phi_m^{(1)}({\bf r},\tau)~\hat\nu_m^{(1)}+\phi_m^{(2)}({\bf r},\tau)~\hat\nu_m^{(2)}
\label{eq_msm}
\end{align}
where $(\phi_m^{(1)}({\bf r},\tau), \phi_m^{(2)}({\bf r},\tau))$ are real amplitudes of the magnetic soft modes.

The Higg's phase obtained by condensation of a combination of the above modes is nothing but the spin ordered phase as we shall see below, while the ``uncondensed" phase represents the $Z_2$ QSL. However, due to the non-trivial projective symmetry group (PSG) transformation of the soft modes under various symmetries of the system and due to the non-trivial mutual semionic statistics between the $e$ and the $m$ excitations, the construction of the critical theory requires careful analysis starting with the PSG analysis of the soft mode amplitudes. To this end, it is useful to define the complex soft mode amplitudes
\begin{align}
\Phi_e=\phi_e^{(1)}+i\phi_e^{(2)}=|\Phi_e|e^{i\theta_e}
\label{eq_ecom}
\end{align}
and
\begin{align}
\Phi_m=\phi_m^{(1)}+i\phi_m^{(2)}=|\Phi_m|e^{i\theta_m}
\label{eq_mcom}
\end{align}
where we have suppressed the arguments for clarity. Now, for the different symmetries considered in Eqs. \ref{eq_gauge_trans_transl}-\ref{eq_gauge_trans_rpi}, we have
\begin{align}
&{\bf T_{d_1}} :\left\{\begin{array}{l}
\Phi_e\rightarrow\Phi_m\\
\Phi_m\rightarrow\Phi_e^*\\
\end{array}\right.~~~
&{\bf T_{d_2}} :\left\{\begin{array}{l}
\Phi_e\rightarrow\Phi_m^*\\
\Phi_m\rightarrow\Phi_e\\
\end{array}\right.\nonumber\\
&{\bf T_{x}} :~\left\{\begin{array}{l}
\Phi_e\rightarrow\Phi_e^*\\
\Phi_m\rightarrow\Phi_m^*\\
\end{array}\right.~~~
&~{\bf T_{y}} :~\left\{\begin{array}{l}
\Phi_e\rightarrow\Phi_e\\
\Phi_m\rightarrow\Phi_m\\
\end{array}\right.\nonumber\\
&{\mathcal{T}} :~~\left\{\begin{array}{l}
\Phi_e\rightarrow-i\Phi_e\\
\Phi_m\rightarrow-i\Phi_m\\
\end{array}\right.~~~
&~{\sigma_v} :~\left\{\begin{array}{l}
\Phi_e\rightarrow i\Phi_e^*\\
\Phi_m\rightarrow i\Phi_m^*\\
\end{array}\right.\nonumber\\
&C_{2z} :\left\{\begin{array}{l}
\Phi_e\rightarrow i\Phi_e^*\\
\Phi_m\rightarrow i\Phi_m^*\\
\end{array}\right.~~~
&R_{\pi} :\left\{\begin{array}{l}
\Phi_e\rightarrow \Phi_e\\
\Phi_m\rightarrow \Phi_m^*\\
\end{array}\right.
\label{eq_emme}
\end{align}
where, we have considered the origin of the coordinates to be centred at the site of the direct lattice. Clearly under ${\bf T_{d_1}}$ and ${\bf T_{d_2}}$ the $e$ and $m$ soft modes transform into each other-- as mentioned above-- due to the fact that the horizontal and vertical bonds interchange under these transformation.  This is an example of anyon permutation symmetry.\cite{PhysRevB.90.115118,PhysRevB.87.104406} Due to this, the mass of the $e$ and $m$ excitations are forced to be same in the critical theory. 

The gauge invariant spin order parameter can be constructed out of the above soft modes\cite{lannert2001quantum,xu2009global,PhysRevB.84.104430} by considering the symmetry transformation, as
\begin{align}
&\ttau_i^z\sim |\Phi_e|^2\cos(2\theta_e)~~~~~~\forall i\in {\rm Horizontal~bonds}\nonumber\\
&\ttau_i^x\sim |\Phi_m|^2\cos(2\theta_m)~~~~~~\forall i\in {\rm Vertical~bonds}
\label{eq_magord}
\end{align} 
Among other transformations, it is clear from the symmetry transformation table that, as expected, the above two spin order parameters are odd under time reversal symmetry, $\mathcal{T}$.

A crucial ingredient missing from the above analysis of the soft modes is the mutual semionic statistics of the electric and the magnetic modes. This can either be implemented using a  $U(1)\times U(1)$ mutual Chern-Simons (CS)theory\cite{freedman2004class,PhysRevB.71.235102,PhysRevB.78.155134,xu2009global} or a slightly more microscopic mutual $Z_2$ CS theory.\cite{PhysRevB.62.7850,PhysRevB.84.104430} Both lead to equivalent results.\cite{ap_sb} Here we shall use the $U(1)\times U(1)$ formalism.

\subsection{Mutual semionic statistics and the $U(1)\times U(1)$ mutual Chern-Simons action}

Within the $U(1)\times U(1)$ mutual CS formalism,\cite{PhysRevB.78.155134,xu2009global,PhysRevB.80.125101} the mutual semionic statistics between the $e$ and $m$ charges is implemented by introducing two internal $U(1)$ gauge fields $A_{\mu}$ and $B_\mu$ that are minimally coupled to the electric ($\Phi_e$) and magnetic ($\Phi_m$) soft modes respectively. The PSG transformation of these fields are obtained from the fact that they are minimally coupled to $\Phi_e$ and $\Phi_m$ respectively.  For the different symmetries in Eqs. \ref{eq_gauge_trans_transl}-\ref{eq_gauge_trans_rpi}, this leads to
\begin{align}
&{\bf T_{d_1}} :\left\{\begin{array}{l}
A_\mu\rightarrow B_\mu\\
B_\mu\rightarrow-A_\mu\\
\end{array}\right.~~~
{\bf T_{d_2}} :\left\{\begin{array}{l}
A_\mu\rightarrow-B_\mu\\
B_\mu\rightarrow A_\mu\\
\end{array}\right.\nonumber\\
&{\bf T_{x}} :~\left\{\begin{array}{l}
A_\mu\rightarrow -A_\mu\\
B_\mu\rightarrow-B_\mu\\
\end{array}\right.~~~
{\bf T_y} :~\left\{\begin{array}{l}
A_\mu\rightarrow A_\mu\\
B_\mu\rightarrow B_\mu\\
\end{array}\right.\nonumber\\
&{\mathcal{T}} :~~\left\{\begin{array}{l}
A_{\mu}\rightarrow -A_{\mu},\\
B_{\mu}\rightarrow -B_{\mu}\\
\end{array}\right.\nonumber\\
&{\sigma_v} :~\left\{\begin{array}{l}
A_{x}\rightarrow -A_{x},~~A_{y}\rightarrow A_{y},~~A_\tau\rightarrow -A_\tau\\
B_{x}\rightarrow -B_{x},~~B_{y}\rightarrow B_{y},~~B_\tau\rightarrow -B_\tau\\
\end{array}\right.\nonumber\\
&{C_{2z}} :\left\{\begin{array}{l}
A_{x}\rightarrow A_{x},~~A_{y}\rightarrow -A_{y},~~A_\tau\rightarrow -A_\tau\\
B_{x}\rightarrow B_{x},~~B_{y}\rightarrow -B_{y},~~B_\tau\rightarrow -B_\tau\\
\end{array}\right.\nonumber\\
&R_\pi :~\left\{\begin{array}{l}
A_{x}\rightarrow -A_{x},~~A_{y}\rightarrow -A_{y},~~A_\tau\rightarrow A_\tau\\
B_{x}\rightarrow B_{x},~~B_{y}\rightarrow B_{y},~~B_\tau\rightarrow -B_\tau\\
\end{array}\right.
\label{eq_abba}
\end{align}

The mutual $U(1)\times U(1)$ CS action in continuum in $(2+1)$ dimensions is then given by\cite{PhysRevB.78.155134,xu2009global} 
\begin{align}
\mathcal{S}_{CS}=\frac{i}{\pi}\int d^2{\bf r}d\tau~\epsilon^{\mu\nu\lambda} A_\mu\partial_\nu B_\lambda
\label{eq_u1cs}
\end{align}
where $\mu,\nu,\lambda=x,y,\tau$. It is easy to see that the above action implements the semionic statistics,\cite{dunne1999aspects} for example, by extremizing $\mathcal{S}_{CS}$ with respect to $A_\mu$ in presence of a static  $e$ charge density, $\rho_e$, which gives
\begin{align}
\rho_e=\frac{1}{\pi}(\partial_xB_y-\partial_yB_x)
\label{eq_emattach}
\end{align} 
Therefore the $m$ charge, $\Phi_m$, sees an odd number of $e$ charge as a source of $\pi$ flux as expected for a $Z_2$ QSL. Note that both $A_\mu$ and $B_\mu$ have their respective Maxwell terms. However such terms are irrelevant in presence of the CS term and the respective photons gain mass.\cite{dunne1999aspects}  Using the symmetry transformation in Eq. \ref{eq_abba}, we find that the CS action (Eq. \ref{eq_u1cs}) is odd under $\mathcal{T}$ and $R_\pi$. However we note that since attachment of $\pi$ and $-\pi$ fluxes are equivalent, the above CS theory is in accordance with these symmetries.\cite{xu2009global} 

\subsection{The Critical Theory}

With this we can now write down the continuum critical action which is given by
\begin{align}
\mathcal{S}_c=\int d^2{\bf r}d\tau~\mathcal{L} +\mathcal{S}_{CS}
\label{eq_softct}
\end{align}
where $\mathcal{S}_{CS}$ is given by Eq. \ref{eq_u1cs} and
\begin{align}
\mathcal{L}=\mathcal{L}_e +\mathcal{L}_m+\mathcal{L}_{em}
\label{eq_spinct}
\end{align}
with
\begin{align}
\mathcal{L}_e=|(\partial_\mu-i A_\mu)\Phi_e|^2+& u|\Phi_e|^2+v|\Phi_e|^4\nonumber\\
&-\lambda\left[(\Phi_e)^4+(\Phi_e^*)^4\right]
\label{eq_e4}
\end{align}
\begin{align}
\mathcal{L}_m=|(\partial_\mu-i B_\mu)\Phi_m|^2+& u|\Phi_m|^2+v|\Phi_m|^4\nonumber\\
&-\lambda\left[(\Phi_m)^4+(\Phi_m^*)^4\right]
\label{eq_m4}
\end{align}
\begin{align}
\mathcal{L}_{em}=w\left[(\Phi_e\Phi_m)^2+(\Phi_e\Phi_m^*)^2+{\rm c.c.}\right]
\label{eq_emint}
\end{align}

At this stage it is useful to draw attention to three important features of the above critical theory. Firstly, at the GMFT level (Eqs. \ref{eq_egmft} and \ref{eq_mgmft}), different horizontal chains are decoupled. Hence the soft modes do not have any rigidity in the vertical direction. However, fluctuations beyond the GMFT level leads to interactions between different horizontal chains. This is clear from Eq. \ref{eq_leading}, where each horizontal chain of $e$ charges are coupled with two $m$ horizontal chains at $Y=y\pm 1/2$. Thus integrating out the high energy $m$ modes generate interaction between neighbouring electric chains and thereby providing effective dispersion to the electric soft mode along the vertical direction. Additional contributions to both horizontal and vertical dispersions are further obtained from higher order corrections of the perturbation theory. However the above mechanism lead to anisotropic dispersion and the couplings for horizontal and vertical directions for the kinetic terms are indeed different. However, such anisotropy can be scaled away by simultaneously re-scaling $y$ (say) and the fields. Such anisotropy would be reflected in terms of correlation functions in terms of lattice unit of length.  

Secondly, due to Eq. \ref{eq_emme} and \ref{eq_abba}, the coupling constants of the $e$ and $m$ modes are equal. In particular the mass is related to the microscopic coupling constants as $u\sim (J_{\rm TC}-J)$ for both the $e$ and $m$ charges. This ensures that both the $e$ and $m$ soft modes condense together unless the translation symmetries, ${\bf T_{d_1}}$ and/or ${\bf T_{d_2}}$ are spontaneously broken.  In terms of the soft modes this is then the continuum version of a $Z_2$ {\it anyon permutation symmetry} which places very strong constraints on the structure of the critical theory and ensures the correct phases as well as phase transitions.

Finally, for $\lambda=w=0$, the system conserves fluxes in both the $e$ and $m$ sectors, $U_e(1)$ and  $U_m(1)$, separately.\cite{xu2009global}  Since, due to the mutual CS term, the fluxes of $A_{\mu} (B_{\mu})$ are attached to $m (e)$ particle densities, the above flux conservation results in charge conservation for both $e$ and $m$ charges. This is broken down when $\lambda\neq 0$ to $Z^e_4$ and $Z^m_4$. Further, $w\neq 0$ indicates short range interaction between the $e$ and $m$ soft modes as expected, say, from Eq. \ref{eq_leading}. Both these terms receive contributions from various terms in the perturbation theory and as such these coupling constants can be both positive or negative. For $w\neq 0$ the symmetry is broken down further to $Z_4$. We note that, in principle, the $\lambda$ term can be generated from the $w$ term at the second order level due to integration of high energy modes with $\lambda\sim w^2/u>0$, but we keep both these symmetry allowed terms as independent for our discussion.

\subsection{The phases}
The critical theory clearly captures the two phases as expected. At the mean field level, for $u>0$, we have
\begin{align}
\langle\Phi_e\rangle=\langle\Phi_m\rangle=0
\end{align}
Therefore both of them can be integrated out and the low energy effective theory is given by $\mathcal{S}_{CS}$ (Eq. \ref{eq_u1cs}) which is the $Z_2$ QSL with the right low energy spectrum consisting of the gapped electric and magnetic charges and a four fold ground state degeneracy in the thermodynamic limit on a two-tori.\cite{PhysRevB.78.155134,xu2009global} 

For $u<0$ both the electric and magnetic modes condense, {\it i.e.},
\begin{align}
\langle\Phi_e\rangle,\langle\Phi_m\rangle\neq 0
\end{align}
Therefore both $A_\mu$ and $B_\mu$ gauge fields acquire mass through Anderson-Higgs mechanism and hence their dynamics can be dropped. To understand the nature of this phase we note that the four fold terms in Eqs. \ref{eq_e4} and \ref{eq_m4} becomes (using Eqs. \ref{eq_ecom} and \ref{eq_mcom})
\begin{align}
\sim -\lambda\left(|\Phi_e|^4\cos(4\theta_e)+|\Phi_m|^4\cos(4\theta_m)\right]
\label{eq_theta4}
\end{align}

Therefore, for $\lambda>0$ the free energy minima occurs for 
\begin{align}
\theta_e,\theta_m=0,\pm \pi/2, \pi
\end{align}
which gives the two possible the symmetry broken partner spin ordered states as is now evident from Eq. \ref{eq_magord} with the spin order parameters being :
\begin{align}
&\langle\ttau_i^z\rangle\sim \langle|\Phi_e|^2\cos(2\theta_e)\rangle\sim \pm 1~~~~~~\forall i\in {\rm Horizontal~bonds}\nonumber\\
&\langle\ttau_i^x\rangle\sim \langle|\Phi_m|^2\cos(2\theta_m)\rangle\sim \pm 1~~~~~~\forall i\in {\rm Vertical~bonds}
\label{eq_magord2}
\end{align} 
Further the state also breaks $\sigma_v$ and  $C_{2z}$. Note that the order parameter is indeed invariant under the $Z_2$ gauge transformations and individual gauge charges are absent in the low energy spectrum in the spin-ordered phases due to the mutual CS term.

In this phase, the interaction between the electric and the magnetic modes (Eq. \ref{eq_emint}) can be written as 
\begin{align}
\mathcal{L}_{em}\sim w|\Phi_e|^2|\Phi_m|^2\cos(2\theta_e)~\cos(2\theta_m)
\label{eq_emint2}
\end{align}
For $w<0(>0)$, this results in ferromagnetic (antiferromagnetic) spin ordering in terms of $\ttau^x$ (on horizontal bonds) and $\ttau^z$ (on the vertical bonds) giving rise to the two states shown in Fig. \ref{fig_stripy_ferro}. The latter choice also breaks translation symmetry under ${\bf T_{d_1}}$ and ${\bf T_{d_2}}$ which interchanges a vertical and horizontal bond. The above phenomenology matches with the underlying microscopics for $w\sim J$. Therefore the above critical theory indeed reproduces the right phases.

It is interesting to note that for $\lambda<0$, Eq. \ref{eq_theta4} shows that the free energy is minimised for 
\begin{align}
\theta_e,\theta_m=\pm \pi/4,\pm 3\pi/4
\end{align}
It is easy to see that this phase is time-reversal symmetric. However, note that in such a state the  order parameters
\begin{align}
\langle|\Phi_e|^2\sin(2\theta_e)\rangle, ~\langle|\Phi_m|^2\sin(2\theta_m)\rangle
\end{align}
are non-zero. These order parameters however break translation symmetry in the horizontal direction, ${\bf T_x}$, and possibly represent some type of bond nematic state. However, for the type of microscopic model that we are concerned with-- as our numerical calculations suggest-- this bond nematic is not relevant and hence we shall not pursue it further. 

\subsection{The critical point}
We now turn to the critical point. It is useful to start with by neglecting the anisotropic terms in the critical theory described by Eq. \ref{eq_spinct} by putting $\lambda=w=0$. The critical action can then be written as
\begin{align}
\mathcal{S}=\int d^2{\bf r}d\tau~\left[\mathcal{L}_e+\mathcal{L}_m\right]+\mathcal{S}_{CS}
\end{align}
where, in this limit 
\begin{align}
\mathcal{L}_e=|(\partial_\mu-i A_\mu)\Phi_e|^2+& u|\Phi_e|^2+v|\Phi_e|^4
\label{eq_le}
\end{align}
\begin{align}
\mathcal{L}_m=|(\partial_\mu-i B_\mu)\Phi_m|^2+& u|\Phi_m|^2+v|\Phi_m|^4
\label{eq_lm}
\end{align}
and $\mathcal{S}_{CS}$ given by Eq. \ref{eq_u1cs}.

This class of mutual $U(1)\times U(1)$ CS theories have been described in a number of different contexts.\cite{xu2009global,PhysRevB.78.155134,PhysRevB.80.125101,PhysRevB.85.045114} Most pertinent to our discussion is Ref. \onlinecite{xu2009global} where such theories were considered in context of transitions out of a $Z_2$ QSL-- similar to the present case. However, there, in absence of the anyon-permutation symmetry that leads to constraint on the masses as given in Eq. \ref{eq_massconst}, the above class of transitions in that case turns out to be fine-tuned and in general separated by an intermediate $e$-Higgs or $m$-Higgs phase each characterised by a distinct spontaneously broken symmetry. Hence the anyon-permutation symmetry due to the microscopic symmetry ${\bf T_{d_{1(2)}}}$ (Eqs. \ref{eq_gauge_trans_transl} and \ref{eq_emme}) is crucial to protect the above critical point facilitating the direct phase transition in the present case. 

Ref. \onlinecite{PhysRevB.85.045114} studied the lattice version of the above model for generic values of the coupling parameters including the self-dual line which is directly relevant to us.  Along the self-dual line, it was found\cite{PhysRevB.85.045114} the $Z_2$ QSL phase gives way to a line of first order transitions (separating the $e$ and $m$ condensates-- not applicable to our work) before it leads to a $e-m$ condensed phase which is characterised exactly through the order parameters as we find here (Eq. \ref{eq_magord}). The meaning of the line of first order phase transition along the self-dual line is not clear in the present context since our severely system size limited numerics did not find any signature of it. 

\begin{figure}
\includegraphics[scale=0.5]{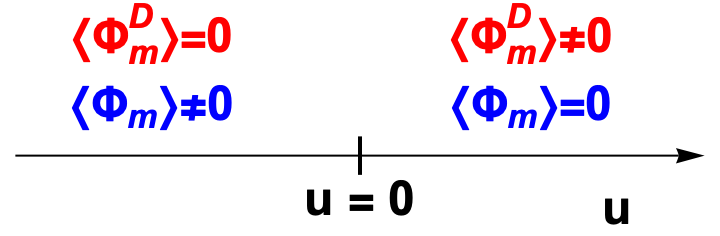}
\caption{The phase diagram corresponding to the critical Lagrangian, $\mathcal{L}_c$ in Eq. \ref{eq_lcem}. At the mean field level the bare $u=0$ corresponds to the synchronised {\it condensing} and {\it gapping out} of $\Phi_m^D$ and $\Phi_e$ respectively as shown. While, this may appear fine-tuned, as explained in the text, this above picture is indeed generic and is protected by symmetry. }
\label{fig_phaseu}
\end{figure}

To gain complementary insights into the critical theory, it is useful to apply particle-vortex duality\cite{peskin1978mandelstam,PhysRevLett.47.1556} for bosons in $(2+1)$ dimensions to the either the $e$ (in Eq. \ref{eq_le}) or $m$ (in Eq. \ref{eq_lm}) sector. Let us choose to dualise the $m$ sector to get the dual of Eq. \ref{eq_lm},
\begin{align}
\mathcal{L}^D_{m}=|(\partial_\mu-ib_\mu)\Phi_m^D|^2+&u_D|\Phi_m^D|^2+v_D|\Phi^D_m|^4\nonumber\\
&+\frac{i}{2\pi}\epsilon^{\mu\nu\lambda}b_\mu\partial_\nu B_\lambda
\label{eq_lmd}
\end{align}
where $\Phi_m^D$ is dual to the $m$ soft mode, $\Phi_m$ which is coupled to the internal gauge field $b_\mu$. In other words, $\Phi_m^D$ is the vortex of the $m$ field, $\Phi_m$. $u_D$ and $v_D$ are the respective couplings. It is clear that when the $m$-vortex condenses, {\it i.e.}, $\langle\Phi_m^D\rangle\neq 0$, then the $m$-charge, $\Phi_m$, is gapped, {\it i.e.} $\langle\Phi_m\rangle=0$, and vice versa. Therefore on general grounds, we expect that at low energies, the dual couplings are given by
\begin{align}
u_D=-\kappa u~~~~~v_D=\eta v
\label{eq_massconst}
\end{align}
where $\kappa,\eta>0$ are proportionality constants and $u, v$ are the coupling constants of the original theory of Eq. \ref{eq_lm}.  This ensures that the $m$-vortex vacuum is mapped to the $m$-charge condensate and vice-versa. From Eq. \ref{eq_u1cs2} and \ref{eq_lmd} we can now integrate out $B_\mu$ to get
\begin{align}
b_\mu+2A_\mu=0
\label{eq_constmd}
\end{align}
which when put back into Eq. \ref{eq_lmd}, gives
\begin{align}
\mathcal{L}^D_{m}=|(\partial_\mu+i2A_\mu)\Phi_m^D|^2-\kappa u|\Phi_m^D|^2+\eta v|\Phi^D_m|^4
\label{eq_lmd2}
\end{align}
The critical action is now given by
\begin{align}
\mathcal{S}=\int d^2{\bf r}d\tau~\mathcal{L}_c
\label{eq_scem}
\end{align}
where 
\begin{align}
\mathcal{L}_c=&|(\partial_\mu-i A_\mu)\Phi_e|^2+|(\partial_\mu+i2 A_\mu)\Phi_m^D|^2\nonumber\\
&+ u(|\Phi_e|^2-\kappa |\Phi_m^D|^2)+v(|\Phi_e|^4+\eta|\Phi_m^D|^4)\nonumber\\
&+g(\epsilon^{\mu\nu\lambda}\partial_\nu A_\lambda)^2
\label{eq_lcem}
\end{align}
where we have now explicityly written the Maxwell term for $A_\mu$ with coupling constant $g(>0)$ in absence of any CS term. The resultant phase diagram is shown in Eq. \ref{fig_phaseu} where $\Phi_e$ ($\Phi_m^D$) are condensed for $u<0(>0)$ with $u=0$ being the critical point where, as an increasing function of $u$ across $u=0$ lead to simultaneous {\it condensation} and {\it un-condensation} (gapping out) of $\Phi_m^D$ and $\Phi_e$ respectively. Clearly this is true irrespective of the renormalisation of the bare mass-scale, $u$, and is not fine-tuned. 

The $\langle\Phi_m^D\rangle\neq 0$ phase represents the QSL where both the $e$ and $m$ charges are gapped. However due to the fact that $\Phi_m^D$ carries charge-2 of the $U(1)$ gauge field $A_\mu$, on condensing $\Phi_m^D$ the gauge group is reduced to $Z_2$ as is appropriate for the $Z_2$ QSL. In this phase both $\Phi_e$ and $\Phi_m$ exist as gapped excitations. To uncover the mutual semionic statistics, we remind ourselves that $\Phi_m$ corresponds to $\pi$-flux of $A_\mu$ due to the mutual CS term (Eq. \ref{eq_u1cs}). In the Higgs phase of $\Phi_m^D$ such fluxes are gapped. However, once excited, the $\Phi_e$ charges are sensitive to it by virtue of their minimal coupling to $A_\mu$ as given by Eq. \ref{eq_lcem}. This description of the $Z_2$ QSL is quite similar to that obtained by disordering a superconductor through condensation of {\it charge-2} vortices.\cite{lannert2001quantum,PhysRevB.62.7850} Indeed in the present case all the {\it even} charges of $A_\mu$ are condensed while the {\it odd} charges are gapped out which is equivalent to the conservation of the $e$ and $m$ charges modulo 2-- as expected in a $Z_2$ QSL. The above description of the $Z_2$ QSL remains unchanged even in the presence of the anisotropy terms in  Eq. \ref{eq_spinct}. 

The spin-ordered state, on the other hand, is obtained for $u<0$ when both $\Phi_e$ and $\Phi_m$ (and hence $\Phi_m^D$ is gapped) are condensed as we described earlier. The photon of $A_\mu$ acquires a gap by Anderson-Higgs mechanism. Indeed all odd charges of $A_\mu$ are condensed in this spin-ordered phase. We, of course could have performed the particle-vortex duality in the $e$ sector in Eq. \ref{eq_le} to obtain an equivalent critical theory in terms of the $m$ charges, $\Phi_m$ and $e$ vortices, $\Phi_e^D$. In particular, we get
\begin{align}
\mathcal{L}_c^{\rm Dual}=&|(\partial_\mu-i B_\mu)\Phi_m|^2+|(\partial_\mu+i2 B_\mu)\Phi_e^D|^2\nonumber\\
&+ u(|\Phi_m|^2-\kappa |\Phi_e^D|^2)+v(|\Phi_m|^4+\eta|\Phi_e^D|^4)\nonumber\\
&+g(\epsilon^{\mu\nu\lambda}\partial_\nu B_\lambda)^2
\label{eq_lcemdual}
\end{align}
which is same as Eq. \ref{eq_lcem} once we identify the following mapping :
\begin{align}
(\Phi_e,\Phi_m^D,A_\mu)\Leftrightarrow (\Phi_m,\Phi_e^D,B_\mu)
\end{align}
which shows that the critical theory is self dual.\cite{PhysRevB.70.075104}

Turning to the anisotropy terms in the critical action in Eq. \ref{eq_spinct} by considering $\lambda,w\neq 0$ in Eqs. \ref{eq_e4}-\ref{eq_emint}. Due to the symmetry under ${\bf T_{d_{1(2)}}}$, we expect that the scaling dimension of the $\Phi_e$ and $\Phi_m$ are equal at the critical point. Hence, in order to judge the relevance of these quartic terms, for our formulation (Eq. \ref{eq_lcem}), it is easiest to start with the $(\Phi_m)^4$ term. From Eq. \ref{eq_lmd} and \ref{eq_constmd}, that the current of $\Phi_m$ is related to the flux of $A_\mu$ as
\begin{align}
j_m^\mu=\frac{1}{2\pi}\epsilon^{\mu\nu\lambda}\partial_\nu b_\lambda=-\frac{1}{\pi}\epsilon^{\mu\nu\lambda}\partial_\nu A_\lambda.
\end{align}
The anisotropic term  breaks the $O(2)$ symmetry of the $m$ sector in Eq. \ref{eq_lcem} down to $Z_4$ since four  $\Phi_m$ charges can be created/annihilated. Each such charge being proportional to $\pi$ flux of $A_\mu$ the $(\Phi_m)^4$ term therefore corresponds to the {\it doubled monopole} operator\cite{PhysRevB.70.220403} of $A_\mu$. For an Abelian Higgs model for a superconductor, such doubled monopoles may be irrelevant in a parameter regime\cite{PhysRevLett.88.232001} raising hope that the present transition may indeed be controlled by the $\lambda=w=0$. However, the critical theory needs to be studied in further detail to settle this issue.

The critical point described by $u=0$ involving the simultaneous condensation and gapping out of the even and odd charges of $A_\mu$ respectively and the gauge flux of $A_\mu$ being conserved is novel and is expected to be different from the transition in Abelian Higgs model\cite{PhysRevLett.47.1556} on one hand and the the transition in the dual Abelian Higgs model describing the condensation of paired vortices\cite{lannert2001quantum,PhysRevB.62.7850} on the other.

The critical theory (Eq. \ref{eq_scem}) therefore suggests that the deconfined quantum phase transition between the  $Z_2$ QSL and the magnetically ordered phase is described by a modified self-dual modified Abelian Higgs model (MAHM) with conserved flux. In absence of the mutual CS term, the critical action in Eq. \ref{eq_softct} describes an easy-plane non-compact projective field theory (easy-plane NCCP$^1$) studied in Ref. \onlinecite{PhysRevB.70.075104} while the one with the mutual CS term was studied in Ref. \onlinecite{PhysRevB.85.045114}. In the second study-- as mentioned before-- it was found that the $QSL$ and the $e-m$ condensate phases are separated by a line of first order transition along the self-dual line. The relevance of this line is not clear in the present context. Hence, at present it is not clear to us whether the present self-dual modified Abelian Higgs model belongs to the same universality class at easy-plane NCCP$^1$.

The transition, therefore, is an example of a deconfined quantum critical point.\cite{senthil2004deconfined} The critical theory is not written in terms of the order parameters but the low energy degrees of freedom of the $Z_2$ QSL. Characteristic to deconfined critical points, the spin order parameter which is a bilinear in terms of the critical field-- the gauge charges. Therefore we expect a large anomalous dimension for the order parameter which naively should be twice that of the critical field. \cite{PhysRevB.70.075104} The above critical theory is expected to be stable in presence of small $\Gamma$ as it does not add any new symmetry allowed terms to the critical theory at the lowest order, thus resulting in the phase diagram as shown in Fig. \ref{fig_pd} for small $\Gamma$.

\subsection{Effect of an external Zeeman field}
\label{subsec_trzee}

So far we have neglected the experimentally relevant possibility of turning on an external magnetic field (we refer to it as a Zeeman field to avoid confusion) on Eq. \ref{eq_hamiltonian} in the anisotropic limit. This perturbation in the isotropic limit is given by
\begin{align}
H_{\rm Zeeman}=-\sum_{i}{\bf h}\cdot{\boldsymbol\sigma}_i
\end{align} 
for the spins, $\sigma^\alpha_i$ on the sites of the honeycomb lattice, where ${\bf h}=(h_x,h_y,h_z)$ is the external Zeeman field. In the anisotropic limit ($|{\bf h}|/K_z$) that we are concerned with, the degenerate perturbation theory (Eq. \ref{eq_deg pert}) gives rise to the following addition in the leading order of ${\bf h}$ to Eq. \ref{full hamiltonian} :
\begin{align}
\mathcal{H}_{\rm z}=-2h_z\sum_{i}\tau^z_i
\label{eq_trzeeunrot}
\end{align}
for the $\tau$-spins on the $z$-bond in the unrotated basis. This is clearly in in agreement with the fact that $\tau^z_i$ is the time reversal odd component of the non-Kramers doublet (Eq. \ref{eq_tausym}).  In the rotated basis (Eq. \ref{eq_rot_ver}) becomes
\begin{align}
\tilde{H}_{\rm z}=-2h_z\left[\sum_{i\in H} \ttau^z_i+\sum_{i\in V}\ttau^x_i\right]
\label{eq_trzee}
\end{align}
where, the first sum is over the horizontal bonds and the second sum is over the vertical bonds.  There are higher order terms in the above Zeeman field including cross terms involving the other perturbing terms in $\mathcal{V}$ of Eq. \ref{eq_pert_hamiltonian}. We neglect the detailed structure of the these higher order time reversal symmetry breaking terms. Notice that the structure of above term is ``oppositte" to that of the leading order pseudo-dipolar term given by Eq. \ref{eq_gammaham_rot} with a crucial difference that unlike Eq. \ref{eq_gammaham_rot}, the present term in Eq. \ref{eq_trzee} is time reversal odd since the Zeeman field breaks the time reversal symmetry. We explore this relation between the Zeeman and the pseudo-dipolar perturbations in the next section in the context of the latter.

In terms of the gauge theory, Eq. \ref{eq_trzee} becomes
\begin{align}
\tilde{\mathcal{H}}_{z}= & -2h_z\sum_{a}\mu^x_a\rho^z_{a,a+\hat x}\mu^x_{a+\hat x}
-J_{TC}\sum_a\mu^z_a\nonumber\\
&-2h_z\sum_{\ba}\tilde\mu^x_{\ba}\tilde\rho^x_{\ba,\ba+\hat x}\tilde\mu^x_{\ba+\hat x} -J_{TC}\sum_{\ba}\tilde\mu^z_{\ba}
\label{eq_trzeedecoup}
\end{align}
This therefore leads to the dispersion of the $e$ and $m$ charges along the horizontal direction which renormalises the results of Eq. \ref{eq_egmft} and \ref{eq_mgmft} for the $e$ and $m$ sectors respectively. 

Crucially, however it lifts the degeneracy of the two time-reversal partner soft modes in Eq. \ref{eq_esm} and \ref{eq_msm}. This allows the following term in addition to the ones already in the soft mode critical action in Eq. \ref{eq_softct}
\begin{align}
\mathcal{S}_{\rm z}=\int d^x d\tau~\mathcal{L}_{\rm z}
\label{eq_trzeesc}
\end{align}
where, 
\begin{align}
\mathcal{L}_{\rm z}&=-\tilde{h}\left[(\Phi_e)^2+(\Phi_e^*)^2+(\Phi_m)^2+(\Phi_m^*)^2\right]\nonumber\\
&= -2\tilde h\left[ |\Phi_e|^2\cos(2\theta_e)+ |\Phi_m|^2\cos(2\theta_m)\right]
\label{eq_trzeesoft}
\end{align}
with $\tilde h\propto h_z$. This clearly lifts the degeneracy between the two time reversal invariant spin states. In presence of this second order term ($\propto \cos2\theta$), the fourth order $Z_4$ anisotropy terms (proportional to $\lambda\cos4\theta$) in Eq. \ref{eq_e4} and \ref{eq_m4} can be neglected. 

The $Z_2$ QSL remains unchanged for $u-\tilde h>0$. However, for $u-\tilde h<0$, inside the spin-ordered phase, for $\tilde h>0(<0)$, we have
\begin{align}
\theta_e,\theta_m=0,\pi (\pm \pi/2)
\end{align}
which is nothing but the polarised phase. This is indeed true for $w<0$ in the $e-m$ coupling term in Eq. \ref{eq_emint} or equivalently Eq. \ref{eq_emint2} where the $\tilde h=0$ state is a ferromagnet. However, for $w>0$, where the $\tilde{h}=0$ ground state is antiferromagnetic, we expect a first order spin-flop transition from the antiferromagnet to a polarised phase within the spin ordered phase. 

The critical point is given for $u-\tilde h=0$. It turns out that for $J<0$, this critical point is similar to that obtained by destabilising the $Z_2$ QSL using the pseudo-dipolar interactions, $\Gamma$. Hence to avoid repetation, we first develop the soft modes of the pseudo-dipolar limit and then return to discuss the critical point for both the Zeeman and the pseudo-dipolar limits together.

\section{Phase Transition between QSL and the trivial paramagnet}
\label{sec_gammact}

Having understood the transition to the spin ordered phase from the QSL we now turn to the transition between the QSL and the trivial paramagnet accessed by tuning the $\Gamma$ term (horizontal axis of the Fig. \ref{fig_pd}). Again a controlled description  of the phase transition is achieved by starting with the QSL and understanding the fate of its excitations-- the gauge charges. The effect of the Zeeman field-- the topic of the last subsection-- sheds crucial insight into this transition. This becomes even more clear by comparing the leading order perturbations due to the $h_z$ (Eq. \ref{eq_trzee}) and $\Gamma$ (Eq. \ref{eq_gammaham_rot}). It is clear that the reflection about the $x=y$ line of the square lattice in Fig. \ref{fig_toric} and passing through the sites maps all the vertical bonds to horizontal bonds and vice versa and thereby mapping Eq. \ref{eq_trzee} to Eq. \ref{eq_gammaham_rot} when we perform a concomitant transformation of $h_z\rightarrow -\Gamma$. We note that under the above reflection the Toric code Hamiltonian (Eq. \ref{eq_toric_code rot}) remains unchanged.  However, now the $\ttau^x (\ttau^z)$ on the horizontal (vertical) bonds are time-reversal odd. This is in accordance with our previous observation that Eq. \ref{eq_gammaham_rot} is time-reversal even. As a consequence a trivial time-reversal symmetric paramagnet realised due to the pseudo-dipolar perturbations, $\Gamma$. 

\subsection{Decoupled vertical Ising chains}

Similar to the case of Heisenberg interactions, we first incorporate  the leading order perturbation in $\Gamma$ given by Eq. \ref{eq_gammaham_rot} to the QSL Hamiltonian in Eq. \ref{eq_toric_code rot} in the rotated basis. In terms of the gauge fields and the gauge charges this becomes 
\begin{equation}\label{eq_gammak_tfim}
\begin{aligned}
\tilde{\mathcal{H}}_{J=0}= & 2\Gamma\sum_{a}\mu^x_a\rho^z_{a,a+\hat y}\mu^x_{a+\hat y}
-J_{TC}\sum_a\mu^z_a\\
&+2\Gamma\sum_{\ba}\tilde\mu^x_{\ba}\tilde\rho^x_{\ba,\ba+\hat y}\tilde\mu^x_{\ba+\hat y} -J_{TC}\sum_{\ba}\tilde\mu^z_{\ba}
\end{aligned}
\end{equation}

Contrast this with the effect of the Zeeman field given in Eq. \ref{eq_trzeedecoup}. Here, evidently, the first (second) line denotes a series of now decoupled {\it vertical}, with respect to Fig. \ref{fig_toric}, transverse field Ising chains\cite{yu2008topological} representing the $e(m)$ charges as opposed to the horizontal ones in Eq. \ref{eq_trzeedecoup}.  This can alternatively be looked up as a consequence of the $x=y$ reflections discussed above which converts the horizontal chains to vertical chains. This leads to important differences, particularly regarding the nature of the phase that the QSL transits into due to the non-trivial projective implementation of the time reversal symmetry (Eq. \ref{eq_gauge_trans_tr}).

Similar to Eq. \ref{eq_gaugechoice}, we can choose the gauge 
\begin{align}
\rho_{a,a+\hat y}^z=\tilde\rho^z_{a,a+\hat y}=+1
\label{eq_gaugechoicem}
\end{align}
as these links do not cross. Thus we get, from Eq. \ref{eq_gammak_tfim}
\begin{align}
\tilde{\mathcal{H}}_{J=0}= & 2\Gamma\sum_{a}\mu^x_a\mu^x_{a+\hat y}
-J_{TC}\sum_a\mu^z_a\nonumber\\
&+2\Gamma\sum_{\ba}\tilde\mu^x_{\ba}\tilde\tilde\mu^x_{\ba+\hat y} -J_{TC}\sum_{\ba}\tilde\mu^z_{\ba}
\end{align}

The paramagnetic phase of each vertical chain-- both for $e$ and $m$ sectors-- for $\Gamma=0$ is clearly the $Z_2$  QSL. On increasing $\Gamma$, the electric and the magnetic charges now develop dispersions along the vertical direction and when the minima of such dispersion hits zero they condense destroying the QSL.  

For $\Gamma<0(>0)$ the soft mode for $e$ and $m$ sectors are given by a charge arrangement similar to Fig. \ref{fig_sm1} (\ref{fig_sm2}). However, very crucially, unlike the Heisenberg case of the previous section these two soft modes are not degenerate because they are not time reversal partners as is evident from Eq. \ref{eq_gauge_trans_tr}. This will become more evident below. Anticipating the difference, we therefore denote these two soft modes as
\begin{align}
\hat\xi_e^{(1)}=1~~~~\hat{\xi}_e^{(2)}=e^{i\pi x}
\end{align}
for the $e$ sector on the direct lattice and
\begin{align}
\hat\xi_m^{(1)}=1~~~~\hat{\xi}_m^{(2)}=e^{i\pi X}
\end{align}
for the $m$ sector on the dual lattice. 

Indeed, as discussed above, for a given choice of $\Gamma$ the two states have different energies and hence we can completely work with one of the soft modes for each sign of $\Gamma$. This becomes more evident in the PSG analysis of the soft modes (see below) where we see that the two soft modes do not mix with each other under symmetry transformations. This, to draw further analogy with the Zeeman perturbation, is like the lifting of the degeneracy of the two soft modes, of Eqs. \ref{eq_esm} and \ref{eq_msm}, as discussed below Eq. \ref{eq_trzeedecoup}. There, the existence of the time reversal symmetric point $h_z=0$ allows us to use a theory with two soft modes each in the $e$ and $m$ sector which is subsequently broken down further by the Zeeman field through Eq. \ref{eq_trzeesoft} in the critical theory.

Similarly, we define a hypothetical $\Gamma=0$ situation where both these modes are degenerate as our starting point and then use a redundant description keeping both the soft modes and obtain the correct description for the phases and phase transition. We note that unlike in the Zeeman case, this hypothetical situation is not realised in our model and indeed is actually an unstable quantum ground state with $\ln 2$ entropy per site as this does not break any symmetry of the anisotropic Hamiltonian. In the passing we note that the soft modes result in two distinct nematic states as given in Eq. \ref{eq_trzeesoft} and hence may be relevant to the isotropic/near isotropic limit of Eq. \ref{eq_hamiltonian}. However, here we use this limit only as a convenient starting point for our analysis of the critical theory.

\subsection{Soft modes}
Similar to Eq. \ref{eq_esm} and \ref{eq_msm}, we expand the gauge charges in terms of the soft modes. This gives 
\begin{align}
\Psi_e({\bf r},\tau)=\tphi_e^{(1)}({\bf r},\tau)~\hat\xi_e^{(1)}+\tphi_e^{(2)}({\bf r},\tau)~\hat\xi_e^{(2)}
\label{eq_game}
\end{align}
for the $e$ sector and 
\begin{align}
\Psi_m({\bf r},\tau)=\tphi_m^{(1)}({\bf r},\tau)~\hat\xi_m^{(1)}+\tphi_m^{(2)}({\bf r},\tau)~\hat\xi_m^{(2)}
\label{eq_gamm}
\end{align}
for the $m$ sector. Here $(\tilde\phi_e^{(1)}, \tilde\phi_e^{(2)})$ and $(\tilde\phi_m^{(1)}, \tilde\phi_m^{(2)})$ are the new $e (m)$ soft mode amplitudes. In order to obtain the PSG transformations we once again combine the two amplitudes of each sector as
\begin{align}
\tPhi_e=\tphi_e^{(1)}+i\tphi_e^{(2)}=|\tPhi_e|e^{i\tilde\theta_e}
\label{eq_tphie}
\end{align}
and
\begin{align}
\tPhi_m=\tphi_m^{(1)}+i\tphi_m^{(2)}=|\tPhi_m|e^{i\tilde\theta_m}
\label{eq_tphim}
\end{align}

The PSG transformations under different symmetries listed in Eqs. \ref{eq_gauge_trans_transl}-\ref{eq_gauge_trans_rpi} are now given by

\begin{align}
&{\bf T_{d_1}} :\left\{\begin{array}{l}
\tPhi_e\rightarrow\tPhi_m\\
\tPhi_m\rightarrow\tPhi_e^*\\
\end{array}\right.~~~
&{\bf T_{d_2}} :\left\{\begin{array}{l}
\tPhi_e\rightarrow\tPhi_m^*\\
\tPhi_m\rightarrow\tPhi_e\\
\end{array}\right.\nonumber\\
&{\bf T_{x}} :~\left\{\begin{array}{l}
\tPhi_e\rightarrow\tPhi_e^*\\
\tPhi_m\rightarrow\tPhi_m^*\\
\end{array}\right.~~~
&{\bf T_{y}} :\left\{\begin{array}{l}
\tPhi_e\rightarrow\tPhi_e\\
\tPhi_m\rightarrow\tPhi_m\\
\end{array}\right.\nonumber\\
&{\mathcal{T}} :~~\left\{\begin{array}{l}
\tPhi_e\rightarrow \tPhi_e^*\\
\tPhi_m\rightarrow \tPhi_m^*\\
\end{array}\right.~~~
&{\sigma_v} :\left\{\begin{array}{l}
\tPhi_e\rightarrow \tPhi_e\\
\tPhi_m\rightarrow \tPhi_m\\
\end{array}\right.\nonumber\\
&C_{2z} :\left\{\begin{array}{l}
\tPhi_e\rightarrow \tPhi_e\\
\tPhi_m\rightarrow \tPhi_m^*\\
\end{array}\right.~~~
&R_{\pi} :\left\{\begin{array}{l}
\tPhi_e\rightarrow \tPhi_e\\
\tPhi_m\rightarrow \tPhi_m^*\\
\end{array}\right.
\label{eq_temme}
\end{align}

We shall find the Higgs phase resulting from condensing the above charges results in a trivial paramagnet. 

\subsection{The mutual $U(1)\times U(1)$ CS critical theory}

Using a redundant description by keeping both the soft modes in each of the electric and magnetic sectors allows us to extend the $U(1)\times U(1)$ mutual CS formalism for this transition. In this case, the CS action is given by
\begin{align}
\tilde{\mathcal{S}}_{CS}=\frac{i}{\pi}\int d^2{\bf r}d\tau~\epsilon^{\mu\nu\lambda} \tA_\mu\partial_\nu \tB_\lambda
\label{eq_u1cs2}
\end{align}
where now $\tA_\mu$ and $\tB_\mu$ are the internal $U(1)$ gauge fields that couple minimally to the soft modes $\tPhi_e$ and $\tPhi_m$ respectively. The PSG of the gauge fields are given by
\begin{align}
&{\bf T_{d_1}} :\left\{\begin{array}{l}
\tA_\mu\rightarrow \tB_\mu\\
\tB_\mu\rightarrow-\tA_\mu\\
\end{array}\right.~~~
{\bf T_{d_2}} :\left\{\begin{array}{l}
\tA_\mu\rightarrow-\tB_\mu\\
\tB_\mu\rightarrow \tA_\mu\\
\end{array}\right.\nonumber\\
\label{eq_tabba}
&{\bf T_{x}} :~\left\{\begin{array}{l}
\tA_\mu\rightarrow -\tA_\mu\\
\tB_\mu\rightarrow-\tB_\mu\\
\end{array}\right.~~~
{\bf T_y} :~\left\{\begin{array}{l}
\tA_\mu\rightarrow \tA_\mu\\
\tB_\mu\rightarrow \tB_\mu\\
\end{array}\right.\nonumber\\
&{\mathcal{T}} :~~\left\{\begin{array}{l}
\tA_\mu\rightarrow \tA_\mu\\
\tB_\mu\rightarrow \tB_\mu\\
\end{array}\right.\\
&{\sigma_v} :~\left\{\begin{array}{l}
\tA_x\rightarrow \tA_x,~~\tA_y\rightarrow-\tA_y,~~\tA_\tau\rightarrow\tA_\tau\\
\tB_x\rightarrow \tB_x,~~\tB_y\rightarrow-\tB_y,~~\tB_\tau\rightarrow\tB_\tau\\
\end{array}\right.\nonumber\\
&{C_{2z}} :\left\{\begin{array}{l}
\tA_x\rightarrow -\tA_x,~~\tA_y\rightarrow \tA_y,~~\tA_\tau\rightarrow\tA_\tau\\
\tB_x\rightarrow \tB_x,~~\tB_y\rightarrow-\tB_y,~~\tB_\tau\rightarrow-\tB_\tau\\
\end{array}\right.\nonumber\\
&{R_\pi} :~\left\{\begin{array}{l}
\tA_x\rightarrow -\tA_x,~~\tA_y\rightarrow -\tA_y,~~\tA_\tau\rightarrow\tA_\tau\\
\tB_x\rightarrow \tB_x,~~\tB_y\rightarrow\tB_y,~~\tB_\tau\rightarrow-\tB_\tau\\
\end{array}\right.\nonumber
\end{align}
which are consistent with the CS action in Eq. \ref{eq_u1cs2} as before upto in inconsequential sign change under $C_{2z}$. The resultant critical field theory is 
\begin{align}
\tilde{\mathcal{S}}_c=\int d^2{\bf r}d\tau \mathcal{L}+\tilde{\mathcal{S}}_{CS}
\label{eq_gammasc}
\end{align}
where $\tilde{\mathcal{S}}_{CS}$ is given by Eq. \ref{eq_u1cs2} and
\begin{align}
\tilde{\mathcal{L}}=\tilde{\mathcal{L}}_e+\tilde{\mathcal{L}}_m+\tilde{\mathcal{L}}_{em}
\end{align}
with
\begin{align}
\tilde{\mathcal{L}}_e=|(\partial_\mu-i \tA_\mu)\tPhi_e|^2+& \tilde{u}|\tPhi_e|^2+\tilde{v}|\tPhi_e|^4\nonumber\\
&-\tilde\lambda\left[(\tPhi_e)^2+(\tPhi_e^*)^2\right]
\label{eq_e2}
\end{align}
\begin{align}
\tilde{\mathcal{L}}_m=|(\partial_\mu-i \tB_\mu)\tPhi_m|^2+& \tilde{u}|\tPhi_m|^2+\tilde{v}|\tPhi_m|^4\nonumber\\
&-\tilde\lambda\left[(\tPhi_m)^2+(\tPhi_m^*)^2\right]
\label{eq_m2}
\end{align}
\begin{align}
\tilde{\mathcal{L}}_{em}=\tilde{w}\left[(\tPhi_e\tPhi_m)^2+(\tPhi_e\tPhi_m^*)^2+{\rm c.c.}\right]
\label{eq_emint2}
\end{align}

Considerations similar to those noted below Eq. \ref{eq_emint} for the case of Heisenberg perturbations, also apply here with an important difference that in the present case, the decoupled limit pertains to vertical Ising chains. In particular the mass term, $u\sim J_{TC}-2\Gamma$.

Note that due to the presence of the second order anisotropic term proportional to $\lambda$ in the critical action given by Eq. \ref{eq_gammasc}, the discussion of the phases and phase transitions has an exact parallel with the case of the Zeeman term (Eq. \ref{eq_trzeesc})-- a consequence of the $x=y$ reflection as discussed above-- with the difference being, in the present case the condensation of the soft modes lead to a time-reversal symmetric paramagnet.

\subsection{The phases}

Turning to the phases, clearly, as before, the un-condensed phase for $\tilde{u}-\tilde{\lambda}>0$, {\it i.e.},
\begin{align}
\langle \tPhi_e\rangle=\langle\tPhi_m\rangle=0
\end{align} 
is the $Z_2$ QSL with gapped $e$ and $m$ charges with mutual semionic statistics. For $\tilde u-\tilde\lambda<0$, the both electric and magnetic charges condense, {\it i.e.}
\begin{align}
\langle \tPhi_e\rangle, \langle\tPhi_m\rangle\neq 0
\end{align} 
This Higgs phase does not break time reversal symmetry. Indeed, in terms of symmetry, the $Z_2$ gauge invariant fields are given by
\begin{align}
&\tPhi_e^2+(\tPhi_e^*)^2\sim |\tPhi_e|^2\cos(2\tilde\theta_e)\sim \ttau^x_i~~~\forall i\in {\rm Horizontal~bonds}\nonumber\\
&\tPhi_m^2+(\tPhi_m^*)^2\sim |\tPhi_m|^2\cos(2\tilde\theta_m)\sim \ttau^z_i~~~\forall i\in {\rm Vertical~bonds}
\label{eq_trivord}
\end{align}

These should be contrasted with  Eq. \ref{eq_magord} which characterises spin ordering. In spite of similar appearences, the above  equations are exactly opposite in terms of the type of the bonds (vertical versus horizontal) with respect to that of Eq. \ref{eq_magord} and this has a central effect in the nature of the resultant phase which for the present case is a symmetric non-degenerate paramagnet. 

For $\tilde\lambda>0(<0)$ in Eq. \ref{eq_e2} and \ref{eq_m2}, the free energy is minimised for 
\begin{align}
\tilde\theta_e, \tilde\theta_m=0,\pi (\pm \pi/2)
\end{align} 
This correspond to ordering (see Eq. \ref{eq_trivord})
\begin{align}
&\ttau^x_i=\pm 1~~~\forall i\in {\rm Horizontal~bonds}\nonumber\\
&\ttau^z_i=\pm 1~~~\forall i\in {\rm Vertical~bonds}
\label{eq_trivord2}
\end{align}
all of which are time reversal symmetric (see Table. \ref{table_tau_symm_fm_h} in Appendix \ref{appen_symm}). For each of these cases, the interaction between the $e$ and $m$ modes is given by Eq. \ref{eq_emint2}, {\it i.e.}
\begin{align}
\tilde{\mathcal{L}}_{em}\sim w|\tPhi_e|^2|\tPhi_m|^2\cos(2\tilde\theta_e)~\cos(2\tilde\theta_m)
\end{align}
For $w<0 (>0)$ the spin components as given in Eq. \ref{eq_trivord2} are parallel (antiparallel). The latter case breaks translation by ${\bf T_{d_1}}$ and ${\bf T_{d_2}}$  which interchanges horizontal and vertical bonds and appears to be not relevant for the present case. The above phenomenology is consistent with the microscopic  of Section \ref{para_pseudo_dipolar_limit} for $\tilde\lambda\sim -\Gamma$. This then completes the discussion of the two trivial paramagnets as shown in Fig. \ref{fig_pd}. We expect that the above theory to be stable to small Heisenberg perturbations, $J$.

\subsection{Effects of Zeeman term and the critical point}

Before discussing the critical point, we would like to understand the effect of the Zeeman term in this pseudo-dipolar limit which would throw critical insight into the nature of the transition both in the Zeeman limit of subsection \ref{subsec_trzee} and the present pseudo-dipolar limit. The starting point is the previously mentioned observation of the leading order Zeeman terms and pseudo-dipolar terms mapping into each other under $x=y$ reflection. This becomes more clear when we consider the leading Zeeman perturbing term given by Eq. \ref{eq_trzee} along with the leading pseudo-dipolar perturbation given by Eq. \ref{eq_gammaham_rot} whence the net leading perturbation is given by
\begin{align}
\mathcal{H}_{\Gamma-z}=\sum_{i\in V}\left[2\Gamma\ttau^z_i-2h_z\ttau^x_i\right]+\sum_{i\in H}\left[2\Gamma\ttau^x_i-2h_z\ttau^z_i\right]
\label{eq_gammazee}
\end{align}
In particular for the limit $h_z=-\Gamma$ the above Hamiltonian becomes
\begin{align}
\mathcal{H}_{\Gamma-z}=2\Gamma\sum_{i}\left[\ttau_i^x+\ttau^z_i\right]
\label{eq_gammazeesp}
\end{align}
In conjunction with the Toric code term in Eq. \ref{eq_toric_code rot}, this is exactly the Hamiltonian studied in Refs. \onlinecite{PhysRevB.82.085114}, \onlinecite{PhysRevB.85.195104} and \onlinecite{PhysRevB.79.033109} for a Toric code QSL in a ``magnetic'' field along the self-dual line. While this inference is drawn on the basis of the leading order perturbations, in the spirit of the discussions presented, it leads to interesting possibilities particularly in the lights of the rich properties of the self-dual line as known from very systematic numerical calculations.\cite{PhysRevB.82.085114} We shall return to this in a moment, but first let us notice that in presence of the Zeeman field, from the symmetry transformations in Eq. \ref{eq_tausym}, it  is clear that the the residual symmetry of the system is generated by only ${\bf T_{d_1}}, {\bf T_{d_2}}$ (and hence ${\bf T_x}, {\bf T_y}$) and $R_\pi$. This is the reason for the exact match between the transformation tables of the soft modes $(\Phi_e, \Phi_m)$ (Eq. \ref{eq_emme}) and $(A_\mu, B_\mu)$ (Eq. \ref{eq_abba}) with $(\tPhi_e, \tPhi_m)$ (Eq. \ref{eq_temme}) and $(\tA_\mu, \tB_\mu)$ (Eq. \ref{eq_tabba}) under ${\bf T_{d_1}}, {\bf T_{d_2}}$ (and hence ${\bf T_x}, {\bf T_y}$) and $R_\pi$. Indeed, we can introduce the following linear superposition of the soft modes
\begin{align}
&\chi_e=\frac{1}{\sqrt{\Gamma^2+h_z^2}}\left[\Gamma\tPhi_e-h_z\Phi_e\right]\\
&\chi_m=\frac{1}{\sqrt{\Gamma^2+h_z^2}}\left[\Gamma\tPhi_m-h_z\Phi_m\right]
\label{eq_zeegamma}
\end{align}
and their corresponding gauge fields
\begin{align}
&C_\mu=\frac{1}{\sqrt{\Gamma^2+h_z^2}}\left[\Gamma\tA_\mu-h_z A_\mu\right]\\
&D_\mu=\frac{1}{\sqrt{\Gamma^2+h_z^2}}\left[\Gamma\tB_\mu-h_z B_\mu\right]
\end{align}
for the perturbation corresponding to Eq. \ref{eq_gammazee} which interpolates between the two limits $\Gamma=0$ and $h_z=0$. The left hand side of the above equation, by construction have the same transformation property as the right hand-side under ${\bf T_{d_1}}, {\bf T_{d_2}}$ (and hence ${\bf T_x}, {\bf T_y}$) and $R_\pi$. Clearly at the special value $h_z=-\Gamma$, corresponding to Eq. \ref{eq_gammazeesp} whence the system enjoys a reflection symmetry to the leading order, stands for an equal superposition of the two sets of soft modes.

Hence following the symmetry arguments as before, we can write a critical action as
\begin{align}
\mathcal{S}^\chi_c=\int d^2{\bf r}~d\tau~\mathcal{L}_c^\chi +\mathcal{S}_{CS}^\chi
\end{align}
where
\begin{align}
\mathcal{S}_{CS}^\chi=\frac{i}{\pi}\int d^2{\bf r}~d\tau~\epsilon^{\mu\nu\lambda}C_\mu\partial_\nu D_\lambda
\end{align}
and 
\begin{align}
{\mathcal{L}}_c^\chi={\mathcal{L}}_e^\chi+{\mathcal{L}}_m^\chi+{\mathcal{L}}_{em}^\chi
\end{align}
with
\begin{align}
{\mathcal{L}}_e^\chi=|(\partial_\mu-i C_\mu)\chi_e|^2+& \tilde{u}|\chi_e|^2+\tilde{v}|\chi_e|^4\nonumber\\
&-\tilde\lambda\left[(\chi_e)^2+(\chi_e^*)^2\right]\nonumber\\
{\mathcal{L}}_m^\chi=|(\partial_\mu-i D_\mu)\chi_m|^2+& \tilde{u}|\chi_m|^2+\tilde{v}|\chi_m|^4\nonumber\\
&-\tilde\lambda\left[(\chi_m)^2+(\chi_m^*)^2\right]
\label{eq_m2chi}
\end{align}
and
\begin{align}
{\mathcal{L}}_{em}^\chi=\tilde{w}\left[(\chi_e\chi_m)^2+(\chi_e\chi_m^*)^2+{\rm c.c.}\right]
\label{eq_emint2}
\end{align}
which clearly interpolates between the two limits in Eqs. \ref{eq_gammasc} (for $h_z=0$) and \ref{eq_trzeesc} (for $\Gamma=0$). 

The discussions of the $Z_2$ QSL for $\tilde u-\tilde\lambda>0$ and the trivial paramagnet for $\tilde u-\tilde\lambda<0$ now directly follows from our previous discussions. Note that the trivial paramagnet continuously deforms from the polarised state (for $\Gamma=0$) to a time-reversal symmetric trivial paramagnet (for $h_z\rightarrow 0^+$) with the $h_z=0$ line being the time reversal symmetric paramagnet.

The two fold anisotropy term proportional to $\lambda$ in given by Eq. \ref{eq_m2chi} carries charge-$2$ under $C_\mu$ and $D_\mu$ respectively. Therefore this is like the pairing term in superconductors, albeit for bosons which breaks down the gauge group from $U(1)$ to $Z_2$.\cite{PhysRevD.19.3682,hansson2004superconductors} However, we note that in our case the $Z_2$ theory in naturally tuned to be along the self-dual line due to the $e\leftrightarrow m$ symmetry. Indeed a similar action was proposed in Ref. \onlinecite{PhysRevB.80.125101} for the transition to from the $Z_2$ QSL to a trivial paramagnet in context of Toric code models. However, there due to the absence of the $e\leftrightarrow m$ symmetry, the masses and other coupling constants of the $e$ sector were different from that of the $m$ sector and thereby away from the self-dual line.

In particular Eq. \ref{eq_gammazeesp} corresponds to the exactly the numerical calculations of Refs. \onlinecite{PhysRevB.82.085114} and \onlinecite{PhysRevB.85.195104} and series expansion techniques of Ref. \onlinecite{PhysRevB.79.033109}. In the lights of our present discussion it is certainly worthwhile to understand if the entire range from $h_z=0$ to $\Gamma=0$ is given by same physics as this would indicate an extremely interesting mapping of the physics under Zeeman perturbations to that of the pseudo-dipolar interactions. Of course this conclusion is tentative at this point as the above mapping is drawn on the basis of the effects of the leading order perturbations and the effects of the higher order terms needs to be taken into account in any systematic numerical calculation. 

Assuming that the physics of the self-dual line is relevant for the present discussion, we now ask what do we know about this line ? From the point of view of $Z_2$ gauge theory with dynamic electric and magnetic charges, the $e$-Higgs and the $m$-Higgs (confined) phases are smoothly connected\cite{PhysRevD.19.3682} with both phase transition belonging to $3D-Z_2$ universality class and meeting at a multicritical point which merges with a line of first order transition along the self-dual line ending in a critical point. Series expansion techniques of Ref. \onlinecite{PhysRevB.79.033109} show that along the self-dual line, the charge gap, $\Delta$, for both the $e$ and $m$ sectors vanishes as $\Delta\sim |h_z-h_z^c|^{z\nu}$ where the critical value of the field is estimated to be $h_z^c\approx0.34$ (for $J_{\rm TC}=1$) and the exponent, $z\nu\approx 0.69-0.70$\cite{PhysRevLett.106.107203,PhysRevB.79.033109}-- different from the $3D-Z_2$ value, $(z\nu)_{3D-Z_2}=0.6301$. The first-order transition, on the other hand ends at $h_c^{1st}\approx 0.42$.\cite{PhysRevB.85.195104} The above picture is confrmed by Monte Carlo calculations.\cite{PhysRevB.82.085114,PhysRevD.21.3360,genovese2003phase,PhysRevB.85.195104}

We end this section with two more comments. First, away from the self-dual line when either $e$ or $m$ charges condense, neglecting the gauge fluctuations of $A_\mu$ and $B_\mu$ at the critical point on the grounds that the CS term makes the respective photons massive results in a transition correctly belonging to the $3D-Z_2$ universality class.\cite{PhysRevB.80.125101} A similar mean-field assumption would lead to $3D-XY$ transition in the present case with $(z\nu)_{3D-XY}\approx 0.67155$.\cite{PhysRevB.63.214503} Second, recent series expansion calculations\cite{PhysRevLett.106.107203} and tensor-network based wave-function analysis\cite{PhysRevLett.122.176401} has been suggested that perturbations on the self-dual line in the vicinity of the multicritical point can open up a gapless phase with power-law correlations for the $e$ and $m$ charges with continuously varying exponents.  However, in our severely finite-size limited ED results, we did not find any signatures of such a phase. The relevance of this physics to the anisotropic Kitaev model in a magnetic field as well as the higher order terms neglected in Eq. \ref{eq_trzee} remains to be understood.

This concludes our discussion of the phases and phase transitions for the ferromagnetic Kitaev-Heisenberg-$\Gamma$ model in the anisotropic limit.

\section{Summary and Outlook}
\label{sec_summ}

We now summarise our results. In this work we have worked out the details for of the phases and phase transitions out of the gapped $Z_2$ QSL in the Toric code limit into a spin-ordered phase (driven by Heisenberg interactions) and a trivial paramagnet (driven by pseudo-dipolar interactions). These interactions are relevant for the material realisation such as $\alpha$-RuCl$_3$, albeit in the isotropic limit. While, even in $\alpha$-RuCl$_3$ and other related materials, the three bonds are not exactly symmetric\cite{ziatdinov2016atomic}, our calculations are clearly do not apply to the known Kitaev materials in its present form but poses important starting point in understanding the unconventional deconfined phase transitions out of the $Z_2$ QSL.

In our present calculation for the anisotropic limit, we find that the transition between the $Z_2$ QSL and the spin ordered phase is given by a self-dual modified Abelian Higgs model whereas that between the QSL and the trivial paramagnet is given by a self-dual $Z_2$ gauge theory. The self duality owes its origin to the anyon permutation symmetry which protects the structure of the critical theory. Would be interesting to understand other examples of such anyon permutation symmetry protected phase transitions. Finally, the phase diagram in Fig. \ref{fig_pd} allows for  interesting multicritical points where all the three phases-- topological, symmetry broken and trivial-- meet. The nature of the multicritical point is worth investigating.

To end, does the present calculations shed any light on the material relevant isotropic Kitaev-Heisenberg-$\Gamma$ model ? Given the correspondence of the QSL and the spin-ordered phases, it is tempting to conclude that the soft modes in the anisotropic limit indeed play an important role-- along with the gapless Majorana-- to determine the critical theory for the isotropic point. Outcomes of calculations along these lines would be interesting.

\acknowledgements
The authors would like to thank A. Agarwala, K. Damle, Y. B. Kim, A. V. Mallik, R. Moessner, A. Prakash, V. B. Shenoy and V. Tripathi for various enlightening discussions and collaborations in related topics. We acknowledge use of open-source QuSpin\cite{weinberg2017quspin,weinberg2019quspin} for exact diagonalisation calculations. KD acknowledges the hospitality of ICTS where a part of this work was done. SB acknowledges financial support through Max Planck partner group on strongly correlated systems at ICTS and SERB-DST (Govt. of India) early career research grant (No. ECR/2017/000504). We acknowledge support of the Department of Atomic Energy, Government of India, under project no.12-R\&D- TFR-5.10-1100. Computations were performed at the ICTS clusters {\it boson} and {\it boson1}.

\appendix
\section{Symmetry transformations}
\label{appen_symm}

\begin{figure}
\centering
\includegraphics[scale=.4]{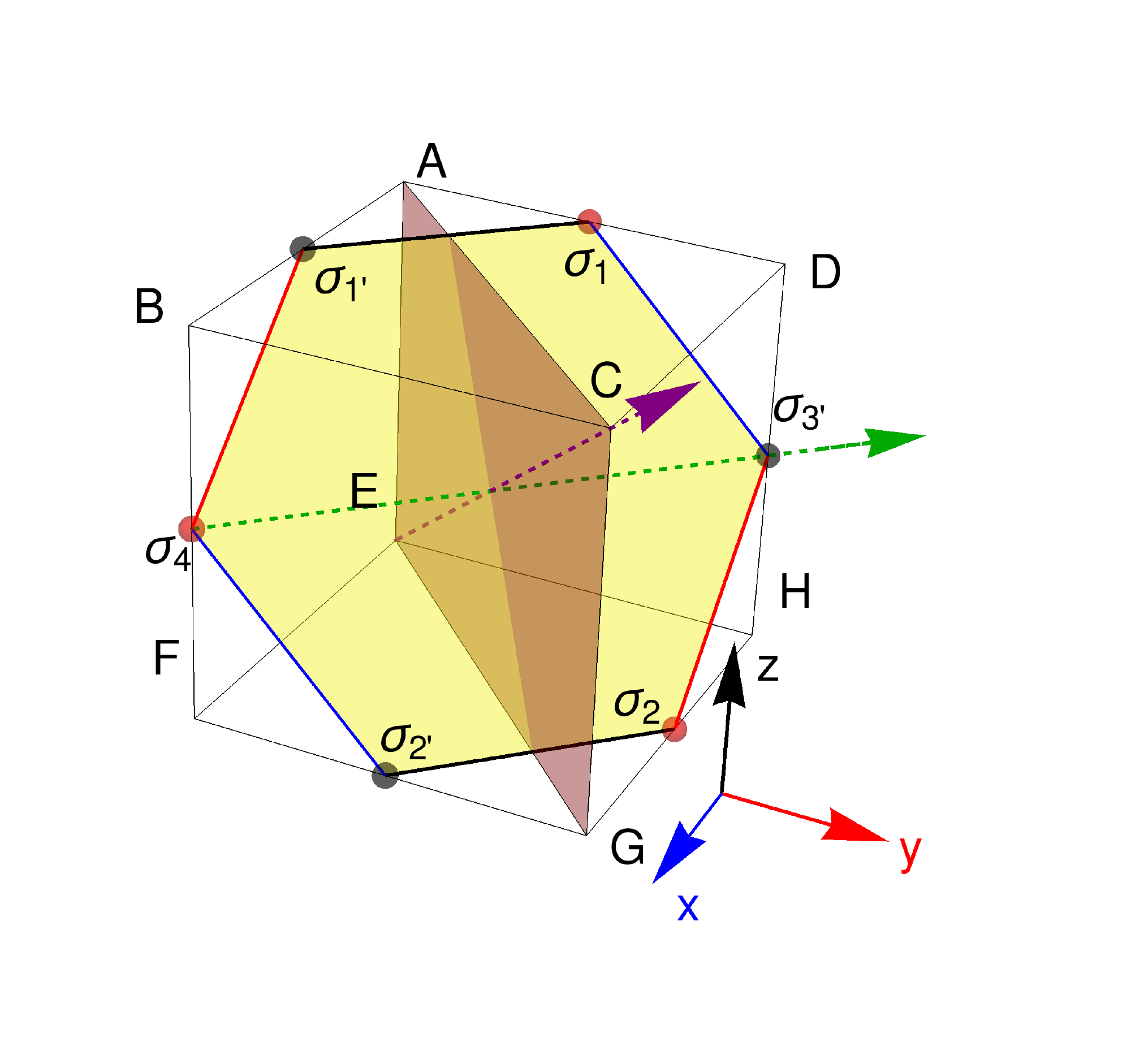}
\caption{The hexagon embedded in a cube with the cartesian $[111]$ direction being normal to the hexagon plane. The red and black sub-lattice structure of the $\sigma$-spins are same as in Fig. \ref{fig_hcomb_symm_2d}. Green dashed arrow along $[\bar110]$ direction is the $\pi$-rotation axis,  $C_{2z}$. Light red plane cutting two of the z-bonds is the mirror reflection, $\sigma_v$. Purple dashed  arrow along $[111]$ direction is the six fold rotation supplemented by the mirror reflection about the honeycomb plane, $\sigma_hC_6$.}
\label{fig_hcomb_symm}
\end{figure}
\begin{figure}
\centering
\includegraphics[scale=.4]{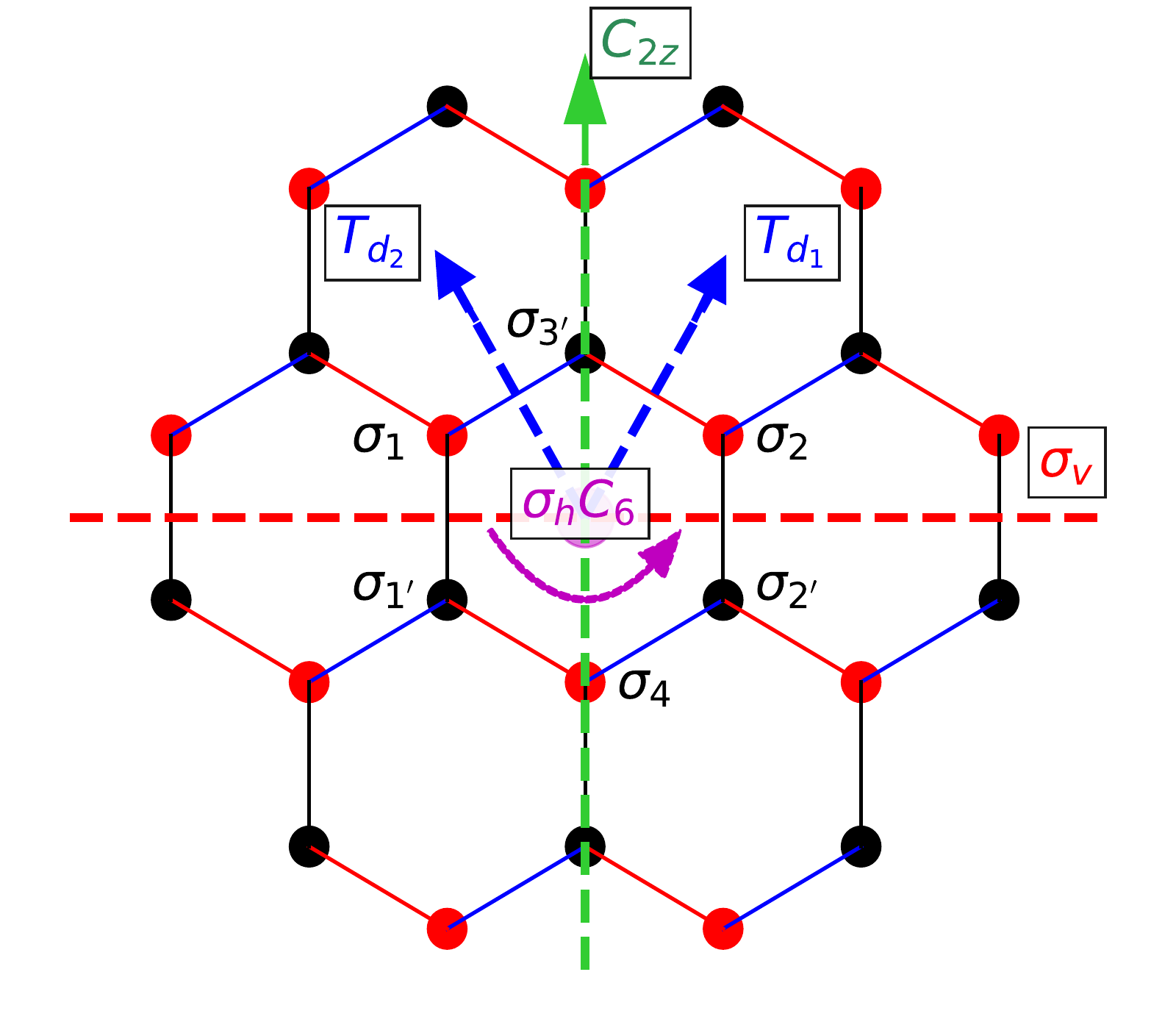}
\caption{Symmetries with respect to the hexagonal plane. The colouring have one to one correspondance with Fig. \ref{fig_hcomb_symm}.}
\label{fig_hcomb_symm_2d}
\end{figure}

Since the Kitaev model is realised in spin-orbit coupled systems, the magnetic moment transforms non-trivially under spin and real space rotations. Following Ref. \onlinecite{PhysRevB.86.085145} we embed the honeycomb in a cube as shown in Fig. \ref{fig_hcomb_symm}. The generators of the symmetries (Fig. \ref{fig_hcomb_symm_2d}) for the underlying honeycomb lattice are given by
\begin{itemize}
\item Time reversal symmetry, $\mathcal{T}$, where the $\sigma^\alpha$ spins transform as Kramers doublets in all the known candidate materials.
\item Translations in the honeycomb plane, $T_{d_1}$ and $T_{d_2}$.
\item $C_6$ rotation about  $[111]$ about the centre of the hexagon, followed by reflection, $\sigma_h$, about the honeycomb plane ($\equiv \sigma_hC_6$).
\item Reflection, $\sigma_v$, about the $x=y$ plane.
\end{itemize}
In the anisotropic limit, $\mathcal{T}, T_{d_1}, T_{d_2}$ and $\sigma_v$ remain intact while $\sigma_hC_6$ is absent. However, the combination $C_{2z}= (\sigma_hC_6)^2\sigma_v(\sigma_hC_6)^{-1}$ is still a symmetry. Further, in addition to the above symmetries we find it useful to consider the $\pi$-rotation about $[111]$ through the center of the hexagon, namely $R_{\pi} =  (\sigma_hC_6)^3 = C_{2z}\sigma_v$. 

\begin{figure}
\centering
\includegraphics[scale=.4]{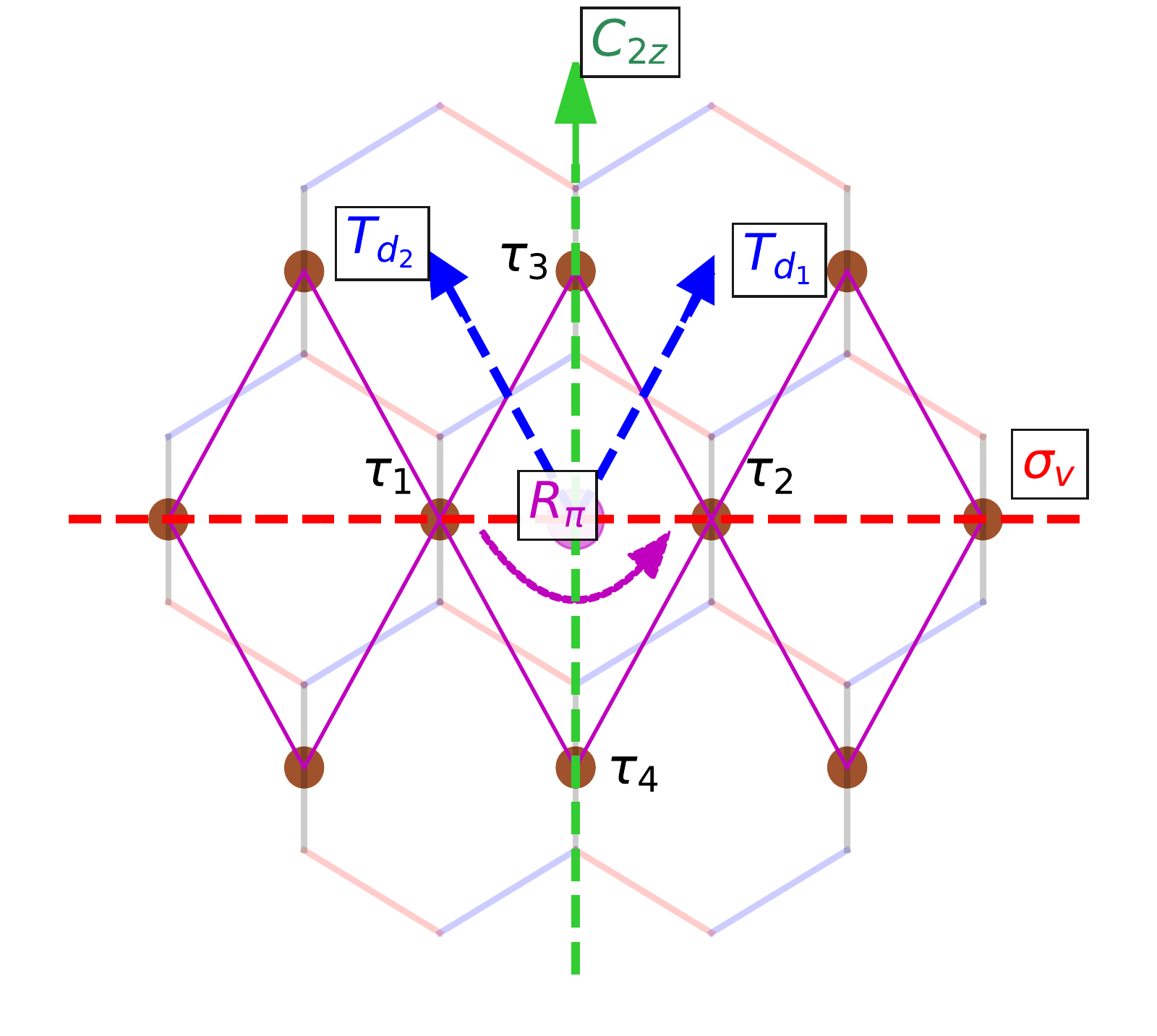}
\caption{In the anisotropic limit the $\sigma$-spin give rise $\tau$-spins, they are labeled by the brown spheres at the vertices of the rhombic lattice (magenta lines). Also the honeycomb lattice is shown in the background. The relevant symmetries are shown following the convention of Fig. \ref{fig_hcomb_symm_2d}.}
\label{fig_hcomb_symm_2d_tau}
\end{figure}

The non-trivial transformation of the $\sigma$ spins under various symmetries (except for the two translations which is rather straight forward) is given in Table \ref{table_sig_symm}. It is now easy to work out the action of the surviving symmetries on the $\tau$ spins as shown in Fig. \ref{fig_hcomb_symm_2d_tau}.

\begin{table}
\begin{center}
 \begin{tabular}{|c|c c|c c|c c|c c| c c c|} 
 \hline
 Symmetry & $1$ & $1^{\prime}$ & $2$ & $2^{\prime}$ & $3$ & $3^{\prime}$ & $4$ & $4^{\prime}$ & $\sigma^x$ & $\sigma^y$ & $\sigma^z$ \\ [0.5ex] 
 \hline\hline
 $\mathcal{T}$ & $1$ & $1^{\prime}$ & $2$ & $2^{\prime}$ & $3$ & $3^{\prime}$ & $4$ & $4^{\prime}$ & -$\sigma^x$ & -$\sigma^y$ & -$\sigma^z$ \\ \hline
 $\sigma_hC_6$ & $1^{\prime}$ & $4$ & $3^{\prime}$ & $2$ & $5^{\prime}$ & ${1}$ & $2^{\prime}$ & ${8}$ & $\sigma^z$ & $\sigma^x$ & $\sigma^y$ \\ \hline
 $\sigma_v$ & ${1^{\prime}}$ & $1$ & ${2^{\prime}}$ & $2$ & ${4^{\prime}}$ & $4$ & ${3^{\prime}}$ & $3$ & -$\sigma^y$ & -$\sigma^x$ & -$\sigma^z$ \\ \hline \hline
 $C_{2z}$ & $2$ & ${2^{\prime}}$ & $1$ & ${1^{\prime}}$ & $3$ & ${3^{\prime}}$ & $4$ & ${4^{\prime}}$ & -$\sigma^y$ & -$\sigma^x$ & -$\sigma^z$ \\ \hline 
 $R_{\pi}$ & ${2^{\prime}}$ & $2$ & ${1^{\prime}}$ & $1$ & ${4^{\prime}}$ & $4$ & ${3^{\prime}}$ & $3$ & $\sigma^x$ & $\sigma^y$ & $\sigma^z$ \\ \hline
\end{tabular}
\caption{The time reversal and point group symmetry operations on the $\sigma$-spin of the central hexagon of the Fig. \ref{fig_hcomb_symm_2d}. Second to fifth columns indicate the transformations of the lattice labels while last column shows the spin space rotations.}

\label{table_sig_symm}
\end{center}
\end{table}

\paragraph{Time reversal Symmetry, $\mathcal{T}$ :}
For the $\sigma$ spins, the time reversal symmetry is implemented by the regular operator $i\sigma^y\mathcal{K}$ where $\mathcal{K}$ is the conjugation operator.  Therefore using Eq. \ref{eq_tau}, under time reversal :
\begin{align}
\mathcal{T}:\{|+\rangle,|-\rangle\}\rightarrow\{|-\rangle,|+\rangle\}
\end{align}
\paragraph{Reflection symmetry, $\sigma_v$ :}
Following Table \ref{table_sig_symm}, we define the symmetry transformation operator for the $\sigma$ spins as 
\begin{align}
 \sigma_v(11^{\prime})&=\hat{\mathbb{E}}(11^{\prime})e^{-i\frac{\hat{n}_v.\vec{\sigma}_{1}\pi}{2}}e^{-i\frac{\hat{n}_v.\vec{\sigma}_{1^{\prime}}\pi}{2}} \\ \nonumber
\end{align}
where $\hat{\mathbb{E}}((11^{\prime}))$ is the exchange operator between the $1\ \&\ 1^{\prime}$ and $\hat{n}_v=\frac{1}{\sqrt{2}}(-1,1,0)$. This gives rise to the following transformation:
\begin{align}
\sigma_v(11^{\prime}):\{\ket{+}_1,\ket{-}_1\}\rightarrow\{i\ket{-}_1,-i\ket{+}_1\}
\end{align}
\paragraph{Rotation about the z-bond, $C_{2z}$ :}
From Table \ref{table_sig_symm} we focus on  the pair $\sigma_{1^{\prime}}$ and $\sigma_{2^{\prime}}$ which are mapped into each other under the symmetry transformation. This gives 
\begin{align}
 C_{2z}(1^{\prime}2^{\prime})&=\hat{\mathbb{E}}(1^{\prime}2^{\prime})e^{-i\frac{\hat{n}_{2z}.\vec{\sigma}_{1^{\prime}}\pi}{2}} e^{-i\frac{\hat{n}_{2z}.\vec{\sigma}_{2^{\prime}}\pi}{2}} 
\end{align}
with $\hat{n}_{2z} = \hat{n}_{v}$. For the $\tau$-spins we therefore have
\begin{align}
C_{2z}(12;1^{\prime}2^{\prime}):\{\ket{+}_{1(2)},\ket{-}_{1(2)}\}\rightarrow\{i\ket{-}_{2(1)},-i\ket{+}_{2(1)}\}
\end{align}
\paragraph{Rotation about the center of the hexagon, $R_{\pi}$ :} From Table \ref{table_sig_symm}, we this transformation does not mix the spin components, but, it introduces lattice transformation including interchange of the two sub-lattices. From Fig. \ref{fig_hcomb_symm_2d}, focusing on the spins $\sigma_{1(1^{\prime})}$ and $\sigma_{2(2^{\prime})}$, the exchange operators $\mathbb{E}(12^{\prime})~\text{and}~\mathbb{E}(1^{\prime}2))$ are defined, they exchange between $\sigma_1~\&~\sigma_{2^{\prime}}$ and $\sigma_2~\&~\sigma_{1^{\prime}}$. The effect of the symmetry operation on the $\tau^z$ basis state is as following
\begin{align}
R_{\pi}(12^{\prime};1^{\prime}2): \{\ket{+}_{1(2)},\ket{-}_{1(2)}\}\rightarrow\{\ket{+}_{2(1)},\ket{-}_{2{1}}\}
\end{align}
The above transformations results in the symmetry table as summarised in Eq. \ref{eq_tausym}. 

Due to the difference in rotation (see Eq. \ref{eq_rot_ver}), the $\ttau$-spins on horizontal and vertical bonds of the square lattice transform differently under the symmetries. This is summarised in Table \ref{table_tau_symm_fm_h}.
\begin{table}
\begin{center}
 \begin{tabular}{|c| c c c|c c c|} 
 \hline
 Symmetry & $\ttau^x_{h}$ & $\ttau^y_{h}$ & $\ttau^z_{h}$  & $-\ttau^x_{v}$ & $\ttau^y_{v}$ & $\ttau^z_{v}$ \\ [0.5ex] 
 \hline\hline
 $\mathcal{T}$ & $\ttau^x_{h}$ & $\ttau^y_{h}$ & $-\ttau^z_{h}$ & $-\ttau^x_{v^{\prime}}$ & $-\ttau^y_{v^{\prime}}$ & $\ttau^z_{v^{\prime}}$ \\  [0.5ex] \hline
 $\sigma_v$ & $\ttau^x_{h^{\prime}}$ & $-\ttau^y_{h^{\prime}}$ & $-\ttau^z_{h^{\prime}}$ & $-\ttau^x_{v^{\prime}}$ & $-\ttau^y_{v^{\prime}}$ & $\ttau^z_{v^{\prime}}$\\ [0.5ex] \hline
 $C_{2z}$ & $\ttau^x_{h^{\prime}}$ & $-\ttau^y_{h^{\prime}}$ & $-\ttau^z_{h^{\prime}}$ & $-\ttau^x_{v^{\prime}}$ & $-\ttau^y_{v^{\prime}}$ & $\ttau^z_{v^{\prime}}$\\ [0.5ex] \hline
 $R_{\pi}$ & $\ttau^x_{h^{\prime}}$ & $\ttau^y_{h^{\prime}}$ & $\ttau^z_{h^{\prime}}$ & $\ttau^x_{v^{\prime}}$ & $\ttau^y_{v^{\prime}}$ & $\ttau^z_{v^{\prime}}$\\ [0.5ex] \hline
 $T_{d_j}$ & $\ttau^z_{v_j}$ & $-\ttau^y_{v_j}$ & $\ttau^x_{v_j}$ & $\ttau^z_{h_j}$ & $-\ttau^y_{h_j}$ & $\ttau^x_{h_j}$ \\ [0.5ex] \hline
  \end{tabular}
\caption{Symmetry transformations of the $\ttau$ spins on the horizontal ($h$) and vertical ($v$) bonds.}
\label{table_tau_symm_fm_h}
\end{center}
\end{table}

\bibliography{biblio}

\begin{thebibliography}{106}%
\makeatletter
\providecommand \@ifxundefined [1]{%
 \@ifx{#1\undefined}
}%
\providecommand \@ifnum [1]{%
 \ifnum #1\expandafter \@firstoftwo
 \else \expandafter \@secondoftwo
 \fi
}%
\providecommand \@ifx [1]{%
 \ifx #1\expandafter \@firstoftwo
 \else \expandafter \@secondoftwo
 \fi
}%
\providecommand \natexlab [1]{#1}%
\providecommand \enquote  [1]{``#1''}%
\providecommand \bibnamefont  [1]{#1}%
\providecommand \bibfnamefont [1]{#1}%
\providecommand \citenamefont [1]{#1}%
\providecommand \href@noop [0]{\@secondoftwo}%
\providecommand \href [0]{\begingroup \@sanitize@url \@href}%
\providecommand \@href[1]{\@@startlink{#1}\@@href}%
\providecommand \@@href[1]{\endgroup#1\@@endlink}%
\providecommand \@sanitize@url [0]{\catcode `\\12\catcode `\$12\catcode
  `\&12\catcode `\#12\catcode `\^12\catcode `\_12\catcode `\%12\relax}%
\providecommand \@@startlink[1]{}%
\providecommand \@@endlink[0]{}%
\providecommand \url  [0]{\begingroup\@sanitize@url \@url }%
\providecommand \@url [1]{\endgroup\@href {#1}{\urlprefix }}%
\providecommand \urlprefix  [0]{URL }%
\providecommand \Eprint [0]{\href }%
\providecommand \doibase [0]{http://dx.doi.org/}%
\providecommand \selectlanguage [0]{\@gobble}%
\providecommand \bibinfo  [0]{\@secondoftwo}%
\providecommand \bibfield  [0]{\@secondoftwo}%
\providecommand \translation [1]{[#1]}%
\providecommand \BibitemOpen [0]{}%
\providecommand \bibitemStop [0]{}%
\providecommand \bibitemNoStop [0]{.\EOS\space}%
\providecommand \EOS [0]{\spacefactor3000\relax}%
\providecommand \BibitemShut  [1]{\csname bibitem#1\endcsname}%
\let\auto@bib@innerbib\@empty
\bibitem [{\citenamefont {Witczak-Krempa}\ \emph {et~al.}(2014)\citenamefont
  {Witczak-Krempa}, \citenamefont {Chen}, \citenamefont {Kim},\ and\
  \citenamefont {Balents}}]{witczak2014correlated}%
  \BibitemOpen
  \bibfield  {author} {\bibinfo {author} {\bibfnamefont {W.}~\bibnamefont
  {Witczak-Krempa}}, \bibinfo {author} {\bibfnamefont {G.}~\bibnamefont
  {Chen}}, \bibinfo {author} {\bibfnamefont {Y.~B.}\ \bibnamefont {Kim}}, \
  and\ \bibinfo {author} {\bibfnamefont {L.}~\bibnamefont {Balents}},\ }\href
  {\doibase https://doi.org/10.1146/annurev-conmatphys-020911-125138}
  {\bibfield  {journal} {\bibinfo  {journal} {Annu. Rev. Condens. Matter
  Phys.}\ }\textbf {\bibinfo {volume} {5}},\ \bibinfo {pages} {57} (\bibinfo
  {year} {2014})}\BibitemShut {NoStop}%
\bibitem [{\citenamefont {Anderson}(1973)}]{anderson1973resonating}%
  \BibitemOpen
  \bibfield  {author} {\bibinfo {author} {\bibfnamefont {P.}~\bibnamefont
  {Anderson}},\ }\href {\doibase https://doi.org/10.1016/0025-5408(73)90167-0}
  {\bibfield  {journal} {\bibinfo  {journal} {Materials Research Bulletin}\
  }\textbf {\bibinfo {volume} {8}},\ \bibinfo {pages} {153} (\bibinfo {year}
  {1973})}\BibitemShut {NoStop}%
\bibitem [{\citenamefont {Anderson}(1987)}]{anderson1987resonating}%
  \BibitemOpen
  \bibfield  {author} {\bibinfo {author} {\bibfnamefont {P.~W.}\ \bibnamefont
  {Anderson}},\ }\href {\doibase 10.1126/science.235.4793.1196} {\bibfield
  {journal} {\bibinfo  {journal} {science}\ }\textbf {\bibinfo {volume}
  {235}},\ \bibinfo {pages} {1196} (\bibinfo {year} {1987})}\BibitemShut
  {NoStop}%
\bibitem [{\citenamefont {Wen}(2002)}]{wen2002quantum}%
  \BibitemOpen
  \bibfield  {author} {\bibinfo {author} {\bibfnamefont {X.-G.}\ \bibnamefont
  {Wen}},\ }\href {\doibase 10.1103/PhysRevB.65.165113} {\bibfield  {journal}
  {\bibinfo  {journal} {Physical Review B}\ }\textbf {\bibinfo {volume} {65}},\
  \bibinfo {pages} {165113} (\bibinfo {year} {2002})}\BibitemShut {NoStop}%
\bibitem [{\citenamefont {Lee}\ \emph {et~al.}(2006)\citenamefont {Lee},
  \citenamefont {Nagaosa},\ and\ \citenamefont {Wen}}]{lee2006doping}%
  \BibitemOpen
  \bibfield  {author} {\bibinfo {author} {\bibfnamefont {P.~A.}\ \bibnamefont
  {Lee}}, \bibinfo {author} {\bibfnamefont {N.}~\bibnamefont {Nagaosa}}, \ and\
  \bibinfo {author} {\bibfnamefont {X.-G.}\ \bibnamefont {Wen}},\ }\href
  {\doibase https://doi.org/10.1103/RevModPhys.78.17} {\bibfield  {journal}
  {\bibinfo  {journal} {Reviews of Modern Physics}\ }\textbf {\bibinfo {volume}
  {78}},\ \bibinfo {pages} {17} (\bibinfo {year} {2006})}\BibitemShut {NoStop}%
\bibitem [{\citenamefont {Moessner}\ and\ \citenamefont
  {Sondhi}(2001{\natexlab{a}})}]{PhysRevLett.86.1881}%
  \BibitemOpen
  \bibfield  {author} {\bibinfo {author} {\bibfnamefont {R.}~\bibnamefont
  {Moessner}}\ and\ \bibinfo {author} {\bibfnamefont {S.~L.}\ \bibnamefont
  {Sondhi}},\ }\href {\doibase 10.1103/PhysRevLett.86.1881} {\bibfield
  {journal} {\bibinfo  {journal} {Phys. Rev. Lett.}\ }\textbf {\bibinfo
  {volume} {86}},\ \bibinfo {pages} {1881} (\bibinfo {year}
  {2001}{\natexlab{a}})}\BibitemShut {NoStop}%
\bibitem [{\citenamefont {Balents}(2010)}]{balents2010spin}%
  \BibitemOpen
  \bibfield  {author} {\bibinfo {author} {\bibfnamefont {L.}~\bibnamefont
  {Balents}},\ }\href {\doibase https://doi.org/10.1038/nature08917} {\bibfield
   {journal} {\bibinfo  {journal} {Nature}\ }\textbf {\bibinfo {volume}
  {464}},\ \bibinfo {pages} {199} (\bibinfo {year} {2010})}\BibitemShut
  {NoStop}%
\bibitem [{\citenamefont {Lee}(2008)}]{lee2008end}%
  \BibitemOpen
  \bibfield  {author} {\bibinfo {author} {\bibfnamefont {P.~A.}\ \bibnamefont
  {Lee}},\ }\href {\doibase 10.1126/science.1163196} {\bibfield  {journal}
  {\bibinfo  {journal} {Science}\ }\textbf {\bibinfo {volume} {321}},\ \bibinfo
  {pages} {1306} (\bibinfo {year} {2008})}\BibitemShut {NoStop}%
\bibitem [{\citenamefont {Savary}\ and\ \citenamefont
  {Balents}(2016)}]{savary2016quantum}%
  \BibitemOpen
  \bibfield  {author} {\bibinfo {author} {\bibfnamefont {L.}~\bibnamefont
  {Savary}}\ and\ \bibinfo {author} {\bibfnamefont {L.}~\bibnamefont
  {Balents}},\ }\href {\doibase https://doi.org/10.1088/0034-4885/80/1/016502}
  {\bibfield  {journal} {\bibinfo  {journal} {Reports on Progress in Physics}\
  }\textbf {\bibinfo {volume} {80}},\ \bibinfo {pages} {016502} (\bibinfo
  {year} {2016})}\BibitemShut {NoStop}%
\bibitem [{\citenamefont {Wen}(2017)}]{wen2017colloquium}%
  \BibitemOpen
  \bibfield  {author} {\bibinfo {author} {\bibfnamefont {X.-G.}\ \bibnamefont
  {Wen}},\ }\href {\doibase https://doi.org/10.1103/RevModPhys.89.041004}
  {\bibfield  {journal} {\bibinfo  {journal} {Reviews of Modern Physics}\
  }\textbf {\bibinfo {volume} {89}},\ \bibinfo {pages} {041004} (\bibinfo
  {year} {2017})}\BibitemShut {NoStop}%
\bibitem [{\citenamefont {Broholm}\ \emph {et~al.}(2020)\citenamefont
  {Broholm}, \citenamefont {Cava}, \citenamefont {Kivelson}, \citenamefont
  {Nocera}, \citenamefont {Norman},\ and\ \citenamefont
  {Senthil}}]{broholm2019quantum}%
  \BibitemOpen
  \bibfield  {author} {\bibinfo {author} {\bibfnamefont {C.}~\bibnamefont
  {Broholm}}, \bibinfo {author} {\bibfnamefont {R.}~\bibnamefont {Cava}},
  \bibinfo {author} {\bibfnamefont {S.}~\bibnamefont {Kivelson}}, \bibinfo
  {author} {\bibfnamefont {D.}~\bibnamefont {Nocera}}, \bibinfo {author}
  {\bibfnamefont {M.}~\bibnamefont {Norman}}, \ and\ \bibinfo {author}
  {\bibfnamefont {T.}~\bibnamefont {Senthil}},\ }\href {\doibase
  10.1126/science.aay0668} {\bibfield  {journal} {\bibinfo  {journal}
  {Science}\ }\textbf {\bibinfo {volume} {367}} (\bibinfo {year} {2020}),\
  10.1126/science.aay0668}\BibitemShut {NoStop}%
\bibitem [{\citenamefont {Cao}\ \emph {et~al.}(2016)\citenamefont {Cao},
  \citenamefont {Banerjee}, \citenamefont {Yan}, \citenamefont {Bridges},
  \citenamefont {Lumsden}, \citenamefont {Mandrus}, \citenamefont {Tennant},
  \citenamefont {Chakoumakos},\ and\ \citenamefont
  {Nagler}}]{PhysRevB.93.134423}%
  \BibitemOpen
  \bibfield  {author} {\bibinfo {author} {\bibfnamefont {H.~B.}\ \bibnamefont
  {Cao}}, \bibinfo {author} {\bibfnamefont {A.}~\bibnamefont {Banerjee}},
  \bibinfo {author} {\bibfnamefont {J.-Q.}\ \bibnamefont {Yan}}, \bibinfo
  {author} {\bibfnamefont {C.~A.}\ \bibnamefont {Bridges}}, \bibinfo {author}
  {\bibfnamefont {M.~D.}\ \bibnamefont {Lumsden}}, \bibinfo {author}
  {\bibfnamefont {D.~G.}\ \bibnamefont {Mandrus}}, \bibinfo {author}
  {\bibfnamefont {D.~A.}\ \bibnamefont {Tennant}}, \bibinfo {author}
  {\bibfnamefont {B.~C.}\ \bibnamefont {Chakoumakos}}, \ and\ \bibinfo {author}
  {\bibfnamefont {S.~E.}\ \bibnamefont {Nagler}},\ }\href {\doibase
  10.1103/PhysRevB.93.134423} {\bibfield  {journal} {\bibinfo  {journal} {Phys.
  Rev. B}\ }\textbf {\bibinfo {volume} {93}},\ \bibinfo {pages} {134423}
  (\bibinfo {year} {2016})}\BibitemShut {NoStop}%
\bibitem [{\citenamefont {Banerjee}\ \emph {et~al.}(2016)\citenamefont
  {Banerjee}, \citenamefont {Bridges}, \citenamefont {Yan}, \citenamefont
  {Aczel}, \citenamefont {Li}, \citenamefont {Stone}, \citenamefont {Granroth},
  \citenamefont {Lumsden}, \citenamefont {Yiu}, \citenamefont {Knolle} \emph
  {et~al.}}]{Banerjee2016}%
  \BibitemOpen
  \bibfield  {author} {\bibinfo {author} {\bibfnamefont {A.}~\bibnamefont
  {Banerjee}}, \bibinfo {author} {\bibfnamefont {C.}~\bibnamefont {Bridges}},
  \bibinfo {author} {\bibfnamefont {J.-Q.}\ \bibnamefont {Yan}}, \bibinfo
  {author} {\bibfnamefont {A.}~\bibnamefont {Aczel}}, \bibinfo {author}
  {\bibfnamefont {L.}~\bibnamefont {Li}}, \bibinfo {author} {\bibfnamefont
  {M.}~\bibnamefont {Stone}}, \bibinfo {author} {\bibfnamefont
  {G.}~\bibnamefont {Granroth}}, \bibinfo {author} {\bibfnamefont
  {M.}~\bibnamefont {Lumsden}}, \bibinfo {author} {\bibfnamefont
  {Y.}~\bibnamefont {Yiu}}, \bibinfo {author} {\bibfnamefont {J.}~\bibnamefont
  {Knolle}},  \emph {et~al.},\ }\href {\doibase
  https://doi.org/10.1038/nmat4604} {\bibfield  {journal} {\bibinfo  {journal}
  {Nature materials}\ }\textbf {\bibinfo {volume} {15}},\ \bibinfo {pages}
  {733} (\bibinfo {year} {2016})}\BibitemShut {NoStop}%
\bibitem [{\citenamefont {Banerjee}\ \emph {et~al.}(2017)\citenamefont
  {Banerjee}, \citenamefont {Yan}, \citenamefont {Knolle}, \citenamefont
  {Bridges}, \citenamefont {Stone}, \citenamefont {Lumsden}, \citenamefont
  {Mandrus}, \citenamefont {Tennant}, \citenamefont {Moessner},\ and\
  \citenamefont {Nagler}}]{Banerjee2017}%
  \BibitemOpen
  \bibfield  {author} {\bibinfo {author} {\bibfnamefont {A.}~\bibnamefont
  {Banerjee}}, \bibinfo {author} {\bibfnamefont {J.}~\bibnamefont {Yan}},
  \bibinfo {author} {\bibfnamefont {J.}~\bibnamefont {Knolle}}, \bibinfo
  {author} {\bibfnamefont {C.~A.}\ \bibnamefont {Bridges}}, \bibinfo {author}
  {\bibfnamefont {M.~B.}\ \bibnamefont {Stone}}, \bibinfo {author}
  {\bibfnamefont {M.~D.}\ \bibnamefont {Lumsden}}, \bibinfo {author}
  {\bibfnamefont {D.~G.}\ \bibnamefont {Mandrus}}, \bibinfo {author}
  {\bibfnamefont {D.~A.}\ \bibnamefont {Tennant}}, \bibinfo {author}
  {\bibfnamefont {R.}~\bibnamefont {Moessner}}, \ and\ \bibinfo {author}
  {\bibfnamefont {S.~E.}\ \bibnamefont {Nagler}},\ }\href {\doibase
  10.1126/science.aah6015} {\bibfield  {journal} {\bibinfo  {journal}
  {Science}\ }\textbf {\bibinfo {volume} {356}},\ \bibinfo {pages} {1055}
  (\bibinfo {year} {2017})}\BibitemShut {NoStop}%
\bibitem [{\citenamefont {Trebst}(2017)}]{trebst2017kitaev}%
  \BibitemOpen
  \bibfield  {author} {\bibinfo {author} {\bibfnamefont {S.}~\bibnamefont
  {Trebst}},\ }\href {https://arxiv.org/abs/1701.07056} {\bibfield  {journal}
  {\bibinfo  {journal} {arXiv preprint arXiv:1701.07056}\ } (\bibinfo {year}
  {2017})}\BibitemShut {NoStop}%
\bibitem [{\citenamefont {Knolle}\ \emph {et~al.}(2018)\citenamefont {Knolle},
  \citenamefont {Bhattacharjee},\ and\ \citenamefont
  {Moessner}}]{PhysRevB.97.134432}%
  \BibitemOpen
  \bibfield  {author} {\bibinfo {author} {\bibfnamefont {J.}~\bibnamefont
  {Knolle}}, \bibinfo {author} {\bibfnamefont {S.}~\bibnamefont
  {Bhattacharjee}}, \ and\ \bibinfo {author} {\bibfnamefont {R.}~\bibnamefont
  {Moessner}},\ }\href {\doibase 10.1103/PhysRevB.97.134432} {\bibfield
  {journal} {\bibinfo  {journal} {Phys. Rev. B}\ }\textbf {\bibinfo {volume}
  {97}},\ \bibinfo {pages} {134432} (\bibinfo {year} {2018})}\BibitemShut
  {NoStop}%
\bibitem [{\citenamefont {Ross}\ \emph {et~al.}(2009)\citenamefont {Ross},
  \citenamefont {Ruff}, \citenamefont {Adams}, \citenamefont {Gardner},
  \citenamefont {Dabkowska}, \citenamefont {Qiu}, \citenamefont {Copley},\ and\
  \citenamefont {Gaulin}}]{PhysRevLett.103.227202}%
  \BibitemOpen
  \bibfield  {author} {\bibinfo {author} {\bibfnamefont {K.~A.}\ \bibnamefont
  {Ross}}, \bibinfo {author} {\bibfnamefont {J.~P.~C.}\ \bibnamefont {Ruff}},
  \bibinfo {author} {\bibfnamefont {C.~P.}\ \bibnamefont {Adams}}, \bibinfo
  {author} {\bibfnamefont {J.~S.}\ \bibnamefont {Gardner}}, \bibinfo {author}
  {\bibfnamefont {H.~A.}\ \bibnamefont {Dabkowska}}, \bibinfo {author}
  {\bibfnamefont {Y.}~\bibnamefont {Qiu}}, \bibinfo {author} {\bibfnamefont
  {J.~R.~D.}\ \bibnamefont {Copley}}, \ and\ \bibinfo {author} {\bibfnamefont
  {B.~D.}\ \bibnamefont {Gaulin}},\ }\href {\doibase
  10.1103/PhysRevLett.103.227202} {\bibfield  {journal} {\bibinfo  {journal}
  {Phys. Rev. Lett.}\ }\textbf {\bibinfo {volume} {103}},\ \bibinfo {pages}
  {227202} (\bibinfo {year} {2009})}\BibitemShut {NoStop}%
\bibitem [{\citenamefont {Ross}\ \emph {et~al.}(2011)\citenamefont {Ross},
  \citenamefont {Savary}, \citenamefont {Gaulin},\ and\ \citenamefont
  {Balents}}]{PhysRevX.1.021002}%
  \BibitemOpen
  \bibfield  {author} {\bibinfo {author} {\bibfnamefont {K.~A.}\ \bibnamefont
  {Ross}}, \bibinfo {author} {\bibfnamefont {L.}~\bibnamefont {Savary}},
  \bibinfo {author} {\bibfnamefont {B.~D.}\ \bibnamefont {Gaulin}}, \ and\
  \bibinfo {author} {\bibfnamefont {L.}~\bibnamefont {Balents}},\ }\href
  {\doibase 10.1103/PhysRevX.1.021002} {\bibfield  {journal} {\bibinfo
  {journal} {Phys. Rev. X}\ }\textbf {\bibinfo {volume} {1}},\ \bibinfo {pages}
  {021002} (\bibinfo {year} {2011})}\BibitemShut {NoStop}%
\bibitem [{\citenamefont {Gaudet}\ \emph {et~al.}(2016)\citenamefont {Gaudet},
  \citenamefont {Ross}, \citenamefont {Kermarrec}, \citenamefont {Butch},
  \citenamefont {Ehlers}, \citenamefont {Dabkowska},\ and\ \citenamefont
  {Gaulin}}]{PhysRevB.93.064406}%
  \BibitemOpen
  \bibfield  {author} {\bibinfo {author} {\bibfnamefont {J.}~\bibnamefont
  {Gaudet}}, \bibinfo {author} {\bibfnamefont {K.~A.}\ \bibnamefont {Ross}},
  \bibinfo {author} {\bibfnamefont {E.}~\bibnamefont {Kermarrec}}, \bibinfo
  {author} {\bibfnamefont {N.~P.}\ \bibnamefont {Butch}}, \bibinfo {author}
  {\bibfnamefont {G.}~\bibnamefont {Ehlers}}, \bibinfo {author} {\bibfnamefont
  {H.~A.}\ \bibnamefont {Dabkowska}}, \ and\ \bibinfo {author} {\bibfnamefont
  {B.~D.}\ \bibnamefont {Gaulin}},\ }\href {\doibase
  10.1103/PhysRevB.93.064406} {\bibfield  {journal} {\bibinfo  {journal} {Phys.
  Rev. B}\ }\textbf {\bibinfo {volume} {93}},\ \bibinfo {pages} {064406}
  (\bibinfo {year} {2016})}\BibitemShut {NoStop}%
\bibitem [{\citenamefont {Thompson}\ \emph {et~al.}(2017)\citenamefont
  {Thompson}, \citenamefont {McClarty}, \citenamefont {Prabhakaran},
  \citenamefont {Cabrera}, \citenamefont {Guidi},\ and\ \citenamefont
  {Coldea}}]{PhysRevLett.119.057203}%
  \BibitemOpen
  \bibfield  {author} {\bibinfo {author} {\bibfnamefont {J.~D.}\ \bibnamefont
  {Thompson}}, \bibinfo {author} {\bibfnamefont {P.~A.}\ \bibnamefont
  {McClarty}}, \bibinfo {author} {\bibfnamefont {D.}~\bibnamefont
  {Prabhakaran}}, \bibinfo {author} {\bibfnamefont {I.}~\bibnamefont
  {Cabrera}}, \bibinfo {author} {\bibfnamefont {T.}~\bibnamefont {Guidi}}, \
  and\ \bibinfo {author} {\bibfnamefont {R.}~\bibnamefont {Coldea}},\ }\href
  {\doibase 10.1103/PhysRevLett.119.057203} {\bibfield  {journal} {\bibinfo
  {journal} {Phys. Rev. Lett.}\ }\textbf {\bibinfo {volume} {119}},\ \bibinfo
  {pages} {057203} (\bibinfo {year} {2017})}\BibitemShut {NoStop}%
\bibitem [{\citenamefont {Scheie}\ \emph {et~al.}(2017)\citenamefont {Scheie},
  \citenamefont {Kindervater}, \citenamefont {S\"aubert}, \citenamefont
  {Duvinage}, \citenamefont {Pfleiderer}, \citenamefont {Changlani},
  \citenamefont {Zhang}, \citenamefont {Harriger}, \citenamefont {Arpino},
  \citenamefont {Koohpayeh}, \citenamefont {Tchernyshyov},\ and\ \citenamefont
  {Broholm}}]{PhysRevLett.119.127201}%
  \BibitemOpen
  \bibfield  {author} {\bibinfo {author} {\bibfnamefont {A.}~\bibnamefont
  {Scheie}}, \bibinfo {author} {\bibfnamefont {J.}~\bibnamefont {Kindervater}},
  \bibinfo {author} {\bibfnamefont {S.}~\bibnamefont {S\"aubert}}, \bibinfo
  {author} {\bibfnamefont {C.}~\bibnamefont {Duvinage}}, \bibinfo {author}
  {\bibfnamefont {C.}~\bibnamefont {Pfleiderer}}, \bibinfo {author}
  {\bibfnamefont {H.~J.}\ \bibnamefont {Changlani}}, \bibinfo {author}
  {\bibfnamefont {S.}~\bibnamefont {Zhang}}, \bibinfo {author} {\bibfnamefont
  {L.}~\bibnamefont {Harriger}}, \bibinfo {author} {\bibfnamefont
  {K.}~\bibnamefont {Arpino}}, \bibinfo {author} {\bibfnamefont {S.~M.}\
  \bibnamefont {Koohpayeh}}, \bibinfo {author} {\bibfnamefont {O.}~\bibnamefont
  {Tchernyshyov}}, \ and\ \bibinfo {author} {\bibfnamefont {C.}~\bibnamefont
  {Broholm}},\ }\href {\doibase 10.1103/PhysRevLett.119.127201} {\bibfield
  {journal} {\bibinfo  {journal} {Phys. Rev. Lett.}\ }\textbf {\bibinfo
  {volume} {119}},\ \bibinfo {pages} {127201} (\bibinfo {year}
  {2017})}\BibitemShut {NoStop}%
\bibitem [{\citenamefont {Hentrich}\ \emph {et~al.}(2018)\citenamefont
  {Hentrich}, \citenamefont {Wolter}, \citenamefont {Zotos}, \citenamefont
  {Brenig}, \citenamefont {Nowak}, \citenamefont {Isaeva}, \citenamefont
  {Doert}, \citenamefont {Banerjee}, \citenamefont {Lampen-Kelley},
  \citenamefont {Mandrus}, \citenamefont {Nagler}, \citenamefont {Sears},
  \citenamefont {Kim}, \citenamefont {B\"uchner},\ and\ \citenamefont
  {Hess}}]{PhysRevLett.120.117204}%
  \BibitemOpen
  \bibfield  {author} {\bibinfo {author} {\bibfnamefont {R.}~\bibnamefont
  {Hentrich}}, \bibinfo {author} {\bibfnamefont {A.~U.~B.}\ \bibnamefont
  {Wolter}}, \bibinfo {author} {\bibfnamefont {X.}~\bibnamefont {Zotos}},
  \bibinfo {author} {\bibfnamefont {W.}~\bibnamefont {Brenig}}, \bibinfo
  {author} {\bibfnamefont {D.}~\bibnamefont {Nowak}}, \bibinfo {author}
  {\bibfnamefont {A.}~\bibnamefont {Isaeva}}, \bibinfo {author} {\bibfnamefont
  {T.}~\bibnamefont {Doert}}, \bibinfo {author} {\bibfnamefont
  {A.}~\bibnamefont {Banerjee}}, \bibinfo {author} {\bibfnamefont
  {P.}~\bibnamefont {Lampen-Kelley}}, \bibinfo {author} {\bibfnamefont {D.~G.}\
  \bibnamefont {Mandrus}}, \bibinfo {author} {\bibfnamefont {S.~E.}\
  \bibnamefont {Nagler}}, \bibinfo {author} {\bibfnamefont {J.}~\bibnamefont
  {Sears}}, \bibinfo {author} {\bibfnamefont {Y.-J.}\ \bibnamefont {Kim}},
  \bibinfo {author} {\bibfnamefont {B.}~\bibnamefont {B\"uchner}}, \ and\
  \bibinfo {author} {\bibfnamefont {C.}~\bibnamefont {Hess}},\ }\href {\doibase
  10.1103/PhysRevLett.120.117204} {\bibfield  {journal} {\bibinfo  {journal}
  {Phys. Rev. Lett.}\ }\textbf {\bibinfo {volume} {120}},\ \bibinfo {pages}
  {117204} (\bibinfo {year} {2018})}\BibitemShut {NoStop}%
\bibitem [{\citenamefont {Banerjee}\ \emph {et~al.}(2018)\citenamefont
  {Banerjee}, \citenamefont {Lampen-Kelley}, \citenamefont {Knolle},
  \citenamefont {Balz}, \citenamefont {Aczel}, \citenamefont {Winn},
  \citenamefont {Liu}, \citenamefont {Pajerowski}, \citenamefont {Yan},
  \citenamefont {Bridges} \emph {et~al.}}]{banerjee2018excitations}%
  \BibitemOpen
  \bibfield  {author} {\bibinfo {author} {\bibfnamefont {A.}~\bibnamefont
  {Banerjee}}, \bibinfo {author} {\bibfnamefont {P.}~\bibnamefont
  {Lampen-Kelley}}, \bibinfo {author} {\bibfnamefont {J.}~\bibnamefont
  {Knolle}}, \bibinfo {author} {\bibfnamefont {C.}~\bibnamefont {Balz}},
  \bibinfo {author} {\bibfnamefont {A.~A.}\ \bibnamefont {Aczel}}, \bibinfo
  {author} {\bibfnamefont {B.}~\bibnamefont {Winn}}, \bibinfo {author}
  {\bibfnamefont {Y.}~\bibnamefont {Liu}}, \bibinfo {author} {\bibfnamefont
  {D.}~\bibnamefont {Pajerowski}}, \bibinfo {author} {\bibfnamefont
  {J.}~\bibnamefont {Yan}}, \bibinfo {author} {\bibfnamefont {C.~A.}\
  \bibnamefont {Bridges}},  \emph {et~al.},\ }\href {\doibase
  https://doi.org/10.1038/s41535-018-0079-2} {\bibfield  {journal} {\bibinfo
  {journal} {npj Quantum Materials}\ }\textbf {\bibinfo {volume} {3}},\
  \bibinfo {pages} {1} (\bibinfo {year} {2018})}\BibitemShut {NoStop}%
\bibitem [{\citenamefont {Ponomaryov}\ \emph {et~al.}(2017)\citenamefont
  {Ponomaryov}, \citenamefont {Schulze}, \citenamefont {Wosnitza},
  \citenamefont {Lampen-Kelley}, \citenamefont {Banerjee}, \citenamefont {Yan},
  \citenamefont {Bridges}, \citenamefont {Mandrus}, \citenamefont {Nagler},
  \citenamefont {Kolezhuk},\ and\ \citenamefont
  {Zvyagin}}]{PhysRevB.96.241107}%
  \BibitemOpen
  \bibfield  {author} {\bibinfo {author} {\bibfnamefont {A.~N.}\ \bibnamefont
  {Ponomaryov}}, \bibinfo {author} {\bibfnamefont {E.}~\bibnamefont {Schulze}},
  \bibinfo {author} {\bibfnamefont {J.}~\bibnamefont {Wosnitza}}, \bibinfo
  {author} {\bibfnamefont {P.}~\bibnamefont {Lampen-Kelley}}, \bibinfo {author}
  {\bibfnamefont {A.}~\bibnamefont {Banerjee}}, \bibinfo {author}
  {\bibfnamefont {J.-Q.}\ \bibnamefont {Yan}}, \bibinfo {author} {\bibfnamefont
  {C.~A.}\ \bibnamefont {Bridges}}, \bibinfo {author} {\bibfnamefont {D.~G.}\
  \bibnamefont {Mandrus}}, \bibinfo {author} {\bibfnamefont {S.~E.}\
  \bibnamefont {Nagler}}, \bibinfo {author} {\bibfnamefont {A.~K.}\
  \bibnamefont {Kolezhuk}}, \ and\ \bibinfo {author} {\bibfnamefont {S.~A.}\
  \bibnamefont {Zvyagin}},\ }\href {\doibase 10.1103/PhysRevB.96.241107}
  {\bibfield  {journal} {\bibinfo  {journal} {Phys. Rev. B}\ }\textbf {\bibinfo
  {volume} {96}},\ \bibinfo {pages} {241107} (\bibinfo {year}
  {2017})}\BibitemShut {NoStop}%
\bibitem [{\citenamefont {Little}\ \emph {et~al.}(2017)\citenamefont {Little},
  \citenamefont {Wu}, \citenamefont {Lampen-Kelley}, \citenamefont {Banerjee},
  \citenamefont {Patankar}, \citenamefont {Rees}, \citenamefont {Bridges},
  \citenamefont {Yan}, \citenamefont {Mandrus}, \citenamefont {Nagler},\ and\
  \citenamefont {Orenstein}}]{PhysRevLett.119.227201}%
  \BibitemOpen
  \bibfield  {author} {\bibinfo {author} {\bibfnamefont {A.}~\bibnamefont
  {Little}}, \bibinfo {author} {\bibfnamefont {L.}~\bibnamefont {Wu}}, \bibinfo
  {author} {\bibfnamefont {P.}~\bibnamefont {Lampen-Kelley}}, \bibinfo {author}
  {\bibfnamefont {A.}~\bibnamefont {Banerjee}}, \bibinfo {author}
  {\bibfnamefont {S.}~\bibnamefont {Patankar}}, \bibinfo {author}
  {\bibfnamefont {D.}~\bibnamefont {Rees}}, \bibinfo {author} {\bibfnamefont
  {C.~A.}\ \bibnamefont {Bridges}}, \bibinfo {author} {\bibfnamefont {J.-Q.}\
  \bibnamefont {Yan}}, \bibinfo {author} {\bibfnamefont {D.}~\bibnamefont
  {Mandrus}}, \bibinfo {author} {\bibfnamefont {S.~E.}\ \bibnamefont {Nagler}},
  \ and\ \bibinfo {author} {\bibfnamefont {J.}~\bibnamefont {Orenstein}},\
  }\href {\doibase 10.1103/PhysRevLett.119.227201} {\bibfield  {journal}
  {\bibinfo  {journal} {Phys. Rev. Lett.}\ }\textbf {\bibinfo {volume} {119}},\
  \bibinfo {pages} {227201} (\bibinfo {year} {2017})}\BibitemShut {NoStop}%
\bibitem [{\citenamefont {Kasahara}\ \emph {et~al.}(2018)\citenamefont
  {Kasahara}, \citenamefont {Ohnishi}, \citenamefont {Mizukami}, \citenamefont
  {Tanaka}, \citenamefont {Ma}, \citenamefont {Sugii}, \citenamefont {Kurita},
  \citenamefont {Tanaka}, \citenamefont {Nasu}, \citenamefont {Motome} \emph
  {et~al.}}]{kasahara2018majorana}%
  \BibitemOpen
  \bibfield  {author} {\bibinfo {author} {\bibfnamefont {Y.}~\bibnamefont
  {Kasahara}}, \bibinfo {author} {\bibfnamefont {T.}~\bibnamefont {Ohnishi}},
  \bibinfo {author} {\bibfnamefont {Y.}~\bibnamefont {Mizukami}}, \bibinfo
  {author} {\bibfnamefont {O.}~\bibnamefont {Tanaka}}, \bibinfo {author}
  {\bibfnamefont {S.}~\bibnamefont {Ma}}, \bibinfo {author} {\bibfnamefont
  {K.}~\bibnamefont {Sugii}}, \bibinfo {author} {\bibfnamefont
  {N.}~\bibnamefont {Kurita}}, \bibinfo {author} {\bibfnamefont
  {H.}~\bibnamefont {Tanaka}}, \bibinfo {author} {\bibfnamefont
  {J.}~\bibnamefont {Nasu}}, \bibinfo {author} {\bibfnamefont {Y.}~\bibnamefont
  {Motome}},  \emph {et~al.},\ }\href {\doibase
  https://doi.org/10.1038/s41586-018-0274-0} {\bibfield  {journal} {\bibinfo
  {journal} {Nature}\ }\textbf {\bibinfo {volume} {559}},\ \bibinfo {pages}
  {227} (\bibinfo {year} {2018})}\BibitemShut {NoStop}%
\bibitem [{\citenamefont {Baskaran}\ \emph {et~al.}(2007)\citenamefont
  {Baskaran}, \citenamefont {Mandal},\ and\ \citenamefont
  {Shankar}}]{PhysRevLett.98.247201}%
  \BibitemOpen
  \bibfield  {author} {\bibinfo {author} {\bibfnamefont {G.}~\bibnamefont
  {Baskaran}}, \bibinfo {author} {\bibfnamefont {S.}~\bibnamefont {Mandal}}, \
  and\ \bibinfo {author} {\bibfnamefont {R.}~\bibnamefont {Shankar}},\ }\href
  {\doibase 10.1103/PhysRevLett.98.247201} {\bibfield  {journal} {\bibinfo
  {journal} {Phys. Rev. Lett.}\ }\textbf {\bibinfo {volume} {98}},\ \bibinfo
  {pages} {247201} (\bibinfo {year} {2007})}\BibitemShut {NoStop}%
\bibitem [{\citenamefont {Kitaev}(2006)}]{kitaev2006anyons}%
  \BibitemOpen
  \bibfield  {author} {\bibinfo {author} {\bibfnamefont {A.}~\bibnamefont
  {Kitaev}},\ }\href {\doibase https://doi.org/10.1016/j.aop.2005.10.005}
  {\bibfield  {journal} {\bibinfo  {journal} {Annals of Physics}\ }\textbf
  {\bibinfo {volume} {321}},\ \bibinfo {pages} {2} (\bibinfo {year}
  {2006})}\BibitemShut {NoStop}%
\bibitem [{\citenamefont {Senthil}(2006)}]{senthil2006quantum}%
  \BibitemOpen
  \bibfield  {author} {\bibinfo {author} {\bibfnamefont {T.}~\bibnamefont
  {Senthil}},\ }\href {\doibase https://doi.org/10.1142/9789812772893_0006}
  {\bibfield  {journal} {\bibinfo  {journal} {International Journal of Modern
  Physics B}\ }\textbf {\bibinfo {volume} {20}},\ \bibinfo {pages} {2603}
  (\bibinfo {year} {2006})}\BibitemShut {NoStop}%
\bibitem [{\citenamefont {Chaikin}\ \emph {et~al.}(1995)\citenamefont
  {Chaikin}, \citenamefont {Lubensky},\ and\ \citenamefont
  {Witten}}]{chaikin1995principles}%
  \BibitemOpen
  \bibfield  {author} {\bibinfo {author} {\bibfnamefont {P.~M.}\ \bibnamefont
  {Chaikin}}, \bibinfo {author} {\bibfnamefont {T.~C.}\ \bibnamefont
  {Lubensky}}, \ and\ \bibinfo {author} {\bibfnamefont {T.~A.}\ \bibnamefont
  {Witten}},\ }\href@noop {} {\emph {\bibinfo {title} {Principles of condensed
  matter physics}}},\ Vol.~\bibinfo {volume} {10}\ (\bibinfo  {publisher}
  {Cambridge university press Cambridge},\ \bibinfo {year} {1995})\BibitemShut
  {NoStop}%
\bibitem [{\citenamefont {Kou}\ \emph {et~al.}(2008)\citenamefont {Kou},
  \citenamefont {Levin},\ and\ \citenamefont {Wen}}]{PhysRevB.78.155134}%
  \BibitemOpen
  \bibfield  {author} {\bibinfo {author} {\bibfnamefont {S.-P.}\ \bibnamefont
  {Kou}}, \bibinfo {author} {\bibfnamefont {M.}~\bibnamefont {Levin}}, \ and\
  \bibinfo {author} {\bibfnamefont {X.-G.}\ \bibnamefont {Wen}},\ }\href
  {\doibase 10.1103/PhysRevB.78.155134} {\bibfield  {journal} {\bibinfo
  {journal} {Phys. Rev. B}\ }\textbf {\bibinfo {volume} {78}},\ \bibinfo
  {pages} {155134} (\bibinfo {year} {2008})}\BibitemShut {NoStop}%
\bibitem [{\citenamefont {Freedman}\ \emph {et~al.}(2004)\citenamefont
  {Freedman}, \citenamefont {Nayak}, \citenamefont {Shtengel}, \citenamefont
  {Walker},\ and\ \citenamefont {Wang}}]{freedman2004class}%
  \BibitemOpen
  \bibfield  {author} {\bibinfo {author} {\bibfnamefont {M.}~\bibnamefont
  {Freedman}}, \bibinfo {author} {\bibfnamefont {C.}~\bibnamefont {Nayak}},
  \bibinfo {author} {\bibfnamefont {K.}~\bibnamefont {Shtengel}}, \bibinfo
  {author} {\bibfnamefont {K.}~\bibnamefont {Walker}}, \ and\ \bibinfo {author}
  {\bibfnamefont {Z.}~\bibnamefont {Wang}},\ }\href {\doibase
  https://doi.org/10.1016/j.aop.2004.01.006} {\bibfield  {journal} {\bibinfo
  {journal} {Annals of Physics}\ }\textbf {\bibinfo {volume} {310}},\ \bibinfo
  {pages} {428} (\bibinfo {year} {2004})}\BibitemShut {NoStop}%
\bibitem [{\citenamefont {Senthil}\ \emph
  {et~al.}(2004{\natexlab{a}})\citenamefont {Senthil}, \citenamefont
  {Vishwanath}, \citenamefont {Balents}, \citenamefont {Sachdev},\ and\
  \citenamefont {Fisher}}]{senthil2004deconfined}%
  \BibitemOpen
  \bibfield  {author} {\bibinfo {author} {\bibfnamefont {T.}~\bibnamefont
  {Senthil}}, \bibinfo {author} {\bibfnamefont {A.}~\bibnamefont {Vishwanath}},
  \bibinfo {author} {\bibfnamefont {L.}~\bibnamefont {Balents}}, \bibinfo
  {author} {\bibfnamefont {S.}~\bibnamefont {Sachdev}}, \ and\ \bibinfo
  {author} {\bibfnamefont {M.~P.}\ \bibnamefont {Fisher}},\ }\href {\doibase
  10.1126/science.1091806} {\bibfield  {journal} {\bibinfo  {journal}
  {Science}\ }\textbf {\bibinfo {volume} {303}},\ \bibinfo {pages} {1490}
  (\bibinfo {year} {2004}{\natexlab{a}})}\BibitemShut {NoStop}%
\bibitem [{\citenamefont {Senthil}\ \emph
  {et~al.}(2004{\natexlab{b}})\citenamefont {Senthil}, \citenamefont {Balents},
  \citenamefont {Sachdev}, \citenamefont {Vishwanath},\ and\ \citenamefont
  {Fisher}}]{senthil2004quantum}%
  \BibitemOpen
  \bibfield  {author} {\bibinfo {author} {\bibfnamefont {T.}~\bibnamefont
  {Senthil}}, \bibinfo {author} {\bibfnamefont {L.}~\bibnamefont {Balents}},
  \bibinfo {author} {\bibfnamefont {S.}~\bibnamefont {Sachdev}}, \bibinfo
  {author} {\bibfnamefont {A.}~\bibnamefont {Vishwanath}}, \ and\ \bibinfo
  {author} {\bibfnamefont {M.~P.}\ \bibnamefont {Fisher}},\ }\href {\doibase
  https://doi.org/10.1103/PhysRevB.70.144407} {\bibfield  {journal} {\bibinfo
  {journal} {Physical Review B}\ }\textbf {\bibinfo {volume} {70}},\ \bibinfo
  {pages} {144407} (\bibinfo {year} {2004}{\natexlab{b}})}\BibitemShut
  {NoStop}%
\bibitem [{\citenamefont {Jackeli}\ and\ \citenamefont
  {Khaliullin}(2009)}]{PhysRevLett.102.017205}%
  \BibitemOpen
  \bibfield  {author} {\bibinfo {author} {\bibfnamefont {G.}~\bibnamefont
  {Jackeli}}\ and\ \bibinfo {author} {\bibfnamefont {G.}~\bibnamefont
  {Khaliullin}},\ }\href {\doibase 10.1103/PhysRevLett.102.017205} {\bibfield
  {journal} {\bibinfo  {journal} {Phys. Rev. Lett.}\ }\textbf {\bibinfo
  {volume} {102}},\ \bibinfo {pages} {017205} (\bibinfo {year}
  {2009})}\BibitemShut {NoStop}%
\bibitem [{\citenamefont {Chaloupka}\ \emph {et~al.}(2010)\citenamefont
  {Chaloupka}, \citenamefont {Jackeli},\ and\ \citenamefont
  {Khaliullin}}]{PhysRevLett.105.027204}%
  \BibitemOpen
  \bibfield  {author} {\bibinfo {author} {\bibfnamefont {J.~c.~v.}\
  \bibnamefont {Chaloupka}}, \bibinfo {author} {\bibfnamefont {G.}~\bibnamefont
  {Jackeli}}, \ and\ \bibinfo {author} {\bibfnamefont {G.}~\bibnamefont
  {Khaliullin}},\ }\href {\doibase 10.1103/PhysRevLett.105.027204} {\bibfield
  {journal} {\bibinfo  {journal} {Phys. Rev. Lett.}\ }\textbf {\bibinfo
  {volume} {105}},\ \bibinfo {pages} {027204} (\bibinfo {year}
  {2010})}\BibitemShut {NoStop}%
\bibitem [{\citenamefont {Rau}\ \emph {et~al.}(2014)\citenamefont {Rau},
  \citenamefont {Lee},\ and\ \citenamefont {Kee}}]{PhysRevLett.112.077204}%
  \BibitemOpen
  \bibfield  {author} {\bibinfo {author} {\bibfnamefont {J.~G.}\ \bibnamefont
  {Rau}}, \bibinfo {author} {\bibfnamefont {E.~K.-H.}\ \bibnamefont {Lee}}, \
  and\ \bibinfo {author} {\bibfnamefont {H.-Y.}\ \bibnamefont {Kee}},\ }\href
  {\doibase 10.1103/PhysRevLett.112.077204} {\bibfield  {journal} {\bibinfo
  {journal} {Phys. Rev. Lett.}\ }\textbf {\bibinfo {volume} {112}},\ \bibinfo
  {pages} {077204} (\bibinfo {year} {2014})}\BibitemShut {NoStop}%
\bibitem [{\citenamefont {Hermanns}\ \emph {et~al.}(2018)\citenamefont
  {Hermanns}, \citenamefont {Kimchi},\ and\ \citenamefont
  {Knolle}}]{Hermanns2018}%
  \BibitemOpen
  \bibfield  {author} {\bibinfo {author} {\bibfnamefont {M.}~\bibnamefont
  {Hermanns}}, \bibinfo {author} {\bibfnamefont {I.}~\bibnamefont {Kimchi}}, \
  and\ \bibinfo {author} {\bibfnamefont {J.}~\bibnamefont {Knolle}},\ }\href
  {\doibase 10.1146/annurev-conmatphys-033117-053934} {\bibfield  {journal}
  {\bibinfo  {journal} {Annual Review of Condensed Matter Physics}\ }\textbf
  {\bibinfo {volume} {9}},\ \bibinfo {pages} {null} (\bibinfo {year} {2018})},\
  \bibinfo {note} {arXiv:1705.01740}\BibitemShut {NoStop}%
\bibitem [{\citenamefont {Singh}\ and\ \citenamefont
  {Gegenwart}(2010)}]{PhysRevB.82.064412}%
  \BibitemOpen
  \bibfield  {author} {\bibinfo {author} {\bibfnamefont {Y.}~\bibnamefont
  {Singh}}\ and\ \bibinfo {author} {\bibfnamefont {P.}~\bibnamefont
  {Gegenwart}},\ }\href {\doibase 10.1103/PhysRevB.82.064412} {\bibfield
  {journal} {\bibinfo  {journal} {Phys. Rev. B}\ }\textbf {\bibinfo {volume}
  {82}},\ \bibinfo {pages} {064412} (\bibinfo {year} {2010})}\BibitemShut
  {NoStop}%
\bibitem [{\citenamefont {Liu}\ \emph {et~al.}(2011)\citenamefont {Liu},
  \citenamefont {Berlijn}, \citenamefont {Yin}, \citenamefont {Ku},
  \citenamefont {Tsvelik}, \citenamefont {Kim}, \citenamefont {Gretarsson},
  \citenamefont {Singh}, \citenamefont {Gegenwart},\ and\ \citenamefont
  {Hill}}]{PhysRevB.83.220403}%
  \BibitemOpen
  \bibfield  {author} {\bibinfo {author} {\bibfnamefont {X.}~\bibnamefont
  {Liu}}, \bibinfo {author} {\bibfnamefont {T.}~\bibnamefont {Berlijn}},
  \bibinfo {author} {\bibfnamefont {W.-G.}\ \bibnamefont {Yin}}, \bibinfo
  {author} {\bibfnamefont {W.}~\bibnamefont {Ku}}, \bibinfo {author}
  {\bibfnamefont {A.}~\bibnamefont {Tsvelik}}, \bibinfo {author} {\bibfnamefont
  {Y.-J.}\ \bibnamefont {Kim}}, \bibinfo {author} {\bibfnamefont
  {H.}~\bibnamefont {Gretarsson}}, \bibinfo {author} {\bibfnamefont
  {Y.}~\bibnamefont {Singh}}, \bibinfo {author} {\bibfnamefont
  {P.}~\bibnamefont {Gegenwart}}, \ and\ \bibinfo {author} {\bibfnamefont
  {J.~P.}\ \bibnamefont {Hill}},\ }\href {\doibase 10.1103/PhysRevB.83.220403}
  {\bibfield  {journal} {\bibinfo  {journal} {Phys. Rev. B}\ }\textbf {\bibinfo
  {volume} {83}},\ \bibinfo {pages} {220403} (\bibinfo {year}
  {2011})}\BibitemShut {NoStop}%
\bibitem [{\citenamefont {Singh}\ \emph {et~al.}(2012)\citenamefont {Singh},
  \citenamefont {Manni}, \citenamefont {Reuther}, \citenamefont {Berlijn},
  \citenamefont {Thomale}, \citenamefont {Ku}, \citenamefont {Trebst},\ and\
  \citenamefont {Gegenwart}}]{PhysRevLett.108.127203}%
  \BibitemOpen
  \bibfield  {author} {\bibinfo {author} {\bibfnamefont {Y.}~\bibnamefont
  {Singh}}, \bibinfo {author} {\bibfnamefont {S.}~\bibnamefont {Manni}},
  \bibinfo {author} {\bibfnamefont {J.}~\bibnamefont {Reuther}}, \bibinfo
  {author} {\bibfnamefont {T.}~\bibnamefont {Berlijn}}, \bibinfo {author}
  {\bibfnamefont {R.}~\bibnamefont {Thomale}}, \bibinfo {author} {\bibfnamefont
  {W.}~\bibnamefont {Ku}}, \bibinfo {author} {\bibfnamefont {S.}~\bibnamefont
  {Trebst}}, \ and\ \bibinfo {author} {\bibfnamefont {P.}~\bibnamefont
  {Gegenwart}},\ }\href {\doibase 10.1103/PhysRevLett.108.127203} {\bibfield
  {journal} {\bibinfo  {journal} {Phys. Rev. Lett.}\ }\textbf {\bibinfo
  {volume} {108}},\ \bibinfo {pages} {127203} (\bibinfo {year}
  {2012})}\BibitemShut {NoStop}%
\bibitem [{\citenamefont {Choi}\ \emph {et~al.}(2012)\citenamefont {Choi},
  \citenamefont {Coldea}, \citenamefont {Kolmogorov}, \citenamefont
  {Lancaster}, \citenamefont {Mazin}, \citenamefont {Blundell}, \citenamefont
  {Radaelli}, \citenamefont {Singh}, \citenamefont {Gegenwart}, \citenamefont
  {Choi}, \citenamefont {Cheong}, \citenamefont {Baker}, \citenamefont
  {Stock},\ and\ \citenamefont {Taylor}}]{PhysRevLett.108.127204}%
  \BibitemOpen
  \bibfield  {author} {\bibinfo {author} {\bibfnamefont {S.~K.}\ \bibnamefont
  {Choi}}, \bibinfo {author} {\bibfnamefont {R.}~\bibnamefont {Coldea}},
  \bibinfo {author} {\bibfnamefont {A.~N.}\ \bibnamefont {Kolmogorov}},
  \bibinfo {author} {\bibfnamefont {T.}~\bibnamefont {Lancaster}}, \bibinfo
  {author} {\bibfnamefont {I.~I.}\ \bibnamefont {Mazin}}, \bibinfo {author}
  {\bibfnamefont {S.~J.}\ \bibnamefont {Blundell}}, \bibinfo {author}
  {\bibfnamefont {P.~G.}\ \bibnamefont {Radaelli}}, \bibinfo {author}
  {\bibfnamefont {Y.}~\bibnamefont {Singh}}, \bibinfo {author} {\bibfnamefont
  {P.}~\bibnamefont {Gegenwart}}, \bibinfo {author} {\bibfnamefont {K.~R.}\
  \bibnamefont {Choi}}, \bibinfo {author} {\bibfnamefont {S.-W.}\ \bibnamefont
  {Cheong}}, \bibinfo {author} {\bibfnamefont {P.~J.}\ \bibnamefont {Baker}},
  \bibinfo {author} {\bibfnamefont {C.}~\bibnamefont {Stock}}, \ and\ \bibinfo
  {author} {\bibfnamefont {J.}~\bibnamefont {Taylor}},\ }\href {\doibase
  10.1103/PhysRevLett.108.127204} {\bibfield  {journal} {\bibinfo  {journal}
  {Phys. Rev. Lett.}\ }\textbf {\bibinfo {volume} {108}},\ \bibinfo {pages}
  {127204} (\bibinfo {year} {2012})}\BibitemShut {NoStop}%
\bibitem [{\citenamefont {Ye}\ \emph {et~al.}(2012)\citenamefont {Ye},
  \citenamefont {Chi}, \citenamefont {Cao}, \citenamefont {Chakoumakos},
  \citenamefont {Fernandez-Baca}, \citenamefont {Custelcean}, \citenamefont
  {Qi}, \citenamefont {Korneta},\ and\ \citenamefont
  {Cao}}]{PhysRevB.85.180403}%
  \BibitemOpen
  \bibfield  {author} {\bibinfo {author} {\bibfnamefont {F.}~\bibnamefont
  {Ye}}, \bibinfo {author} {\bibfnamefont {S.}~\bibnamefont {Chi}}, \bibinfo
  {author} {\bibfnamefont {H.}~\bibnamefont {Cao}}, \bibinfo {author}
  {\bibfnamefont {B.~C.}\ \bibnamefont {Chakoumakos}}, \bibinfo {author}
  {\bibfnamefont {J.~A.}\ \bibnamefont {Fernandez-Baca}}, \bibinfo {author}
  {\bibfnamefont {R.}~\bibnamefont {Custelcean}}, \bibinfo {author}
  {\bibfnamefont {T.~F.}\ \bibnamefont {Qi}}, \bibinfo {author} {\bibfnamefont
  {O.~B.}\ \bibnamefont {Korneta}}, \ and\ \bibinfo {author} {\bibfnamefont
  {G.}~\bibnamefont {Cao}},\ }\href {\doibase 10.1103/PhysRevB.85.180403}
  {\bibfield  {journal} {\bibinfo  {journal} {Phys. Rev. B}\ }\textbf {\bibinfo
  {volume} {85}},\ \bibinfo {pages} {180403} (\bibinfo {year}
  {2012})}\BibitemShut {NoStop}%
\bibitem [{\citenamefont {Biffin}\ \emph
  {et~al.}(2014{\natexlab{a}})\citenamefont {Biffin}, \citenamefont {Johnson},
  \citenamefont {Choi}, \citenamefont {Freund}, \citenamefont {Manni},
  \citenamefont {Bombardi}, \citenamefont {Manuel}, \citenamefont {Gegenwart},\
  and\ \citenamefont {Coldea}}]{PhysRevB.90.205116}%
  \BibitemOpen
  \bibfield  {author} {\bibinfo {author} {\bibfnamefont {A.}~\bibnamefont
  {Biffin}}, \bibinfo {author} {\bibfnamefont {R.~D.}\ \bibnamefont {Johnson}},
  \bibinfo {author} {\bibfnamefont {S.}~\bibnamefont {Choi}}, \bibinfo {author}
  {\bibfnamefont {F.}~\bibnamefont {Freund}}, \bibinfo {author} {\bibfnamefont
  {S.}~\bibnamefont {Manni}}, \bibinfo {author} {\bibfnamefont
  {A.}~\bibnamefont {Bombardi}}, \bibinfo {author} {\bibfnamefont
  {P.}~\bibnamefont {Manuel}}, \bibinfo {author} {\bibfnamefont
  {P.}~\bibnamefont {Gegenwart}}, \ and\ \bibinfo {author} {\bibfnamefont
  {R.}~\bibnamefont {Coldea}},\ }\href {\doibase 10.1103/PhysRevB.90.205116}
  {\bibfield  {journal} {\bibinfo  {journal} {Phys. Rev. B}\ }\textbf {\bibinfo
  {volume} {90}},\ \bibinfo {pages} {205116} (\bibinfo {year}
  {2014}{\natexlab{a}})}\BibitemShut {NoStop}%
\bibitem [{\citenamefont {Biffin}\ \emph
  {et~al.}(2014{\natexlab{b}})\citenamefont {Biffin}, \citenamefont {Johnson},
  \citenamefont {Kimchi}, \citenamefont {Morris}, \citenamefont {Bombardi},
  \citenamefont {Analytis}, \citenamefont {Vishwanath},\ and\ \citenamefont
  {Coldea}}]{PhysRevLett.113.197201}%
  \BibitemOpen
  \bibfield  {author} {\bibinfo {author} {\bibfnamefont {A.}~\bibnamefont
  {Biffin}}, \bibinfo {author} {\bibfnamefont {R.~D.}\ \bibnamefont {Johnson}},
  \bibinfo {author} {\bibfnamefont {I.}~\bibnamefont {Kimchi}}, \bibinfo
  {author} {\bibfnamefont {R.}~\bibnamefont {Morris}}, \bibinfo {author}
  {\bibfnamefont {A.}~\bibnamefont {Bombardi}}, \bibinfo {author}
  {\bibfnamefont {J.~G.}\ \bibnamefont {Analytis}}, \bibinfo {author}
  {\bibfnamefont {A.}~\bibnamefont {Vishwanath}}, \ and\ \bibinfo {author}
  {\bibfnamefont {R.}~\bibnamefont {Coldea}},\ }\href {\doibase
  10.1103/PhysRevLett.113.197201} {\bibfield  {journal} {\bibinfo  {journal}
  {Phys. Rev. Lett.}\ }\textbf {\bibinfo {volume} {113}},\ \bibinfo {pages}
  {197201} (\bibinfo {year} {2014}{\natexlab{b}})}\BibitemShut {NoStop}%
\bibitem [{\citenamefont {Nussinov}\ and\ \citenamefont
  {Brink}(2013)}]{nussinov2013compass}%
  \BibitemOpen
  \bibfield  {author} {\bibinfo {author} {\bibfnamefont {Z.}~\bibnamefont
  {Nussinov}}\ and\ \bibinfo {author} {\bibfnamefont {J.~v.~d.}\ \bibnamefont
  {Brink}},\ }\href {https://arxiv.org/abs/1303.5922} {\bibfield  {journal}
  {\bibinfo  {journal} {arXiv preprint arXiv:1303.5922}\ } (\bibinfo {year}
  {2013})}\BibitemShut {NoStop}%
\bibitem [{\citenamefont {Kimchi}\ and\ \citenamefont
  {Vishwanath}(2014)}]{PhysRevB.89.014414}%
  \BibitemOpen
  \bibfield  {author} {\bibinfo {author} {\bibfnamefont {I.}~\bibnamefont
  {Kimchi}}\ and\ \bibinfo {author} {\bibfnamefont {A.}~\bibnamefont
  {Vishwanath}},\ }\href {\doibase 10.1103/PhysRevB.89.014414} {\bibfield
  {journal} {\bibinfo  {journal} {Phys. Rev. B}\ }\textbf {\bibinfo {volume}
  {89}},\ \bibinfo {pages} {014414} (\bibinfo {year} {2014})}\BibitemShut
  {NoStop}%
\bibitem [{\citenamefont {Lee}\ \emph {et~al.}(2014)\citenamefont {Lee},
  \citenamefont {Schaffer}, \citenamefont {Bhattacharjee},\ and\ \citenamefont
  {Kim}}]{lee2014heisenberg}%
  \BibitemOpen
  \bibfield  {author} {\bibinfo {author} {\bibfnamefont {E.~K.-H.}\
  \bibnamefont {Lee}}, \bibinfo {author} {\bibfnamefont {R.}~\bibnamefont
  {Schaffer}}, \bibinfo {author} {\bibfnamefont {S.}~\bibnamefont
  {Bhattacharjee}}, \ and\ \bibinfo {author} {\bibfnamefont {Y.~B.}\
  \bibnamefont {Kim}},\ }\href {\doibase
  https://doi.org/10.1103/PhysRevB.89.045117} {\bibfield  {journal} {\bibinfo
  {journal} {Physical Review B}\ }\textbf {\bibinfo {volume} {89}},\ \bibinfo
  {pages} {045117} (\bibinfo {year} {2014})}\BibitemShut {NoStop}%
\bibitem [{\citenamefont {Reuther}\ \emph {et~al.}(2011)\citenamefont
  {Reuther}, \citenamefont {Thomale},\ and\ \citenamefont
  {Trebst}}]{PhysRevB.84.100406}%
  \BibitemOpen
  \bibfield  {author} {\bibinfo {author} {\bibfnamefont {J.}~\bibnamefont
  {Reuther}}, \bibinfo {author} {\bibfnamefont {R.}~\bibnamefont {Thomale}}, \
  and\ \bibinfo {author} {\bibfnamefont {S.}~\bibnamefont {Trebst}},\ }\href
  {\doibase 10.1103/PhysRevB.84.100406} {\bibfield  {journal} {\bibinfo
  {journal} {Phys. Rev. B}\ }\textbf {\bibinfo {volume} {84}},\ \bibinfo
  {pages} {100406} (\bibinfo {year} {2011})}\BibitemShut {NoStop}%
\bibitem [{\citenamefont {Jiang}\ \emph {et~al.}(2011)\citenamefont {Jiang},
  \citenamefont {Gu}, \citenamefont {Qi},\ and\ \citenamefont
  {Trebst}}]{PhysRevB.83.245104}%
  \BibitemOpen
  \bibfield  {author} {\bibinfo {author} {\bibfnamefont {H.-C.}\ \bibnamefont
  {Jiang}}, \bibinfo {author} {\bibfnamefont {Z.-C.}\ \bibnamefont {Gu}},
  \bibinfo {author} {\bibfnamefont {X.-L.}\ \bibnamefont {Qi}}, \ and\ \bibinfo
  {author} {\bibfnamefont {S.}~\bibnamefont {Trebst}},\ }\href {\doibase
  10.1103/PhysRevB.83.245104} {\bibfield  {journal} {\bibinfo  {journal} {Phys.
  Rev. B}\ }\textbf {\bibinfo {volume} {83}},\ \bibinfo {pages} {245104}
  (\bibinfo {year} {2011})}\BibitemShut {NoStop}%
\bibitem [{\citenamefont {Oitmaa}(2015)}]{PhysRevB.92.020405}%
  \BibitemOpen
  \bibfield  {author} {\bibinfo {author} {\bibfnamefont {J.}~\bibnamefont
  {Oitmaa}},\ }\href {\doibase 10.1103/PhysRevB.92.020405} {\bibfield
  {journal} {\bibinfo  {journal} {Phys. Rev. B}\ }\textbf {\bibinfo {volume}
  {92}},\ \bibinfo {pages} {020405} (\bibinfo {year} {2015})}\BibitemShut
  {NoStop}%
\bibitem [{\citenamefont {Schaffer}\ \emph {et~al.}(2012)\citenamefont
  {Schaffer}, \citenamefont {Bhattacharjee},\ and\ \citenamefont
  {Kim}}]{PhysRevB.86.224417}%
  \BibitemOpen
  \bibfield  {author} {\bibinfo {author} {\bibfnamefont {R.}~\bibnamefont
  {Schaffer}}, \bibinfo {author} {\bibfnamefont {S.}~\bibnamefont
  {Bhattacharjee}}, \ and\ \bibinfo {author} {\bibfnamefont {Y.~B.}\
  \bibnamefont {Kim}},\ }\href {\doibase 10.1103/PhysRevB.86.224417} {\bibfield
   {journal} {\bibinfo  {journal} {Phys. Rev. B}\ }\textbf {\bibinfo {volume}
  {86}},\ \bibinfo {pages} {224417} (\bibinfo {year} {2012})}\BibitemShut
  {NoStop}%
\bibitem [{\citenamefont {Price}\ and\ \citenamefont
  {Perkins}(2012)}]{PhysRevLett.109.187201}%
  \BibitemOpen
  \bibfield  {author} {\bibinfo {author} {\bibfnamefont {C.~C.}\ \bibnamefont
  {Price}}\ and\ \bibinfo {author} {\bibfnamefont {N.~B.}\ \bibnamefont
  {Perkins}},\ }\href {\doibase 10.1103/PhysRevLett.109.187201} {\bibfield
  {journal} {\bibinfo  {journal} {Phys. Rev. Lett.}\ }\textbf {\bibinfo
  {volume} {109}},\ \bibinfo {pages} {187201} (\bibinfo {year}
  {2012})}\BibitemShut {NoStop}%
\bibitem [{\citenamefont {Gohlke}\ \emph {et~al.}(2018)\citenamefont {Gohlke},
  \citenamefont {Wachtel}, \citenamefont {Yamaji}, \citenamefont {Pollmann},\
  and\ \citenamefont {Kim}}]{PhysRevB.97.075126}%
  \BibitemOpen
  \bibfield  {author} {\bibinfo {author} {\bibfnamefont {M.}~\bibnamefont
  {Gohlke}}, \bibinfo {author} {\bibfnamefont {G.}~\bibnamefont {Wachtel}},
  \bibinfo {author} {\bibfnamefont {Y.}~\bibnamefont {Yamaji}}, \bibinfo
  {author} {\bibfnamefont {F.}~\bibnamefont {Pollmann}}, \ and\ \bibinfo
  {author} {\bibfnamefont {Y.~B.}\ \bibnamefont {Kim}},\ }\href {\doibase
  10.1103/PhysRevB.97.075126} {\bibfield  {journal} {\bibinfo  {journal} {Phys.
  Rev. B}\ }\textbf {\bibinfo {volume} {97}},\ \bibinfo {pages} {075126}
  (\bibinfo {year} {2018})}\BibitemShut {NoStop}%
\bibitem [{\citenamefont {Lee}\ \emph {et~al.}(2020)\citenamefont {Lee},
  \citenamefont {Kaneko}, \citenamefont {Chern}, \citenamefont {Okubo},
  \citenamefont {Yamaji}, \citenamefont {Kawashima},\ and\ \citenamefont
  {Kim}}]{lee2019magnetic}%
  \BibitemOpen
  \bibfield  {author} {\bibinfo {author} {\bibfnamefont {H.-Y.}\ \bibnamefont
  {Lee}}, \bibinfo {author} {\bibfnamefont {R.}~\bibnamefont {Kaneko}},
  \bibinfo {author} {\bibfnamefont {L.~E.}\ \bibnamefont {Chern}}, \bibinfo
  {author} {\bibfnamefont {T.}~\bibnamefont {Okubo}}, \bibinfo {author}
  {\bibfnamefont {Y.}~\bibnamefont {Yamaji}}, \bibinfo {author} {\bibfnamefont
  {N.}~\bibnamefont {Kawashima}}, \ and\ \bibinfo {author} {\bibfnamefont
  {Y.~B.}\ \bibnamefont {Kim}},\ }\href {\doibase
  https://doi.org/10.1038/s41467-020-15320-x} {\bibfield  {journal} {\bibinfo
  {journal} {Nature communications}\ }\textbf {\bibinfo {volume} {11}},\
  \bibinfo {pages} {1} (\bibinfo {year} {2020})}\BibitemShut {NoStop}%
\bibitem [{\citenamefont {Kitaev}(2003)}]{kitaev2003fault}%
  \BibitemOpen
  \bibfield  {author} {\bibinfo {author} {\bibfnamefont {A.~Y.}\ \bibnamefont
  {Kitaev}},\ }\href {\doibase https://doi.org/10.1016/S0003-4916(02)00018-0}
  {\bibfield  {journal} {\bibinfo  {journal} {Annals of Physics}\ }\textbf
  {\bibinfo {volume} {303}},\ \bibinfo {pages} {2} (\bibinfo {year}
  {2003})}\BibitemShut {NoStop}%
\bibitem [{\citenamefont {Trebst}\ \emph {et~al.}(2007)\citenamefont {Trebst},
  \citenamefont {Werner}, \citenamefont {Troyer}, \citenamefont {Shtengel},\
  and\ \citenamefont {Nayak}}]{PhysRevLett.98.070602}%
  \BibitemOpen
  \bibfield  {author} {\bibinfo {author} {\bibfnamefont {S.}~\bibnamefont
  {Trebst}}, \bibinfo {author} {\bibfnamefont {P.}~\bibnamefont {Werner}},
  \bibinfo {author} {\bibfnamefont {M.}~\bibnamefont {Troyer}}, \bibinfo
  {author} {\bibfnamefont {K.}~\bibnamefont {Shtengel}}, \ and\ \bibinfo
  {author} {\bibfnamefont {C.}~\bibnamefont {Nayak}},\ }\href {\doibase
  10.1103/PhysRevLett.98.070602} {\bibfield  {journal} {\bibinfo  {journal}
  {Phys. Rev. Lett.}\ }\textbf {\bibinfo {volume} {98}},\ \bibinfo {pages}
  {070602} (\bibinfo {year} {2007})}\BibitemShut {NoStop}%
\bibitem [{\citenamefont {Moon}\ and\ \citenamefont
  {Xu}(2012)}]{PhysRevB.86.214414}%
  \BibitemOpen
  \bibfield  {author} {\bibinfo {author} {\bibfnamefont {E.-G.}\ \bibnamefont
  {Moon}}\ and\ \bibinfo {author} {\bibfnamefont {C.}~\bibnamefont {Xu}},\
  }\href {\doibase 10.1103/PhysRevB.86.214414} {\bibfield  {journal} {\bibinfo
  {journal} {Phys. Rev. B}\ }\textbf {\bibinfo {volume} {86}},\ \bibinfo
  {pages} {214414} (\bibinfo {year} {2012})}\BibitemShut {NoStop}%
\bibitem [{\citenamefont {Qi}\ and\ \citenamefont
  {Gu}(2014)}]{PhysRevB.89.235122}%
  \BibitemOpen
  \bibfield  {author} {\bibinfo {author} {\bibfnamefont {Y.}~\bibnamefont
  {Qi}}\ and\ \bibinfo {author} {\bibfnamefont {Z.-C.}\ \bibnamefont {Gu}},\
  }\href {\doibase 10.1103/PhysRevB.89.235122} {\bibfield  {journal} {\bibinfo
  {journal} {Phys. Rev. B}\ }\textbf {\bibinfo {volume} {89}},\ \bibinfo
  {pages} {235122} (\bibinfo {year} {2014})}\BibitemShut {NoStop}%
\bibitem [{\citenamefont {Quinn}\ \emph {et~al.}(2015)\citenamefont {Quinn},
  \citenamefont {Bhattacharjee},\ and\ \citenamefont
  {Moessner}}]{PhysRevB.91.134419}%
  \BibitemOpen
  \bibfield  {author} {\bibinfo {author} {\bibfnamefont {E.}~\bibnamefont
  {Quinn}}, \bibinfo {author} {\bibfnamefont {S.}~\bibnamefont
  {Bhattacharjee}}, \ and\ \bibinfo {author} {\bibfnamefont {R.}~\bibnamefont
  {Moessner}},\ }\href {\doibase 10.1103/PhysRevB.91.134419} {\bibfield
  {journal} {\bibinfo  {journal} {Phys. Rev. B}\ }\textbf {\bibinfo {volume}
  {91}},\ \bibinfo {pages} {134419} (\bibinfo {year} {2015})}\BibitemShut
  {NoStop}%
\bibitem [{\citenamefont {Sela}\ \emph {et~al.}(2014)\citenamefont {Sela},
  \citenamefont {Jiang}, \citenamefont {Gerlach},\ and\ \citenamefont
  {Trebst}}]{PhysRevB.90.035113}%
  \BibitemOpen
  \bibfield  {author} {\bibinfo {author} {\bibfnamefont {E.}~\bibnamefont
  {Sela}}, \bibinfo {author} {\bibfnamefont {H.-C.}\ \bibnamefont {Jiang}},
  \bibinfo {author} {\bibfnamefont {M.~H.}\ \bibnamefont {Gerlach}}, \ and\
  \bibinfo {author} {\bibfnamefont {S.}~\bibnamefont {Trebst}},\ }\href
  {\doibase 10.1103/PhysRevB.90.035113} {\bibfield  {journal} {\bibinfo
  {journal} {Phys. Rev. B}\ }\textbf {\bibinfo {volume} {90}},\ \bibinfo
  {pages} {035113} (\bibinfo {year} {2014})}\BibitemShut {NoStop}%
\bibitem [{\citenamefont {Wachtel}\ and\ \citenamefont
  {Orgad}(2019)}]{PhysRevB.99.115104}%
  \BibitemOpen
  \bibfield  {author} {\bibinfo {author} {\bibfnamefont {G.}~\bibnamefont
  {Wachtel}}\ and\ \bibinfo {author} {\bibfnamefont {D.}~\bibnamefont
  {Orgad}},\ }\href {\doibase 10.1103/PhysRevB.99.115104} {\bibfield  {journal}
  {\bibinfo  {journal} {Phys. Rev. B}\ }\textbf {\bibinfo {volume} {99}},\
  \bibinfo {pages} {115104} (\bibinfo {year} {2019})}\BibitemShut {NoStop}%
\bibitem [{\citenamefont {Nanda}\ and\ \citenamefont
  {Bhattacharjee}(2020)}]{animesh_afm}%
  \BibitemOpen
  \bibfield  {author} {\bibinfo {author} {\bibfnamefont {A.}~\bibnamefont
  {Nanda}}\ and\ \bibinfo {author} {\bibfnamefont {S.}~\bibnamefont
  {Bhattacharjee}},\ }\href@noop {} {\bibfield  {journal} {\bibinfo  {journal}
  {(unpublished)}\ } (\bibinfo {year} {2020})}\BibitemShut {NoStop}%
\bibitem [{\citenamefont {Essin}\ and\ \citenamefont
  {Hermele}(2013)}]{PhysRevB.87.104406}%
  \BibitemOpen
  \bibfield  {author} {\bibinfo {author} {\bibfnamefont {A.~M.}\ \bibnamefont
  {Essin}}\ and\ \bibinfo {author} {\bibfnamefont {M.}~\bibnamefont
  {Hermele}},\ }\href {\doibase 10.1103/PhysRevB.87.104406} {\bibfield
  {journal} {\bibinfo  {journal} {Phys. Rev. B}\ }\textbf {\bibinfo {volume}
  {87}},\ \bibinfo {pages} {104406} (\bibinfo {year} {2013})}\BibitemShut
  {NoStop}%
\bibitem [{\citenamefont {Schaffer}\ \emph {et~al.}(2013)\citenamefont
  {Schaffer}, \citenamefont {Bhattacharjee},\ and\ \citenamefont
  {Kim}}]{PhysRevB.88.174405}%
  \BibitemOpen
  \bibfield  {author} {\bibinfo {author} {\bibfnamefont {R.}~\bibnamefont
  {Schaffer}}, \bibinfo {author} {\bibfnamefont {S.}~\bibnamefont
  {Bhattacharjee}}, \ and\ \bibinfo {author} {\bibfnamefont {Y.~B.}\
  \bibnamefont {Kim}},\ }\href {\doibase 10.1103/PhysRevB.88.174405} {\bibfield
   {journal} {\bibinfo  {journal} {Phys. Rev. B}\ }\textbf {\bibinfo {volume}
  {88}},\ \bibinfo {pages} {174405} (\bibinfo {year} {2013})}\BibitemShut
  {NoStop}%
\bibitem [{\citenamefont {Teo}\ \emph {et~al.}(2014)\citenamefont {Teo},
  \citenamefont {Roy},\ and\ \citenamefont {Chen}}]{PhysRevB.90.115118}%
  \BibitemOpen
  \bibfield  {author} {\bibinfo {author} {\bibfnamefont {J.~C.~Y.}\
  \bibnamefont {Teo}}, \bibinfo {author} {\bibfnamefont {A.}~\bibnamefont
  {Roy}}, \ and\ \bibinfo {author} {\bibfnamefont {X.}~\bibnamefont {Chen}},\
  }\href {\doibase 10.1103/PhysRevB.90.115118} {\bibfield  {journal} {\bibinfo
  {journal} {Phys. Rev. B}\ }\textbf {\bibinfo {volume} {90}},\ \bibinfo
  {pages} {115118} (\bibinfo {year} {2014})}\BibitemShut {NoStop}%
\bibitem [{\citenamefont {Fradkin}\ and\ \citenamefont
  {Shenker}(1979)}]{PhysRevD.19.3682}%
  \BibitemOpen
  \bibfield  {author} {\bibinfo {author} {\bibfnamefont {E.}~\bibnamefont
  {Fradkin}}\ and\ \bibinfo {author} {\bibfnamefont {S.~H.}\ \bibnamefont
  {Shenker}},\ }\href {\doibase 10.1103/PhysRevD.19.3682} {\bibfield  {journal}
  {\bibinfo  {journal} {Phys. Rev. D}\ }\textbf {\bibinfo {volume} {19}},\
  \bibinfo {pages} {3682} (\bibinfo {year} {1979})}\BibitemShut {NoStop}%
\bibitem [{\citenamefont {{Villain, J.}}\ \emph {et~al.}(1980)\citenamefont
  {{Villain, J.}}, \citenamefont {{Bidaux, R.}}, \citenamefont {{Carton,
  J.-P.}},\ and\ \citenamefont {{Conte, R.}}}]{refId0}%
  \BibitemOpen
  \bibfield  {author} {\bibinfo {author} {\bibnamefont {{Villain, J.}}},
  \bibinfo {author} {\bibnamefont {{Bidaux, R.}}}, \bibinfo {author}
  {\bibnamefont {{Carton, J.-P.}}}, \ and\ \bibinfo {author} {\bibnamefont
  {{Conte, R.}}},\ }\href {\doibase 10.1051/jphys:0198000410110126300}
  {\bibfield  {journal} {\bibinfo  {journal} {J. Phys. France}\ }\textbf
  {\bibinfo {volume} {41}},\ \bibinfo {pages} {1263} (\bibinfo {year}
  {1980})}\BibitemShut {NoStop}%
\bibitem [{\citenamefont {Moessner}\ and\ \citenamefont
  {Sondhi}(2001{\natexlab{b}})}]{moessner2001ising}%
  \BibitemOpen
  \bibfield  {author} {\bibinfo {author} {\bibfnamefont {R.}~\bibnamefont
  {Moessner}}\ and\ \bibinfo {author} {\bibfnamefont {S.~L.}\ \bibnamefont
  {Sondhi}},\ }\href {\doibase https://doi.org/10.1103/PhysRevB.63.224401}
  {\bibfield  {journal} {\bibinfo  {journal} {Physical Review B}\ }\textbf
  {\bibinfo {volume} {63}},\ \bibinfo {pages} {224401} (\bibinfo {year}
  {2001}{\natexlab{b}})}\BibitemShut {NoStop}%
\bibitem [{\citenamefont {You}\ \emph {et~al.}(2012)\citenamefont {You},
  \citenamefont {Kimchi},\ and\ \citenamefont
  {Vishwanath}}]{PhysRevB.86.085145}%
  \BibitemOpen
  \bibfield  {author} {\bibinfo {author} {\bibfnamefont {Y.-Z.}\ \bibnamefont
  {You}}, \bibinfo {author} {\bibfnamefont {I.}~\bibnamefont {Kimchi}}, \ and\
  \bibinfo {author} {\bibfnamefont {A.}~\bibnamefont {Vishwanath}},\ }\href
  {\doibase 10.1103/PhysRevB.86.085145} {\bibfield  {journal} {\bibinfo
  {journal} {Phys. Rev. B}\ }\textbf {\bibinfo {volume} {86}},\ \bibinfo
  {pages} {085145} (\bibinfo {year} {2012})}\BibitemShut {NoStop}%
\bibitem [{\citenamefont {Wen}(2003)}]{PhysRevLett.90.016803}%
  \BibitemOpen
  \bibfield  {author} {\bibinfo {author} {\bibfnamefont {X.-G.}\ \bibnamefont
  {Wen}},\ }\href {\doibase 10.1103/PhysRevLett.90.016803} {\bibfield
  {journal} {\bibinfo  {journal} {Phys. Rev. Lett.}\ }\textbf {\bibinfo
  {volume} {90}},\ \bibinfo {pages} {016803} (\bibinfo {year}
  {2003})}\BibitemShut {NoStop}%
\bibitem [{\citenamefont {Kou}\ \emph {et~al.}(2005)\citenamefont {Kou},
  \citenamefont {Qi},\ and\ \citenamefont {Weng}}]{PhysRevB.71.235102}%
  \BibitemOpen
  \bibfield  {author} {\bibinfo {author} {\bibfnamefont {S.-P.}\ \bibnamefont
  {Kou}}, \bibinfo {author} {\bibfnamefont {X.-L.}\ \bibnamefont {Qi}}, \ and\
  \bibinfo {author} {\bibfnamefont {Z.-Y.}\ \bibnamefont {Weng}},\ }\href
  {\doibase 10.1103/PhysRevB.71.235102} {\bibfield  {journal} {\bibinfo
  {journal} {Phys. Rev. B}\ }\textbf {\bibinfo {volume} {71}},\ \bibinfo
  {pages} {235102} (\bibinfo {year} {2005})}\BibitemShut {NoStop}%
\bibitem [{\citenamefont {Xu}\ and\ \citenamefont
  {Sachdev}(2009)}]{xu2009global}%
  \BibitemOpen
  \bibfield  {author} {\bibinfo {author} {\bibfnamefont {C.}~\bibnamefont
  {Xu}}\ and\ \bibinfo {author} {\bibfnamefont {S.}~\bibnamefont {Sachdev}},\
  }\href {\doibase https://doi.org/10.1103/PhysRevB.79.064405} {\bibfield
  {journal} {\bibinfo  {journal} {Physical Review B}\ }\textbf {\bibinfo
  {volume} {79}},\ \bibinfo {pages} {064405} (\bibinfo {year}
  {2009})}\BibitemShut {NoStop}%
\bibitem [{\citenamefont {Motrunich}\ and\ \citenamefont
  {Senthil}(2005)}]{PhysRevB.71.125102}%
  \BibitemOpen
  \bibfield  {author} {\bibinfo {author} {\bibfnamefont {O.~I.}\ \bibnamefont
  {Motrunich}}\ and\ \bibinfo {author} {\bibfnamefont {T.}~\bibnamefont
  {Senthil}},\ }\href {\doibase 10.1103/PhysRevB.71.125102} {\bibfield
  {journal} {\bibinfo  {journal} {Phys. Rev. B}\ }\textbf {\bibinfo {volume}
  {71}},\ \bibinfo {pages} {125102} (\bibinfo {year} {2005})}\BibitemShut
  {NoStop}%
\bibitem [{\citenamefont {Kamiya}\ \emph {et~al.}(2015)\citenamefont {Kamiya},
  \citenamefont {Kato}, \citenamefont {Nasu},\ and\ \citenamefont
  {Motome}}]{PhysRevB.92.100403}%
  \BibitemOpen
  \bibfield  {author} {\bibinfo {author} {\bibfnamefont {Y.}~\bibnamefont
  {Kamiya}}, \bibinfo {author} {\bibfnamefont {Y.}~\bibnamefont {Kato}},
  \bibinfo {author} {\bibfnamefont {J.}~\bibnamefont {Nasu}}, \ and\ \bibinfo
  {author} {\bibfnamefont {Y.}~\bibnamefont {Motome}},\ }\href {\doibase
  10.1103/PhysRevB.92.100403} {\bibfield  {journal} {\bibinfo  {journal} {Phys.
  Rev. B}\ }\textbf {\bibinfo {volume} {92}},\ \bibinfo {pages} {100403}
  (\bibinfo {year} {2015})}\BibitemShut {NoStop}%
\bibitem [{\citenamefont {Weinberg}\ and\ \citenamefont
  {Bukov}(2017)}]{weinberg2017quspin}%
  \BibitemOpen
  \bibfield  {author} {\bibinfo {author} {\bibfnamefont {P.}~\bibnamefont
  {Weinberg}}\ and\ \bibinfo {author} {\bibfnamefont {M.}~\bibnamefont
  {Bukov}},\ }\href {\doibase 10.21468/SciPostPhys.2.1.003} {\bibfield
  {journal} {\bibinfo  {journal} {SciPost Phys}\ }\textbf {\bibinfo {volume}
  {2}} (\bibinfo {year} {2017}),\ 10.21468/SciPostPhys.2.1.003}\BibitemShut
  {NoStop}%
\bibitem [{\citenamefont {Weinberg}\ and\ \citenamefont
  {Bukov}(2019)}]{weinberg2019quspin}%
  \BibitemOpen
  \bibfield  {author} {\bibinfo {author} {\bibfnamefont {P.}~\bibnamefont
  {Weinberg}}\ and\ \bibinfo {author} {\bibfnamefont {M.}~\bibnamefont
  {Bukov}},\ }\href {\doibase 10.21468/SciPostPhys.7.2.020} {\bibfield
  {journal} {\bibinfo  {journal} {SciPost Phys.}\ }\textbf {\bibinfo {volume}
  {7}},\ \bibinfo {pages} {020} (\bibinfo {year} {2019})}\BibitemShut {NoStop}%
\bibitem [{\citenamefont {Kitaev}\ and\ \citenamefont
  {Preskill}(2006)}]{kitaev2006topological}%
  \BibitemOpen
  \bibfield  {author} {\bibinfo {author} {\bibfnamefont {A.}~\bibnamefont
  {Kitaev}}\ and\ \bibinfo {author} {\bibfnamefont {J.}~\bibnamefont
  {Preskill}},\ }\href {\doibase 10.1103/PhysRevLett.96.110404} {\bibfield
  {journal} {\bibinfo  {journal} {Physical review letters}\ }\textbf {\bibinfo
  {volume} {96}},\ \bibinfo {pages} {110404} (\bibinfo {year}
  {2006})}\BibitemShut {NoStop}%
\bibitem [{\citenamefont {Levin}\ and\ \citenamefont
  {Wen}(2006)}]{levin2006detecting}%
  \BibitemOpen
  \bibfield  {author} {\bibinfo {author} {\bibfnamefont {M.}~\bibnamefont
  {Levin}}\ and\ \bibinfo {author} {\bibfnamefont {X.-G.}\ \bibnamefont
  {Wen}},\ }\href {\doibase 10.1103/PhysRevLett.96.110405} {\bibfield
  {journal} {\bibinfo  {journal} {Physical review letters}\ }\textbf {\bibinfo
  {volume} {96}},\ \bibinfo {pages} {110405} (\bibinfo {year}
  {2006})}\BibitemShut {NoStop}%
\bibitem [{\citenamefont {Yu}\ \emph {et~al.}(2009)\citenamefont {Yu},
  \citenamefont {Kwok}, \citenamefont {Cao},\ and\ \citenamefont
  {Gu}}]{yu2009fidelity}%
  \BibitemOpen
  \bibfield  {author} {\bibinfo {author} {\bibfnamefont {W.-C.}\ \bibnamefont
  {Yu}}, \bibinfo {author} {\bibfnamefont {H.-M.}\ \bibnamefont {Kwok}},
  \bibinfo {author} {\bibfnamefont {J.}~\bibnamefont {Cao}}, \ and\ \bibinfo
  {author} {\bibfnamefont {S.-J.}\ \bibnamefont {Gu}},\ }\href {\doibase
  https://doi.org/10.1103/PhysRevE.80.021108} {\bibfield  {journal} {\bibinfo
  {journal} {Physical Review E}\ }\textbf {\bibinfo {volume} {80}},\ \bibinfo
  {pages} {021108} (\bibinfo {year} {2009})}\BibitemShut {NoStop}%
\bibitem [{\citenamefont {Hansson}\ \emph {et~al.}(2004)\citenamefont
  {Hansson}, \citenamefont {Oganesyan},\ and\ \citenamefont
  {Sondhi}}]{hansson2004superconductors}%
  \BibitemOpen
  \bibfield  {author} {\bibinfo {author} {\bibfnamefont {T.}~\bibnamefont
  {Hansson}}, \bibinfo {author} {\bibfnamefont {V.}~\bibnamefont {Oganesyan}},
  \ and\ \bibinfo {author} {\bibfnamefont {S.~L.}\ \bibnamefont {Sondhi}},\
  }\href {\doibase https://doi.org/10.1016/j.aop.2004.05.006} {\bibfield
  {journal} {\bibinfo  {journal} {Annals of Physics}\ }\textbf {\bibinfo
  {volume} {313}},\ \bibinfo {pages} {497} (\bibinfo {year}
  {2004})}\BibitemShut {NoStop}%
\bibitem [{\citenamefont {Senthil}\ and\ \citenamefont
  {Fisher}(2000)}]{PhysRevB.62.7850}%
  \BibitemOpen
  \bibfield  {author} {\bibinfo {author} {\bibfnamefont {T.}~\bibnamefont
  {Senthil}}\ and\ \bibinfo {author} {\bibfnamefont {M.~P.~A.}\ \bibnamefont
  {Fisher}},\ }\href {\doibase 10.1103/PhysRevB.62.7850} {\bibfield  {journal}
  {\bibinfo  {journal} {Phys. Rev. B}\ }\textbf {\bibinfo {volume} {62}},\
  \bibinfo {pages} {7850} (\bibinfo {year} {2000})}\BibitemShut {NoStop}%
\bibitem [{\citenamefont {Bhattacharjee}(2011)}]{PhysRevB.84.104430}%
  \BibitemOpen
  \bibfield  {author} {\bibinfo {author} {\bibfnamefont {S.}~\bibnamefont
  {Bhattacharjee}},\ }\href {\doibase 10.1103/PhysRevB.84.104430} {\bibfield
  {journal} {\bibinfo  {journal} {Phys. Rev. B}\ }\textbf {\bibinfo {volume}
  {84}},\ \bibinfo {pages} {104430} (\bibinfo {year} {2011})}\BibitemShut
  {NoStop}%
\bibitem [{\citenamefont {Sachdev}(2018)}]{sachdev2018topological}%
  \BibitemOpen
  \bibfield  {author} {\bibinfo {author} {\bibfnamefont {S.}~\bibnamefont
  {Sachdev}},\ }\href {\doibase 10.1088/1361-6633/aae110} {\bibfield  {journal}
  {\bibinfo  {journal} {Reports on Progress in Physics}\ }\textbf {\bibinfo
  {volume} {82}},\ \bibinfo {pages} {014001} (\bibinfo {year}
  {2018})}\BibitemShut {NoStop}%
\bibitem [{\citenamefont {Hal\'asz}\ \emph {et~al.}(2014)\citenamefont
  {Hal\'asz}, \citenamefont {Chalker},\ and\ \citenamefont
  {Moessner}}]{PhysRevB.90.035145}%
  \BibitemOpen
  \bibfield  {author} {\bibinfo {author} {\bibfnamefont {G.~B.}\ \bibnamefont
  {Hal\'asz}}, \bibinfo {author} {\bibfnamefont {J.~T.}\ \bibnamefont
  {Chalker}}, \ and\ \bibinfo {author} {\bibfnamefont {R.}~\bibnamefont
  {Moessner}},\ }\href {\doibase 10.1103/PhysRevB.90.035145} {\bibfield
  {journal} {\bibinfo  {journal} {Phys. Rev. B}\ }\textbf {\bibinfo {volume}
  {90}},\ \bibinfo {pages} {035145} (\bibinfo {year} {2014})}\BibitemShut
  {NoStop}%
\bibitem [{\citenamefont {Savary}\ and\ \citenamefont
  {Balents}(2012)}]{PhysRevLett.108.037202}%
  \BibitemOpen
  \bibfield  {author} {\bibinfo {author} {\bibfnamefont {L.}~\bibnamefont
  {Savary}}\ and\ \bibinfo {author} {\bibfnamefont {L.}~\bibnamefont
  {Balents}},\ }\href {\doibase 10.1103/PhysRevLett.108.037202} {\bibfield
  {journal} {\bibinfo  {journal} {Phys. Rev. Lett.}\ }\textbf {\bibinfo
  {volume} {108}},\ \bibinfo {pages} {037202} (\bibinfo {year}
  {2012})}\BibitemShut {NoStop}%
\bibitem [{\citenamefont {Lannert}\ \emph {et~al.}(2001)\citenamefont
  {Lannert}, \citenamefont {Fisher},\ and\ \citenamefont
  {Senthil}}]{lannert2001quantum}%
  \BibitemOpen
  \bibfield  {author} {\bibinfo {author} {\bibfnamefont {C.}~\bibnamefont
  {Lannert}}, \bibinfo {author} {\bibfnamefont {M.~P.}\ \bibnamefont {Fisher}},
  \ and\ \bibinfo {author} {\bibfnamefont {T.}~\bibnamefont {Senthil}},\ }\href
  {\doibase https://doi.org/10.1103/PhysRevB.63.134510} {\bibfield  {journal}
  {\bibinfo  {journal} {Physical Review B}\ }\textbf {\bibinfo {volume} {63}},\
  \bibinfo {pages} {134510} (\bibinfo {year} {2001})}\BibitemShut {NoStop}%
\bibitem [{\citenamefont {Prakash}\ and\ \citenamefont
  {Bhattacharjee}(2020)}]{ap_sb}%
  \BibitemOpen
  \bibfield  {author} {\bibinfo {author} {\bibfnamefont {A.}~\bibnamefont
  {Prakash}}\ and\ \bibinfo {author} {\bibfnamefont {S.}~\bibnamefont
  {Bhattacharjee}},\ }\href@noop {} {\bibfield  {journal} {\bibinfo  {journal}
  {(unpublished)}\ } (\bibinfo {year} {2020})}\BibitemShut {NoStop}%
\bibitem [{\citenamefont {Kou}\ \emph {et~al.}(2009)\citenamefont {Kou},
  \citenamefont {Yu},\ and\ \citenamefont {Wen}}]{PhysRevB.80.125101}%
  \BibitemOpen
  \bibfield  {author} {\bibinfo {author} {\bibfnamefont {S.-P.}\ \bibnamefont
  {Kou}}, \bibinfo {author} {\bibfnamefont {J.}~\bibnamefont {Yu}}, \ and\
  \bibinfo {author} {\bibfnamefont {X.-G.}\ \bibnamefont {Wen}},\ }\href
  {\doibase 10.1103/PhysRevB.80.125101} {\bibfield  {journal} {\bibinfo
  {journal} {Phys. Rev. B}\ }\textbf {\bibinfo {volume} {80}},\ \bibinfo
  {pages} {125101} (\bibinfo {year} {2009})}\BibitemShut {NoStop}%
\bibitem [{\citenamefont {Dunne}(1999)}]{dunne1999aspects}%
  \BibitemOpen
  \bibfield  {author} {\bibinfo {author} {\bibfnamefont {G.~V.}\ \bibnamefont
  {Dunne}},\ }in\ \href@noop {} {\emph {\bibinfo {booktitle} {Aspects
  topologiques de la physique en basse dimension. Topological aspects of low
  dimensional systems}}}\ (\bibinfo  {publisher} {Springer},\ \bibinfo {year}
  {1999})\ pp.\ \bibinfo {pages} {177--263}\BibitemShut {NoStop}%
\bibitem [{\citenamefont {Geraedts}\ and\ \citenamefont
  {Motrunich}(2012)}]{PhysRevB.85.045114}%
  \BibitemOpen
  \bibfield  {author} {\bibinfo {author} {\bibfnamefont {S.~D.}\ \bibnamefont
  {Geraedts}}\ and\ \bibinfo {author} {\bibfnamefont {O.~I.}\ \bibnamefont
  {Motrunich}},\ }\href {\doibase 10.1103/PhysRevB.85.045114} {\bibfield
  {journal} {\bibinfo  {journal} {Phys. Rev. B}\ }\textbf {\bibinfo {volume}
  {85}},\ \bibinfo {pages} {045114} (\bibinfo {year} {2012})}\BibitemShut
  {NoStop}%
\bibitem [{\citenamefont {Peskin}(1978)}]{peskin1978mandelstam}%
  \BibitemOpen
  \bibfield  {author} {\bibinfo {author} {\bibfnamefont {M.~E.}\ \bibnamefont
  {Peskin}},\ }\href {\doibase https://doi.org/10.1016/0003-4916(78)90252-X}
  {\bibfield  {journal} {\bibinfo  {journal} {Annals of Physics}\ }\textbf
  {\bibinfo {volume} {113}},\ \bibinfo {pages} {122} (\bibinfo {year}
  {1978})}\BibitemShut {NoStop}%
\bibitem [{\citenamefont {Dasgupta}\ and\ \citenamefont
  {Halperin}(1981)}]{PhysRevLett.47.1556}%
  \BibitemOpen
  \bibfield  {author} {\bibinfo {author} {\bibfnamefont {C.}~\bibnamefont
  {Dasgupta}}\ and\ \bibinfo {author} {\bibfnamefont {B.~I.}\ \bibnamefont
  {Halperin}},\ }\href {\doibase 10.1103/PhysRevLett.47.1556} {\bibfield
  {journal} {\bibinfo  {journal} {Phys. Rev. Lett.}\ }\textbf {\bibinfo
  {volume} {47}},\ \bibinfo {pages} {1556} (\bibinfo {year}
  {1981})}\BibitemShut {NoStop}%
\bibitem [{\citenamefont {Motrunich}\ and\ \citenamefont
  {Vishwanath}(2004)}]{PhysRevB.70.075104}%
  \BibitemOpen
  \bibfield  {author} {\bibinfo {author} {\bibfnamefont {O.~I.}\ \bibnamefont
  {Motrunich}}\ and\ \bibinfo {author} {\bibfnamefont {A.}~\bibnamefont
  {Vishwanath}},\ }\href {\doibase 10.1103/PhysRevB.70.075104} {\bibfield
  {journal} {\bibinfo  {journal} {Phys. Rev. B}\ }\textbf {\bibinfo {volume}
  {70}},\ \bibinfo {pages} {075104} (\bibinfo {year} {2004})}\BibitemShut
  {NoStop}%
\bibitem [{\citenamefont {Levin}\ and\ \citenamefont
  {Senthil}(2004)}]{PhysRevB.70.220403}%
  \BibitemOpen
  \bibfield  {author} {\bibinfo {author} {\bibfnamefont {M.}~\bibnamefont
  {Levin}}\ and\ \bibinfo {author} {\bibfnamefont {T.}~\bibnamefont
  {Senthil}},\ }\href {\doibase 10.1103/PhysRevB.70.220403} {\bibfield
  {journal} {\bibinfo  {journal} {Phys. Rev. B}\ }\textbf {\bibinfo {volume}
  {70}},\ \bibinfo {pages} {220403} (\bibinfo {year} {2004})}\BibitemShut
  {NoStop}%
\bibitem [{\citenamefont {Kleinert}\ \emph {et~al.}(2002)\citenamefont
  {Kleinert}, \citenamefont {Nogueira},\ and\ \citenamefont
  {Sudb\o{}}}]{PhysRevLett.88.232001}%
  \BibitemOpen
  \bibfield  {author} {\bibinfo {author} {\bibfnamefont {H.}~\bibnamefont
  {Kleinert}}, \bibinfo {author} {\bibfnamefont {F.~S.}\ \bibnamefont
  {Nogueira}}, \ and\ \bibinfo {author} {\bibfnamefont {A.}~\bibnamefont
  {Sudb\o{}}},\ }\href {\doibase 10.1103/PhysRevLett.88.232001} {\bibfield
  {journal} {\bibinfo  {journal} {Phys. Rev. Lett.}\ }\textbf {\bibinfo
  {volume} {88}},\ \bibinfo {pages} {232001} (\bibinfo {year}
  {2002})}\BibitemShut {NoStop}%
\bibitem [{\citenamefont {Yu}\ \emph {et~al.}(2008)\citenamefont {Yu},
  \citenamefont {Kou},\ and\ \citenamefont {Wen}}]{yu2008topological}%
  \BibitemOpen
  \bibfield  {author} {\bibinfo {author} {\bibfnamefont {J.}~\bibnamefont
  {Yu}}, \bibinfo {author} {\bibfnamefont {S.-P.}\ \bibnamefont {Kou}}, \ and\
  \bibinfo {author} {\bibfnamefont {X.-G.}\ \bibnamefont {Wen}},\ }\href
  {\doibase https://doi.org/10.1209/0295-5075/84/17004} {\bibfield  {journal}
  {\bibinfo  {journal} {EPL (Europhysics Letters)}\ }\textbf {\bibinfo {volume}
  {84}},\ \bibinfo {pages} {17004} (\bibinfo {year} {2008})}\BibitemShut
  {NoStop}%
\bibitem [{\citenamefont {Tupitsyn}\ \emph {et~al.}(2010)\citenamefont
  {Tupitsyn}, \citenamefont {Kitaev}, \citenamefont {Prokof'ev},\ and\
  \citenamefont {Stamp}}]{PhysRevB.82.085114}%
  \BibitemOpen
  \bibfield  {author} {\bibinfo {author} {\bibfnamefont {I.~S.}\ \bibnamefont
  {Tupitsyn}}, \bibinfo {author} {\bibfnamefont {A.}~\bibnamefont {Kitaev}},
  \bibinfo {author} {\bibfnamefont {N.~V.}\ \bibnamefont {Prokof'ev}}, \ and\
  \bibinfo {author} {\bibfnamefont {P.~C.~E.}\ \bibnamefont {Stamp}},\ }\href
  {\doibase 10.1103/PhysRevB.82.085114} {\bibfield  {journal} {\bibinfo
  {journal} {Phys. Rev. B}\ }\textbf {\bibinfo {volume} {82}},\ \bibinfo
  {pages} {085114} (\bibinfo {year} {2010})}\BibitemShut {NoStop}%
\bibitem [{\citenamefont {Wu}\ \emph {et~al.}(2012)\citenamefont {Wu},
  \citenamefont {Deng},\ and\ \citenamefont {Prokof'ev}}]{PhysRevB.85.195104}%
  \BibitemOpen
  \bibfield  {author} {\bibinfo {author} {\bibfnamefont {F.}~\bibnamefont
  {Wu}}, \bibinfo {author} {\bibfnamefont {Y.}~\bibnamefont {Deng}}, \ and\
  \bibinfo {author} {\bibfnamefont {N.}~\bibnamefont {Prokof'ev}},\ }\href
  {\doibase 10.1103/PhysRevB.85.195104} {\bibfield  {journal} {\bibinfo
  {journal} {Phys. Rev. B}\ }\textbf {\bibinfo {volume} {85}},\ \bibinfo
  {pages} {195104} (\bibinfo {year} {2012})}\BibitemShut {NoStop}%
\bibitem [{\citenamefont {Vidal}\ \emph {et~al.}(2009)\citenamefont {Vidal},
  \citenamefont {Dusuel},\ and\ \citenamefont {Schmidt}}]{PhysRevB.79.033109}%
  \BibitemOpen
  \bibfield  {author} {\bibinfo {author} {\bibfnamefont {J.}~\bibnamefont
  {Vidal}}, \bibinfo {author} {\bibfnamefont {S.}~\bibnamefont {Dusuel}}, \
  and\ \bibinfo {author} {\bibfnamefont {K.~P.}\ \bibnamefont {Schmidt}},\
  }\href {\doibase 10.1103/PhysRevB.79.033109} {\bibfield  {journal} {\bibinfo
  {journal} {Phys. Rev. B}\ }\textbf {\bibinfo {volume} {79}},\ \bibinfo
  {pages} {033109} (\bibinfo {year} {2009})}\BibitemShut {NoStop}%
\bibitem [{\citenamefont {Dusuel}\ \emph {et~al.}(2011)\citenamefont {Dusuel},
  \citenamefont {Kamfor}, \citenamefont {Or\'us}, \citenamefont {Schmidt},\
  and\ \citenamefont {Vidal}}]{PhysRevLett.106.107203}%
  \BibitemOpen
  \bibfield  {author} {\bibinfo {author} {\bibfnamefont {S.}~\bibnamefont
  {Dusuel}}, \bibinfo {author} {\bibfnamefont {M.}~\bibnamefont {Kamfor}},
  \bibinfo {author} {\bibfnamefont {R.}~\bibnamefont {Or\'us}}, \bibinfo
  {author} {\bibfnamefont {K.~P.}\ \bibnamefont {Schmidt}}, \ and\ \bibinfo
  {author} {\bibfnamefont {J.}~\bibnamefont {Vidal}},\ }\href {\doibase
  10.1103/PhysRevLett.106.107203} {\bibfield  {journal} {\bibinfo  {journal}
  {Phys. Rev. Lett.}\ }\textbf {\bibinfo {volume} {106}},\ \bibinfo {pages}
  {107203} (\bibinfo {year} {2011})}\BibitemShut {NoStop}%
\bibitem [{\citenamefont {Jongeward}\ \emph {et~al.}(1980)\citenamefont
  {Jongeward}, \citenamefont {Stack},\ and\ \citenamefont
  {Jayaprakash}}]{PhysRevD.21.3360}%
  \BibitemOpen
  \bibfield  {author} {\bibinfo {author} {\bibfnamefont {G.~A.}\ \bibnamefont
  {Jongeward}}, \bibinfo {author} {\bibfnamefont {J.~D.}\ \bibnamefont
  {Stack}}, \ and\ \bibinfo {author} {\bibfnamefont {C.}~\bibnamefont
  {Jayaprakash}},\ }\href {\doibase 10.1103/PhysRevD.21.3360} {\bibfield
  {journal} {\bibinfo  {journal} {Phys. Rev. D}\ }\textbf {\bibinfo {volume}
  {21}},\ \bibinfo {pages} {3360} (\bibinfo {year} {1980})}\BibitemShut
  {NoStop}%
\bibitem [{\citenamefont {Genovese}\ \emph {et~al.}(2003)\citenamefont
  {Genovese}, \citenamefont {Gliozzi}, \citenamefont {Rago},\ and\
  \citenamefont {Torrero}}]{genovese2003phase}%
  \BibitemOpen
  \bibfield  {author} {\bibinfo {author} {\bibfnamefont {L.}~\bibnamefont
  {Genovese}}, \bibinfo {author} {\bibfnamefont {F.}~\bibnamefont {Gliozzi}},
  \bibinfo {author} {\bibfnamefont {A.}~\bibnamefont {Rago}}, \ and\ \bibinfo
  {author} {\bibfnamefont {C.}~\bibnamefont {Torrero}},\ }\href {\doibase
  https://doi.org/10.1016/S0920-5632(03)01713-4} {\bibfield  {journal}
  {\bibinfo  {journal} {Nuclear Physics B-Proceedings Supplements}\ }\textbf
  {\bibinfo {volume} {119}},\ \bibinfo {pages} {894} (\bibinfo {year}
  {2003})}\BibitemShut {NoStop}%
\bibitem [{\citenamefont {Campostrini}\ \emph {et~al.}(2001)\citenamefont
  {Campostrini}, \citenamefont {Hasenbusch}, \citenamefont {Pelissetto},
  \citenamefont {Rossi},\ and\ \citenamefont {Vicari}}]{PhysRevB.63.214503}%
  \BibitemOpen
  \bibfield  {author} {\bibinfo {author} {\bibfnamefont {M.}~\bibnamefont
  {Campostrini}}, \bibinfo {author} {\bibfnamefont {M.}~\bibnamefont
  {Hasenbusch}}, \bibinfo {author} {\bibfnamefont {A.}~\bibnamefont
  {Pelissetto}}, \bibinfo {author} {\bibfnamefont {P.}~\bibnamefont {Rossi}}, \
  and\ \bibinfo {author} {\bibfnamefont {E.}~\bibnamefont {Vicari}},\ }\href
  {\doibase 10.1103/PhysRevB.63.214503} {\bibfield  {journal} {\bibinfo
  {journal} {Phys. Rev. B}\ }\textbf {\bibinfo {volume} {63}},\ \bibinfo
  {pages} {214503} (\bibinfo {year} {2001})}\BibitemShut {NoStop}%
\bibitem [{\citenamefont {Zhu}\ and\ \citenamefont
  {Zhang}(2019)}]{PhysRevLett.122.176401}%
  \BibitemOpen
  \bibfield  {author} {\bibinfo {author} {\bibfnamefont {G.-Y.}\ \bibnamefont
  {Zhu}}\ and\ \bibinfo {author} {\bibfnamefont {G.-M.}\ \bibnamefont
  {Zhang}},\ }\href {\doibase 10.1103/PhysRevLett.122.176401} {\bibfield
  {journal} {\bibinfo  {journal} {Phys. Rev. Lett.}\ }\textbf {\bibinfo
  {volume} {122}},\ \bibinfo {pages} {176401} (\bibinfo {year}
  {2019})}\BibitemShut {NoStop}%
\bibitem [{\citenamefont {Ziatdinov}\ \emph {et~al.}(2016)\citenamefont
  {Ziatdinov}, \citenamefont {Banerjee}, \citenamefont {Maksov}, \citenamefont
  {Berlijn}, \citenamefont {Zhou}, \citenamefont {Cao}, \citenamefont {Yan},
  \citenamefont {Bridges}, \citenamefont {Mandrus}, \citenamefont {Nagler}
  \emph {et~al.}}]{ziatdinov2016atomic}%
  \BibitemOpen
  \bibfield  {author} {\bibinfo {author} {\bibfnamefont {M.}~\bibnamefont
  {Ziatdinov}}, \bibinfo {author} {\bibfnamefont {A.}~\bibnamefont {Banerjee}},
  \bibinfo {author} {\bibfnamefont {A.}~\bibnamefont {Maksov}}, \bibinfo
  {author} {\bibfnamefont {T.}~\bibnamefont {Berlijn}}, \bibinfo {author}
  {\bibfnamefont {W.}~\bibnamefont {Zhou}}, \bibinfo {author} {\bibfnamefont
  {H.}~\bibnamefont {Cao}}, \bibinfo {author} {\bibfnamefont {J.-Q.}\
  \bibnamefont {Yan}}, \bibinfo {author} {\bibfnamefont {C.~A.}\ \bibnamefont
  {Bridges}}, \bibinfo {author} {\bibfnamefont {D.}~\bibnamefont {Mandrus}},
  \bibinfo {author} {\bibfnamefont {S.~E.}\ \bibnamefont {Nagler}},  \emph
  {et~al.},\ }\href {\doibase https://doi.org/10.1038/ncomms13774} {\bibfield
  {journal} {\bibinfo  {journal} {Nature communications}\ }\textbf {\bibinfo
  {volume} {7}},\ \bibinfo {pages} {13774} (\bibinfo {year}
  {2016})}\BibitemShut {NoStop}%
\end{thebibliography}%
\end{document}